\pdfoutput=1

\documentclass[11pt,twoside,a4paper,cmspaper,final,collab]{cms-tdr}
\def\svnVersion{cf095ca-D}\def\svnDate{2019/10/15}\def\cmsCernNoTag{CERN-EP-2019-180}\def\cmsCernDate{\today}\def\cmsMessage{"Published in the European Physical Journal C as \href{http://dx.doi.org/10.1140/epjc/s10052-019-7493-x}{\doi{10.1140/epjc/s10052-019-7493-x}.}"}

\begin{document}\cmsNoteHeader{SUS-19-005}

\hyphenation{had-ron-i-za-tion}
\hyphenation{cal-or-i-me-ter}
\hyphenation{de-vices}
\newlength\cmsFigWidth
\newlength\cmsTabSkip\setlength{\cmsTabSkip}{1ex}
\ifthenelse{\boolean{cms@external}}{\setlength\cmsFigWidth{0.40\textwidth}}{\setlength\cmsFigWidth{0.49\textwidth}}
\ifthenelse{\boolean{cms@external}}{\providecommand{\cmsLeft}{upper\xspace}}{\providecommand{\cmsLeft}{left\xspace}}
\ifthenelse{\boolean{cms@external}}{\providecommand{\cmsRight}{lower\xspace}}{\providecommand{\cmsRight}{right\xspace}}
\ifthenelse{\boolean{cms@external}}{\providecommand{\cmsLLeft}{Upper\xspace}}{\providecommand{\cmsLLeft}{Left\xspace}}
\ifthenelse{\boolean{cms@external}}{\providecommand{\cmsRRight}{Lower\xspace}}{\providecommand{\cmsRRight}{Right\xspace}}
\ifthenelse{\boolean{cms@external}}{\providecommand{\cmsMiddle}{Lower\xspace}}{\providecommand{\cmsMiddle}{Right\xspace}}
\providecommand{\cmsTable}[1]{\resizebox{\textwidth}{!}{#1}}
\ifthenelse{\boolean{cms@external}}{\providecommand{\cmsTableAlt}[1]{#1}}{\providecommand{\cmsTableAlt}[1]{\resizebox{\textwidth}{!}{#1}}}
\ifthenelse{\boolean{cms@external}}{\providecommand{\cmsTableTwo}[1]{\resizebox{\columnwidth}{!}{#1}}}{\providecommand{\cmsTableTwo}[1]{#1}}
\DeclareGraphicsExtensions{.png,.PNG,.pdf,.PDF}
\cmsNoteHeader{SUS-19-005}

\title{Searches for physics beyond the standard model with the \mttwo variable in hadronic final states with and without disappearing tracks in proton-proton collisions at $\sqrt{s}=13\TeV$}
\titlerunning{Searches for physics beyond the standard model with the \mttwo variable at 13\TeV}

\date{\today}

\newcommand{\Lint}{137\fbinv}
\newcommand{\mttwo}{\ensuremath{M_{\mathrm{T2}}}\xspace}
\newcommand{\Met}{\ptmiss}
\newcommand{\vMet}{\ptvecmiss}
\newcommand{\dpmin}{\ensuremath{\Delta\phi_{\text{min}}}\xspace}
\newcommand{\njets}{\ensuremath{N_{\mathrm{j}}}\xspace}
\newcommand{\nbtags}{\ensuremath{N_{\PQb}}\xspace}
\newcommand{\Mt}{\ensuremath{M_{\mathrm{T}}}\xspace}
\newcommand{\wjets}{\ensuremath{\PW\text{+jets}}\xspace}
\newcommand{\zjets}{\ensuremath{\PZ\text{+jets}}\xspace}
\newcommand{\vjets}{\ensuremath{\PV\text{+jets}}\xspace}
\newcommand{\vzw}{\ensuremath{(\PV=\PW,\PZ)}\xspace}
\newcommand{\ttjets}{\ensuremath{\ttbar \text{+jets}}\xspace}
\newcommand{\ttV}{\ensuremath{\ttbar\PV}\xspace}
\newcommand{\qqbarpr}{\ensuremath{\qqbar'}\xspace}
\newcommand{\znunu}{\ensuremath{\PZ\to\PGn\PAGn}\xspace}
\newcommand{\ptj}{\ensuremath{\pt^{\text{jet}}}\xspace}
\newcommand{\fshort}{\ensuremath{f_{\text{short}}}\xspace}
\newcommand{\gluino}{\PSg}
\newcommand{\lsp}{\PSGczDo}
\newcommand{\astop}{\PASQt}
\newcommand{\chizz}{\PSGczDt}
\newcommand{\chargino}{\PSGcpmDo}
\newcommand{\charginoplus}{\PSGcpDo}
\newcommand{\charginominus}{\PSGcmDo}
\providecommand{\PSGcmpDo}{\HepParticle{\PSGc}{1}{\mp}\Xspace}
\newcommand{\PQtL}{\ensuremath{\PQt_{\text{L}}}\Xspace}
\newcommand{\PQbL}{\ensuremath{\PQb_{\text{L}}}\Xspace}
\newcommand{\ttbb}{\ensuremath{\ttbar \PQb\PAQb}\xspace}
\newcommand{\ttjj}{\ensuremath{\ttbar \text{jj}}\xspace}
\newcommand{\lqs}{\ensuremath{\mathrm{LQ}_{\mathrm{S}}}\xspace}
\newcommand{\lqv}{\ensuremath{\mathrm{LQ}_{\mathrm{V}}}\xspace}
\providecommand{\NA}{\text{---}\xspace}

\abstract{
Two related searches for phenomena beyond the standard model (BSM) are performed using events with hadronic jets and significant transverse momentum imbalance. The results are based on a sample of proton-proton collisions at a center-of-mass energy of $13\TeV$, collected by the CMS experiment at the LHC in 2016--2018 and corresponding to an integrated luminosity of \Lint. The first search is inclusive, based on signal regions defined by the hadronic energy in the event, the jet multiplicity, the number of jets identified as originating from bottom quarks, and the value of the kinematic variable \mttwo for events with at least two jets. For events with exactly one jet, the transverse momentum of the jet is used instead. The second search looks in addition for disappearing tracks produced by BSM long-lived charged particles that decay within the volume of the tracking detector. No excess event yield is observed above the predicted standard model background. This is used to constrain a range of BSM models that predict the following: the pair production of gluinos and squarks in the context of supersymmetry models conserving $R$-parity, with or without intermediate long-lived charginos produced in the decay chain; the resonant production of a colored scalar state decaying to a massive Dirac fermion and a quark; or the pair production of scalar and vector leptoquarks each decaying to a neutrino and a top, bottom, or light-flavor quark. In most of the cases, the results obtained are the most stringent constraints to date.
}

\hypersetup{
pdfauthor={CMS Collaboration},
pdftitle={Searches for physics beyond the standard model with the MT2 variable in hadronic final states with and without disappearing tracks in proton-proton collisions at sqrt(s)=13 TeV},
pdfsubject={CMS},
pdfkeywords={CMS, physics, supersymmetry}}

\maketitle

\section{Introduction}
\label{sec:intro}

{\tolerance=1200
We present results of two related searches for physics beyond the standard model (BSM) in events with jets and significant transverse momentum imbalance.
These are based on a data set of proton-proton ($\Pp\Pp$) collisions at $\sqrt{s} = 13\TeV$, collected with the CMS detector at the CERN LHC in 2016--2018,
and corresponding to an integrated luminosity of \Lint.
\par}

The first is an inclusive search that exploits the transverse momentum imbalance as inferred from the kinematic variable \mttwo~\cite{MT2variable},
defined in Section~\ref{sec:evtsel:evtsel}, in events with at least two hadronic jets, or the transverse momentum (\pt) of the jet in events with just one jet.
Similar searches were previously conducted by both the ATLAS~\cite{ATLASat2016:manyjets,ATLASat2016:sbottom,ATLASat2016:stop,ATLASat2016:manyb,ATLASat2016:monojet,ATLASat2016:hadronic}
and CMS ~\cite{RA2b_2016,MT2_2016,AlphaT_2016,MT2_2016LQ,Razor_2016} Collaborations.
Our analysis builds on the work presented in Refs.~\cite{MT2_2016,MT2_2016LQ}, using improved methods to estimate the background from standard model (SM) processes,
in particular the multijet background arising from instrumental effects.
Event counts in bins of the number of jets (\njets), the number of jets identified as originating from the fragmentation of a bottom quark (\PQb-tagged jets, \nbtags), the scalar \pt sum of all selected jets (\HT), and
the \mttwo variable or the \pt of the single jet,
are compared against estimates of the background from SM processes, as derived from dedicated data control samples.

The second search aims at extending the sensitivity of the inclusive search for scenarios where the mass spectrum of potential new particles is compressed.
In such scenarios, some theoretical models~\cite{AMSB_1,AMSB_2} predict the existence of
long-lived charged particles that can be identified as
disappearing tracks,
when they decay within the volume of the tracking detector
and their charged decay products are below the \pt detection threshold. Such signatures are rare in the SM and are often dominated by instrumental effects.
The presence of disappearing tracks is exploited in order to suppress the background from SM processes, and to enhance the sensitivity towards these scenarios.
Similar analyses were previously conducted by both the ATLAS~\cite{ATLAS8TeVDisTracks,ATLASat2016:distracks}
and CMS~\cite{EXO8TeV_distracks,HSCP_2,EXO_distracks,HSCP_1} Collaborations.
We use events with at least two jets, and the \mttwo variable to further suppress the background from SM processes.
Event counts in bins of \njets, \HT, disappearing track length, and disappearing track \pt are compared against
estimates of the background from SM processes derived from dedicated data control samples.

The results are interpreted
in the context of
simplified models~\cite{sms1,sms2,sms3,sms4,sms5} of $R$-parity~\cite{lqtheory_11} conserving supersymmetry
(SUSY)~\cite{Ramond:1971gb,Golfand:1971iw,Neveu:1971rx,Volkov:1972jx,Wess:1973kz,Wess:1974tw,Fayet:1974pd,Nilles:1983ge}
where gluinos and squarks are pair-produced and the lightest SUSY particle is a neutralino.

The results of the inclusive \mttwo search are also
interpreted in the context of
a BSM scenario
where a colored scalar state $\phi$ is resonantly produced through coupling to quarks, and decays to
an invisible massive Dirac fermion $\psi$ and an SM quark. This is referred to as the mono-$\phi$ model.
It has been recently proposed as an explanation of an excess in data in regions with low jet multiplicities, identified
in the context of a reinterpretation~\cite{monojetexcess1,monojetexcess2} of the results of
the previous inclusive \mttwo search~\cite{MT2_2016}
as well as of other similar searches by both the ATLAS~\cite{ATLASat2016:monojet,ATLASat2016:hadronic} and CMS~\cite{RA2b_2016,CMSmonojet} Collaborations.

{\tolerance=1200
Finally, the inclusive \mttwo search is interpreted using
models of leptoquark (LQ) pair production, similarly to Ref.~\cite{MT2_2016LQ}.
Leptoquarks are hypothetical particles with quantum numbers of both quarks and leptons~\cite{lqtheory}.
The spin of an LQ state is either 0 (scalar LQ or $\mathrm{LQ_{S}}$) or 1 (vector LQ or $\mathrm{LQ_{V}}$).
Leptoquarks appear in BSM theories such as grand unified theories~\cite{lqtheory,lqtheory_2,lqtheory_3,lqtheory_4},
technicolor models~\cite{lqtheory_5,lqtheory_6,lqtheory_7,lqtheory_8},
compositeness scenarios~\cite{lqtheory_9,lqtheory_10},
and $R$-parity violating SUSY~\cite{Ramond:1971gb,Golfand:1971iw,Neveu:1971rx,Volkov:1972jx,Wess:1973kz,Wess:1974tw,Fayet:1974pd,Nilles:1983ge,rpvsusy_9},
 and have been suggested as an explanation of the anomalies observed in flavor physics~\cite{lqfp1,lqfp2,lqfp3,lqfp4,lqfp5,lqmodel,lqtoolbox} by the BaBar~\cite{babar1,babar2},
Belle~\cite{belle1,belle2,belle3,belle4,bellenew},
and LHCb~\cite{lhcb1,lhcb2,lhcb3,lhcb4,lhcb5,lhcbnew} Collaborations.
The best fit model of Refs.~\cite{lqmodel,lqtoolbox} predicts an $\mathrm{LQ_{V}}$ with a mass of $\mathcal{O}\left(\TeVns{}\right)$
decaying with 50\% branching fraction to either a top quark and a neutrino ($\PQt\PGn$) or a bottom quark and a \PGt lepton ($\PQb\PGt$),
which would be expected to be visible at the LHC.
The final states and kinematic variables resulting from the pair production of $\mathrm{LQ_{S}}$, each decaying to a quark and a neutrino,
are the same as those considered in searches for squark pair production in $R$-parity conserving SUSY, assuming that the squark decays directly to
a quark and a massless neutralino~\cite{lqfpmt2,MT2_2016LQ}. The decay products of $\mathrm{LQ_{V}}$ are also found to have similar kinematic properties~\cite{lqfpmt2,MT2_2016LQ}.
Therefore, as the search presented in this paper is already optimized for squark pair production, it is also sensitive to LQ pair production.
The LQ production with decays to a quark and a neutrino has been constrained using LHC data by both the ATLAS~\cite{atlas1,atlas2,atlas3}
and CMS~\cite{cms1,cms2,cms3,MT2_2016LQ,cms4,cms5} Collaborations,
either by reinterpreting the existing squark searches, or considering scenarios with mixed branching fractions where an LQ also decays to a quark and a charged lepton.
The same signatures have been previously covered at the Fermilab Tevatron by the CDF (e.g., in Refs.~\cite{cdf2,cdf3,cdf1}) and \DZERO (e.g., in Refs.~\cite{d01,d03,d02}) Collaborations.
Constraints have been placed by direct searches for single LQ production performed at HERA by the H1~\cite{h11} and ZEUS~\cite{zeus1} Collaborations.
Finally, searches for LQs decaying to $\PQb\PGt$ have been performed by the ATLAS~\cite{atlas11}, CMS~\cite{cms11,cms22}, CDF~\cite{cdf22,cdf11}, and \DZERO~\cite{d011} Collaborations.
\par}

After a brief description of the CMS detector in Section~\ref{sec:cms}, the event selection and categorization as well as details of the Monte Carlo (MC) simulation
are presented in Section~\ref{sec:evtsel}.
Section~\ref{sec:bkgds} describes the SM background estimation.
Results and their interpretations are presented in Sections~\ref{sec:results}~and~\ref{sec:interpretation}, respectively.
Finally, a summary is provided in Section~\ref{sec:conclusions}.

\section{The CMS detector}
\label{sec:cms}

The central feature of the CMS apparatus is a superconducting solenoid of 6\unit{m} internal diameter
providing a magnetic field of 3.8\unit{T}. Within the solenoid volume are a silicon pixel and strip tracker, a
lead tungstate crystal electromagnetic calorimeter, and a brass and scintillator hadron calorimeter, each composed of
a barrel and two endcap sections. Forward calorimeters extend the pseudorapidity ($\eta$) coverage provided by the
 barrel and endcap detectors. Muons are measured in gas-ionization detectors embedded in the steel flux-return yoke outside
the solenoid. The first level of the CMS trigger system, composed of custom hardware processors, uses information from the
calorimeters and muon detectors to select the most interesting events in a fixed time interval of less than 4\mus. The high-level
 trigger processor farm further decreases the event rate from around 100\unit{kHz} to about 1\unit{kHz}, before data storage.
A more detailed description of the CMS detector and trigger system, together with a definition of the coordinate system used and the relevant
kinematic variables, can be found in Refs.~\cite{Chatrchyan:2008zzk,CMStrigger}.
The pixel tracker was upgraded before the start of the data taking period in 2017,
providing one additional layer of measurements compared to the older tracker~\cite{phase1trackerTDR}.

\section{Event selection and Monte Carlo simulation}
\label{sec:evtsel}

\subsection{Event selection}
\label{sec:evtsel:evtsel}

Events
are processed using the particle-flow (PF)
algorithm~\cite{pflowNEW},
which aims at reconstructing and identifying each individual particle in an event,
with an optimal combination of information from the elements of the CMS detector.
The particles reconstructed with this algorithm are hereafter referred to as PF candidates.
The physics objects and the event preselection are similar to those described in Ref.~\cite{MT2_2016}; they are summarized in Table~\ref{tab:evtsel}, and described in detail below.
We select events with at least one reconstructed vertex and at least one jet, and veto events with an isolated lepton (\Pe or \PGm) or an isolated charged PF candidate.
The isolated charged PF candidate veto is designed to provide additional rejection against events with electrons and muons, as well as to reject hadronic \PGt decays.

\begin{table*}[tbhp!]
  \setlength{\extrarowheight}{.45em}
  \centering
    \topcaption{\label{tab:evtsel}
      Summary of the trigger requirements and the kinematic offline event preselection requirements
      on the reconstructed physics objects, for both the inclusive \mttwo search and the search for disappearing tracks.
      Here
$R$ is the distance parameter of the anti-\kt algorithm.
      To veto leptons and tracks, the transverse mass \Mt is determined using the veto object and the \vMet.
      The variable $\pt^{\text{sum}}$ is a measure of object isolation and it denotes the \pt sum of all additional PF candidates in a cone around the lepton or the track.
The size of the cone is listed in the table in units of $\Delta R \equiv \sqrt{\smash[b]{(\Delta \phi)^2
  + (\Delta \eta)^2}}$. The lepton (track) \pt is denoted as $\pt^{\text{lep}}$ ($\pt^{\text{track}}$).
      Further details of the lepton selection are given in Refs.~\cite{MT2_2016,MT2at13TeV}.
      The $i$th-highest \pt jet is denoted as $j_\mathrm{i}$.}
    \cmsTableAlt{
    \begin{tabular}{ l  l }
      \multirow{10}{*}{Trigger} & 2016: \\
      & $\Met>120\GeV$ and $\mht>120\GeV$, or \\
      & $\HT>300\GeV$ and $\Met>110\GeV$, or \\
      & $\HT>900\GeV$, or jet $\pt>450\GeV$ \\ [\cmsTabSkip]
      & 2017 and 2018: \\
      & $\Met>120\GeV$ and $\mht>120\GeV$, or \\
      & $\HT>60\GeV$ and $\Met>120\GeV$ and $\mht>120\GeV$, or \\
      & $\HT>500\GeV$ and $\Met>100\GeV$ and $\mht>100\GeV$, or \\
      & $\HT>800\GeV$ and $\Met>75\GeV$ and $\mht>75\GeV$, or \\
      & $\HT>1050\GeV$, or jet $\pt>500\GeV$ \\  [\cmsTabSkip]
      Jet selection & $R=0.4$, $\pt>30\GeV$, $\abs{\eta}<2.4$ \\ [\cmsTabSkip]
      \PQb-tagged jet selection & $\pt>20\GeV$, $\abs{\eta}<2.4$ and \PQb tag \\  [\cmsTabSkip]
      $\HT$ & $\HT>250\GeV$ \\ [\cmsTabSkip]
      \multirow{3}{*}{$\Met$} & $\Met>250\GeV$ for $\HT<1200\GeV$ or $\njets=1$,
      else $\Met>30\GeV$\\
      & $\dpmin = \Delta\phi\left(\vMet, j_{\mathrm{1,2,3,4}}\right)>0.3$ \\
      & $\abs{\vMet-\htvecmiss}/\Met<0.5$ \\ [\cmsTabSkip]
      \multirow{5}{*}{\mttwo (if $\njets \geq 2$)} & Inclusive \mttwo search: \\
      & $\mttwo>200\GeV$ for $\HT<1500\GeV$, else $\mttwo>400\GeV$ \\ [\cmsTabSkip]
      & Disappearing tracks search: \\
      & $\mttwo>200\GeV$ \\ [\cmsTabSkip]
      \multirow{2}{*}{$\pt^{\text{sum}} $ cone (isolation)} & Veto \Pe or \Pgm: $\Delta R= \min(0.2, \max(10\GeV/\pt^{\text{lep}},0.05)) $ \\
      & Veto track: $\Delta R=0.3$ \\ [\cmsTabSkip]
      Veto electron & $\pt>10\GeV$, $\abs{\eta}<2.4$, $\pt^{\text{sum}} < 0.1 \, \pt^{\text{lep}}$  \\
      Veto electron track & $\pt>5\GeV$, $\abs{\eta}<2.4$, $\Mt<100\GeV$, $\pt^{\text{sum}}
      < 0.2 \, \pt^{\text{lep}}$ \\ [\cmsTabSkip]
      Veto muon & $\pt>10\GeV$, $\abs{\eta}<2.4$, $\pt^{\text{sum}} < 0.2 \,\pt^{\text{lep}}$ \\
      Veto muon track & $\pt>5\GeV$, $\abs{\eta}<2.4$, $\Mt<100\GeV$, $\pt^{\text{sum}}
      < 0.2 \, \pt^{\text{lep}}$ \\ [\cmsTabSkip]
      Veto track & $\pt>10\GeV$, $\abs{\eta}<2.4$, $\Mt<100\GeV$,
      $\pt^{\text{sum}} < 0.1 \, \pt^{\text{track}}$ \\
    \end{tabular}
    }
\end{table*}

Jets are formed by clustering PF candidates using the
anti-\kt algorithm~\cite{Cacciari:2008gp, Cacciari:2011ma}
and are
corrected for contributions from event pileup~\cite{cacciari-2008-659}
and the effects of nonuniform detector response~\cite{JER8TeV,JER}.
Only jets passing the selection criteria in Table~\ref{tab:evtsel} are used for counting and for the determination of kinematic variables.
In particular, we consider jets with $\pt>30\GeV$ and $\abs{\eta}<2.4$, unless otherwise stated.
Jets that contain the decay of a bottom-flavored hadron
are identified using a deep neural network algorithm~\cite{btagDeepCSV}
with a working point chosen such that the efficiency to identify a bottom
quark jet is in the range 55--70\% for jet \pt
between 20 and 400\GeV.
The misidentification rate is approximately 1--2\% for light-flavor or gluon jets, and 10--15\% for charm jets.
We count \PQb-tagged jets with $\pt>20\GeV$ and $\abs{\eta}<2.4$.
The minimum \pt threshold used for counting \PQb-tagged jets is lowered to 20\GeV instead of 30, as used for \njets,
in order to maximize the sensitivity towards BSM scenarios with bottom quarks.

The negative of the vector \pt sum of all selected jets is denoted
by \htvecmiss, while the missing transverse momentum \vMet is defined as the negative of the vector \pt sum of
all reconstructed PF candidates. Their magnitudes are referred to as \mht and \Met, respectively.
The \vMet is further adjusted to reflect the jet energy corrections~\cite{JER8TeV,JER}.
Events with possible contributions from beam halo processes or anomalous noise in the calorimeter
are rejected using dedicated filters~\cite{Chatrchyan:2011tn,MetFiltersRun2}.
For events with at least two jets, we start with the pair having the largest dijet invariant mass
and iteratively cluster all selected jets using an
algorithm that minimizes the Lund
distance measure~\cite{LundDistRef1,Phythia64} until two stable pseudo-jets are obtained.
The resulting pseudo-jets together with the \vMet are used to calculate the kinematic variable \mttwo~\cite{MT2variable} as:
\begin{linenomath}
\begin{equation}
\mttwo = \min_{\vMet{}^{ \mathrm{X}(1)} + \vMet{}^{ \mathrm{X}(2)} = \vMet}
  \left[ \max \left( \Mt^{(1)} , \Mt^{(2)} \right) \right],
\label{eq.MT2.definition}
\end{equation}
\end{linenomath}
where $\vMet{}^{ \mathrm{X}(i)}$ ($i=1,2$) are trial vectors
obtained by decomposing \vMet,  and
$\Mt^{(i)}$ are the transverse masses~\cite{MT}
obtained by pairing either of the trial vectors with one of the two pseudo-jets.
The minimization is performed over all trial momenta satisfying the \vMet constraint.
The background from multijet events (discussed in Section~\ref{sec:bkgds}) is characterized by small values of
\mttwo, while processes with significant genuine \vMet yield larger values of \mttwo.
More detailed discussions of the \mttwo variable properties are given in Refs.~\cite{MT2at13TeV,MT2at8TeV,MT2at7TeV}.

In both the inclusive \mttwo search and the search for disappearing tracks,
collision events are selected using triggers with requirements on \HT, \Met, \mht, and jet \pt.
The combined trigger efficiency, as measured in an orthogonal data sample of events with an isolated electron,
is found to be $>$97\% across the full kinematic range of the search.
To suppress background from multijet production, we require $\mttwo > 200\GeV$ in events with $\njets \geq 2$.
In the inclusive \mttwo search, this \mttwo threshold is increased to 400\GeV for events with $\HT > 1500\GeV$ to maintain
multijet processes as a subdominant background in all search regions.
In events with $\njets = 1$, where \mttwo is not defined, we require $\ptj>250\GeV$ and $\Met>250\GeV$.
As a protection against jet mismeasurement, we require the minimum difference in the azimuthal angle between the \vMet  vector and
the direction of each of the four \pt-leading jets, \dpmin, to be greater than 0.3 radians, and the magnitude of the difference between \vMet  and \htvecmiss  to be less than half of \Met.
For the determination of \dpmin, we consider jets with $\abs{\eta}<4.7$.
If fewer than four such jets are found, all are considered in the \dpmin calculation.

In the search for disappearing tracks, events are selected requiring in addition the presence of at least one disappearing track.
These are defined as well-reconstructed isolated tracks with no measurement points
in at least two of the outermost layers of the tracker and no associated energy deposits in the
calorimeter.  These tracks are
predominantly not considered as candidates by the PF algorithm;
as a result
they are not included in the calculation of \vMet.

\subsection{Event categorization}
\label{sec:evtsel:evtcat}

\subsubsection{Inclusive \texorpdfstring{\mttwo}{MT2} search}
\label{sec:evtsel:evtcat:mt2}

Events containing at least two jets are categorized by the values of \njets, \nbtags, and \HT.
Each category is referred to as a
topological region.
Signal regions are defined by further dividing topological regions into bins of \mttwo.
Events with only one jet are selected if the jet \pt is at least 250\GeV,
and are classified according to the \pt of this jet and whether the event contains a \PQb-tagged jet.
The 282 search regions are summarized in Tables~\ref{tab:yieldsmonojet}--\ref{tab:yieldsUHh} in Appendix~\ref{app:srs:mt2}.
We also define \emph{super signal regions}, covering a subset of the kinematic space of the full analysis with simpler inclusive selection criteria.
The super signal regions can be used to obtain approximate interpretations of our
result, as discussed in Section~\ref{sec:results}, where these regions are defined.

\subsubsection{Search for disappearing tracks}
\label{sec:evtsel:evtcat:distracks}

In the following, the selected disappearing tracks are called
short tracks (STs).
We also define short track candidates (STCs) as disappearing tracks that are required to satisfy relaxed selection criteria on the track quality and isolation compared to an ST,
but not the tight ones required for STs.
Both STs and STCs are required to have no measurement points
in at least two of the outermost layers of the tracker and no associated energy deposits in the
calorimeter.

We select events with at least one ST and at least two jets, and we categorize them by the values of \njets and \HT.
Disappearing tracks are categorized according to their length and \pt,
in order to maximize the sensitivity to a range of lifetimes of potential BSM long-lived charged particles,
and to distinguish tracks reconstructed with different precision.
Two bins of \pt are defined as:
\begin{itemize}
\item $15<\pt<50\GeV$,
\item $\pt>50\GeV$.
\end{itemize}
Additionally, four track length categories are defined, depending on the number of layers of the tracking detector with a measurement:
\begin{itemize}
\item pixel tracks (P), having at least three layers with a measurement in the pixel tracking detector, and none in the strip tracking detector,
\item medium length tracks (M), having less than seven layers with a measurement, and at least one outside of the pixel tracking detector,
\item long tracks (L), having at least seven layers with a measurement.
\end{itemize}
For 2017--2018 data, we further split the P tracks into two categories:
\begin{itemize}
\item pixel tracks having three layers with a measurement (P3),
\item pixel tracks having at least four layers with a measurement (P4).
\end{itemize}
For long (L) tracks, no categorization in bins of \pt is applied.

The full track selection requirements for both STs and STCs are listed in Table~\ref{tab:stsel} of Appendix~\ref{app:stsel}, together with the track length categories they belong to.
For signal STs, the track reconstruction and selection efficiency ranges from 50 to 65\%, depending on the track length and the data taking period.

The 68 search regions (28 used for the categorization of the 2016 data set, and 40 for the 2017--2018 data set)
are summarized in Tables~\ref{tab:sr1_distracks}--\ref{tab:sr2_distracks} in Appendix~\ref{app:srs:distr}.

\subsection{Monte Carlo simulation}
\label{sec:evtsel:mc}

{\tolerance=2000
The MC simulation is used to design the search,
to help estimate SM backgrounds, and to evaluate the sensitivity
to simplified models of BSM physics.

The main background samples (\zjets, \wjets, \ttjets, and multijet),
as well as BSM signal samples,
are generated at leading order (LO) precision with the \MGvATNLO~2 (2.2.2, or 2.4.2) generator~\cite{mg5amcnlo}.
Up to four, three, or two additional partons are considered in the matrix element calculations for the generation of the \vjets~\vzw,
\ttjets, and signal samples, respectively.
Other background processes are also considered: \ttV samples with up to two additional partons in the matrix element calculations are generated at LO precision
with the \MGvATNLO~2 generator,
while single top quark samples are generated at next-to-leading order (NLO) precision with the \MGvATNLO~2 or \POWHEG~($\mathrm{v}1.0$, or $\mathrm{v}2.0$)~\cite{Nason:2004rx,Frixione:2007vw,Alioli:2009je,Alioli:2010xd,Re:2010bp} generators.
Finally, contributions from rarer processes such as diboson, triboson, and four top quark production, are also considered and found to be negligible.
The expected yields of all samples are normalized using the most precise available cross section calculations,
typically corresponding to
NLO or next-to-NLO (NNLO) accuracy~\cite{Gavin:2010az, Gavin:2012sy, Czakon:2011xx, mg5amcnlo, Alioli:2009je, Re:2010bp, Borschensky:2014cia}.

The detector response of SM samples and 2016 signal samples containing long-lived objects is modeled with the \GEANTfour~\cite{geant4} program,
while the CMS fast simulation framework~\cite{fastsim,fastsimrun2} is used for other signal samples, and
uncertainties are derived to account for the potential mismodeling of the event kinematics.

For all simulated samples, generators are interfaced with \PYTHIA~8.2~(8.205, 8.212, 8.226, or 8.230)~\cite{pythia82} for fragmentation and parton showering.
For samples simulated at LO (NLO) precision, the MLM~\cite{mlm} (FxFx~\cite{fxfx}) prescription is used to match partons from the matrix element calculation to those from the parton showers.
The CUETP8M1~\cite{CUETP8M1bib} \PYTHIA~8.2 tune is used for the 2016 SM background and signal samples.
For 2017 and 2018, the CP5 and CP2 tunes~\cite{CP15bib} are used for the SM background and signal samples, respectively.
The NNPDF2.3LO (NNPDF2.3NLO)~\cite{nnpdf23bib} parton distribution functions (PDFs) are used to generate the 2016 LO (NLO) samples,
while the NNPDF3.1LO (NNPDF3.1NNLO)~\cite{nnpdf3bib} PDFs are used for the 2017 and 2018 samples.
\par}

The output of the detector simulation is processed using the same chain of reconstruction algorithms as for collision data.

To improve on the \MGvATNLO modeling of the
multiplicity of additional jets from initial-state radiation (ISR) in the 2016 sample,
\MGvATNLO~\ttbar MC events are weighted based on the
number of ISR jets ($N_\mathrm{j}^\mathrm{ISR}$) so as to make the jet
multiplicity agree with data.
The same reweighting procedure is applied to BSM MC events.
The weighting factors are obtained from a control region enriched in \ttbar,
defined as 
events with two leptons and exactly two \PQb-tagged jets, and vary between 0.92 for $N_\mathrm{j}^\mathrm{ISR}=1$ and 0.51 for $N_\mathrm{j}^\mathrm{ISR}\geq 6$.
We take one half of the deviation from unity as the systematic uncertainty in these reweighting factors, to cover 
for the experimental uncertainties in their derivation and 
for differences between \ttbar and BSM production.
Owing to a better tuning of the MC generators, this reweighting procedure is not necessary for 2017 and 2018 \MGvATNLO~\ttbar MC samples,
while it is still applied to BSM MC events.

{\tolerance=1000
To improve the modeling of the flavor of additional jets, the simulation of \ttbar and \ttV events is
corrected to account for the measured
ratio of \ttbb/\ttjj cross sections reported in Ref.~\cite{ttbbweight}.
Specifically, simulated \ttbar and \ttV events with two \PQb quarks not originating from top quark decay are weighted
to account for the CMS measurement of the ratio of cross sections $\sigma(\ttbb)/\sigma(\ttjj)$, which was
found to be a factor of $1.7 \pm 0.5$ larger than the MC prediction~\cite{ttbbweight}.
\par}

\section{Background estimation}
\label{sec:bkgds}

\subsection{Inclusive \texorpdfstring{\mttwo}{MT2} search}
\label{sec:mt2bkgds}

The backgrounds in jets-plus-\Met final states arise from three categories of SM processes.

\begin{itemize}
\item The lost-lepton (LL) background: events with a lepton from a \PW\ boson decay where the lepton is either out of acceptance, not reconstructed, not identified, or not isolated.
  This background originates mostly from  \wjets and \ttjets events, with smaller contributions from more rare processes, such as diboson or \ttV production.
\item The irreducible background: \zjets events, where the \PZ\ boson decays to neutrinos.  This background is the most difficult to distinguish from the final states arising from potential signals.
  It is a major background in nearly all search regions, its
  importance decreasing with
increasing \nbtags.
\item The instrumental background: mostly multijet events with no genuine \Met.
  These events enter a search region due to either significant jet momentum mismeasurements or sources of anomalous noise.
  This is a subdominant background compared to others, after events are selected, as described in Section~\ref{sec:evtsel:evtsel}.
\end{itemize}

The backgrounds are estimated from data control regions.
In the presence of BSM physics, these control regions could be affected by signal contamination.
Although the expected signal contamination is typically negligible,
its potential impact is accounted for in the interpretation of the results, as further described in Section~\ref{sec:interpretation}.

\subsubsection{Estimation of the background from events with leptonic \PW boson decays}
\label{sec:bkgds:ll}

The LL background is estimated from control regions with exactly one lepton candidate (\Pe or \PGm) selected using
the same triggers
and preselection criteria used for the signal regions,
with the exception of the lepton veto, which is inverted.
The transverse mass \Mt determined using the lepton candidate and the \vMet is required to satisfy $\Mt<100\GeV$,
in order to suppress the potential signal contamination of the control regions. 
Selected events are binned according to the same criteria as the search regions.
The background in each signal bin,
$N^{\mathrm{SR}}_{\mathrm{LL}}$, is obtained
by scaling the number of events in the control region,
$N^{\mathrm{CR}}_{1\ell}$,
using transfer factors $R^{0\ell/1\ell}_{\mathrm{MC}}$, as detailed below:
\begin{itemize}
\item For events with $\njets=1$:
\ifthenelse{\boolean{cms@external}}{
\begin{linenomath}
\begin{multline}
\label{eq:ll1j}
  N^{\mathrm{SR}}_{\mathrm{LL}} \left(\ptj,\nbtags\right) \\=
  N^{\mathrm{CR}}_{1\ell} \left(\ptj,\nbtags\right) R^{0\ell/1\ell}_{\mathrm{MC}} \left(\ptj,\nbtags\right).
\end{multline}
\end{linenomath}
}{
\begin{linenomath}
\begin{equation}
\label{eq:ll1j}
  N^{\mathrm{SR}}_{\mathrm{LL}} \left(\ptj,\nbtags\right) = N^{\mathrm{CR}}_{1\ell} \left(\ptj,\nbtags\right) \, R^{0\ell/1\ell}_{\mathrm{MC}} \left(\ptj,\nbtags\right).
\end{equation}
\end{linenomath}
}
\item For events with $\njets\geq2$:
\ifthenelse{\boolean{cms@external}}{
\begin{linenomath}
\begin{multline}
\label{eq:ll2j}
  N^{\mathrm{SR}}_{\mathrm{LL}} \left(\Omega,\mttwo\right) = N^{\mathrm{CR}}_{1\ell} \left(\Omega,\mttwo\right) \\
  \times R^{0\ell/1\ell}_{\mathrm{MC}} \left(\Omega,\mttwo\right) \, k_{\mathrm{LL}} \left(\mttwo|\Omega\right),
\end{multline}
\end{linenomath}
}{
\begin{linenomath}
\begin{equation}
\label{eq:ll2j}
  N^{\mathrm{SR}}_{\mathrm{LL}} \left(\Omega,\mttwo\right) = N^{\mathrm{CR}}_{1\ell} \left(\Omega,\mttwo\right) \, R^{0\ell/1\ell}_{\mathrm{MC}} \left(\Omega,\mttwo\right) \, k_{\mathrm{LL}} \left(\mttwo|\Omega\right),
\end{equation}
\end{linenomath}
}
where:
\ifthenelse{\boolean{cms@external}}{
\begin{linenomath}
\begin{equation}
\label{eq:omegall}
        \Omega \equiv \\
        \left(\HT,\njets,\nbtags\right).
\end{equation}
\end{linenomath}
}{
\begin{linenomath}
\begin{equation}
\label{eq:omegall}
        \Omega \equiv \left(\HT,\njets,\nbtags\right).
\end{equation}
\end{linenomath}
}
\end{itemize}
{\tolerance=1000
The single-lepton control regions have 1--2 times as many events as the corresponding signal regions.
The factor $R^{0\ell/1\ell}_{\mathrm{MC}}$ accounts for lepton acceptance and efficiency,
as well as the expected contribution from the decay of \PW\ bosons to hadrons through an intermediate \Pgt\ lepton.
It is obtained from MC simulation, and corrected for the measured
differences in the lepton efficiencies between data and simulation.
\par}

For events with $\njets\geq2$,
the factor $k_{\mathrm{LL}}$ is one, except at high \mttwo values,
where the single-lepton control sample
has insufficient data to allow $N^{\mathrm{CR}}_{1\ell}$ to be measured in each (\HT, \njets, \nbtags, \mttwo) bin.
In such cases, $N^{\mathrm{CR}}_{1\ell}$ is integrated over the remaining \mttwo bins of the same (\HT, \njets, \nbtags) region, and the distribution
in \mttwo across these bins is taken from simulation and applied through the factor $k_{\mathrm{LL}}$.

\begin{figure*}[htb]
  \centering
    \includegraphics[width=0.48\textwidth]{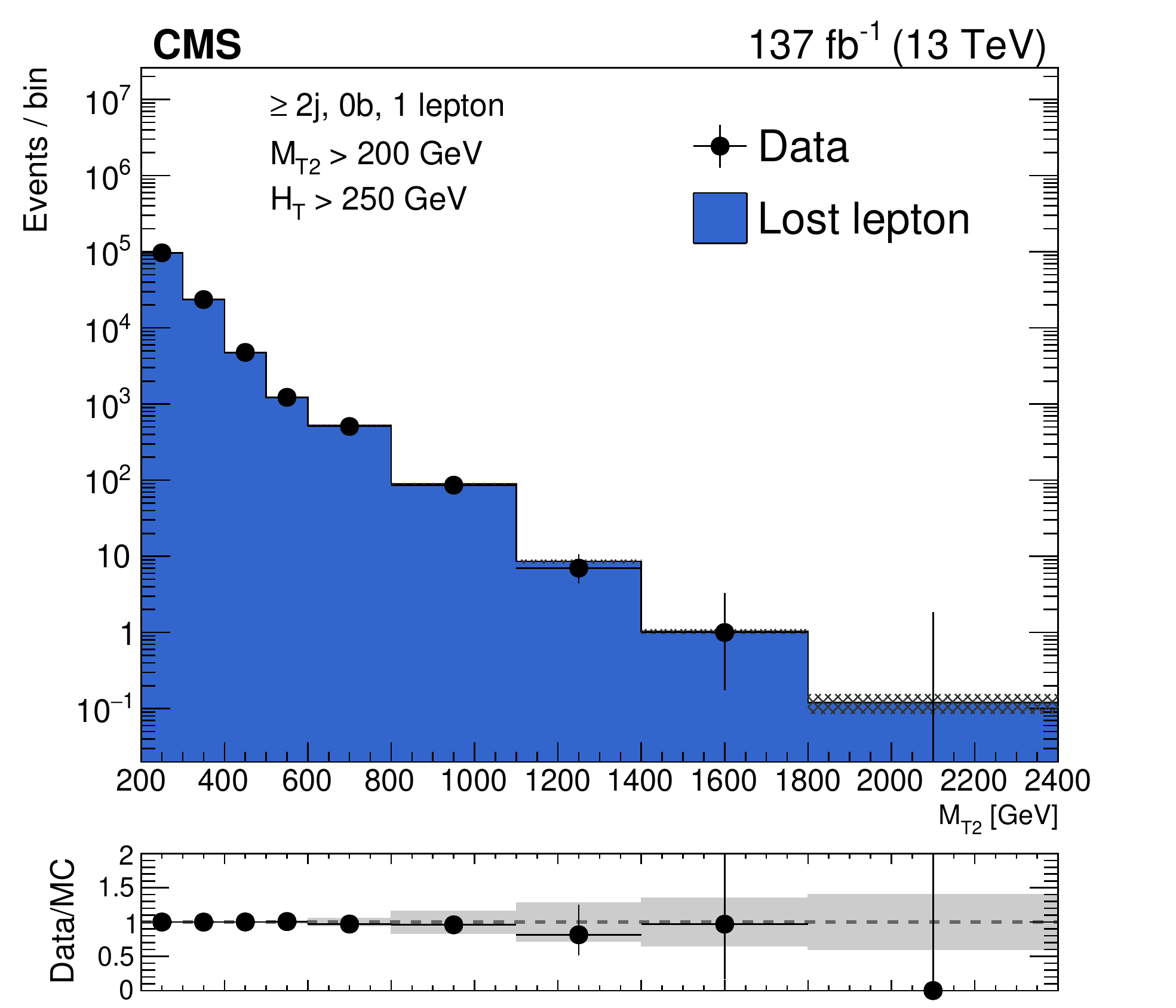}
    \includegraphics[width=0.48\textwidth]{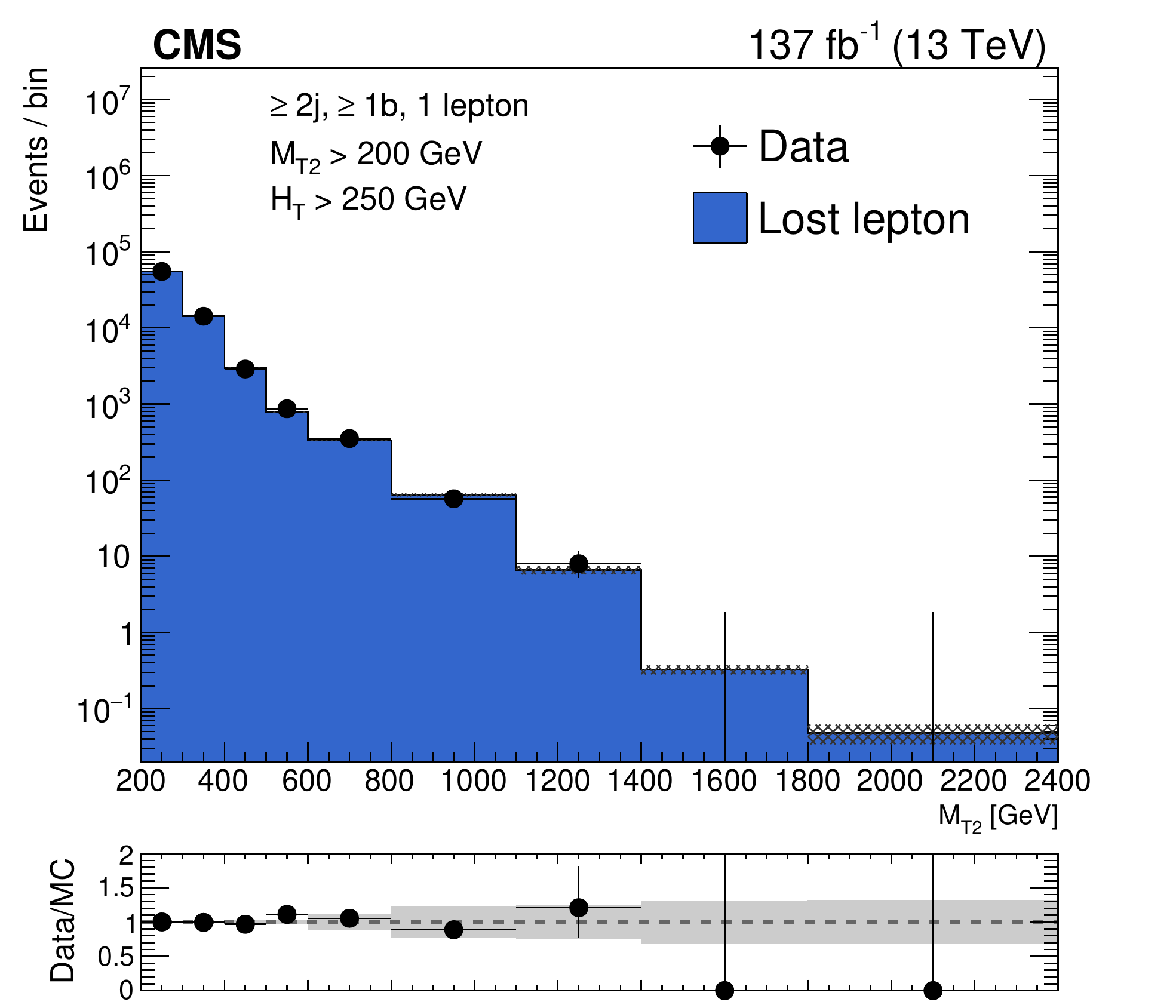}
    \caption{Distributions of the \mttwo variable in data and simulation for the single-lepton control region, after
      normalizing the simulation to data in bins of \HT, \njets, and \nbtags,
      for events with no \PQb-tagged jets (left), and events with at
      least one \PQb-tagged jet (right).
      The hatched bands on the top panels show the MC statistical uncertainty,
      while the solid gray bands in the ratio plots show the systematic uncertainty in the \mttwo shape.
      The bins have different widths, denoted by the horizontal bars.}
    \label{fig:ll_mt2}
\end{figure*}

The MC modeling of \mttwo is checked in data, in single-lepton events
with either $\nbtags=0$ or $\nbtags\geq1$,
as shown in the left and right panels of Fig.~\ref{fig:ll_mt2}, respectively.
The predicted distributions in the comparison are obtained by summing all the relevant regions, after normalizing MC event yields to data and distributing events among
the \mttwo bins according to the expectation from simulation.

Uncertainties arising from the limited size of the control samples and
from theoretical and experimental
considerations
are evaluated and propagated to the final estimate.
The dominant uncertainty in $R^{0\ell/1\ell}_{\mathrm{MC}}$ is due to the modeling
of the lepton efficiency (for electrons, muons, and hadronically decaying \PGt leptons) and jet energy scale (JES),
and is of order 15--20\%.
The uncertainty in the \mttwo extrapolation via $k_{\mathrm{LL}}$, which is as large as 40\%,
arises primarily from the JES,
the relative fractions of \wjets\ and \ttjets events,
and the choice of the renormalization ($\mu_{\mathrm{R}}$) and factorization ($\mu_{\mathrm{F}}$) scales used in the event generation.

The uncertainties in the LL background prediction are summarized in Table~\ref{tab:llepsyst} together with their typical size ranges across the search bins.
\begin{table}[tbhp]
    \centering
    \topcaption{\label{tab:llepsyst}
    Summary of systematic uncertainties in the lost-lepton background prediction, together with their typical size ranges across the search bins.
      }
    \begin{tabular}{ l  c }
      \hline
      Source & Range [\%] \\
      \hline
      Limited size of data control samples & 5--100\\
      Limited size of MC samples & 0--50\\
      $\Pe/\PGm$ efficiency & 0--10\\
      \PGt efficiency & 0--3\\
      \PQb tagging efficiency & 0--3\\
      Jet energy scale & 0--5\\
      $\Mt\left(\text{lepton},~\vMet\right)$ selection efficiency & 0--3\\
      \mttwo shape uncertainty (if $k_{\mathrm{LL}}\neq1$) & 0--40\\
      $\mu_{\mathrm{R}}$ and $\mu_{\mathrm{F}}$ variation & 0--5\\
      $\ttbb/\ttjj$ weight & 0--25\\
      \hline
    \end{tabular}
\end{table}

\subsubsection{Estimation of the background from \texorpdfstring{$\PZ(\PGn\PAGn)+\text{jets}$}{PZ nu nu-bar + jets}}
\label{sec:bkgds:zinv}

The $\PZ\to\PGn\PAGn$ background is estimated from a
$\PZ\to\ell^{+}\ell^{-}$ ($\ell = \Pe,\Pgm$) control sample
selected using dilepton triggers.
The trigger efficiency, measured from a sample of events in data with large \HT, is found to be greater than 97\% in the selected kinematic range.

The leptons in the control sample are required to
be of the same flavor and have opposite charge. The \pt of the leading and trailing leptons
must be at least 100 and 30\GeV, respectively.
Finally, the invariant mass of the lepton pair must be within 20\GeV of the \PZ\ boson mass.

After requiring that the \pt of the dilepton system is at least 200\GeV (corresponding to the $\mttwo>200\GeV$ requirement), the preselection requirements are applied based on kinematic variables
recalculated after removing the dilepton system from the event to replicate the $\PZ\to\PGn\PAGn$ kinematic properties.
For events with $\njets = 1$, one control region is defined for each bin of jet \pt.
For events with at least two jets, the selected events are binned in \HT, \njets, and \nbtags,
but not in \mttwo, to increase the dilepton event yield in each control region.

The contribution to each control region from flavor-symmetric processes, most importantly \ttbar production, is estimated
using different-flavor (DF) $\Pe\Pgm$ events obtained with the same selection criteria
as same-flavor (SF) $\Pe\Pe$ and  $\Pgm\Pgm$ events.
The background in each signal bin is then obtained using transfer factors.
\begin{itemize}
\item For events with $\njets=1$, according to:
\ifthenelse{\boolean{cms@external}}{
\begin{linenomath}
\begin{multline}
  \label{eq:zinv1j}
N^{\mathrm{SR}}_{\PZ\to\PGn\PAGn} \left(\ptj,\nbtags\right) = \Bigl[N^{\mathrm{CRSF}}_{\ell\ell} \left(\ptj,\nbtags\right)\\
 - N^{\mathrm{CRDF}}_{\ell\ell} \left(\ptj,\nbtags\right) \, R^{\mathrm{SF}/\mathrm{DF}} \Bigr] \\
  \times R^{\PZ\to\PGn\PAGn/Z\to\ell^{+}\ell^{-}}_{\mathrm{MC}} \left(\ptj,\nbtags\right).
\end{multline}
\end{linenomath}
}{
\begin{linenomath}
\begin{multline}
  \label{eq:zinv1j}
N^{\mathrm{SR}}_{\PZ\to\PGn\PAGn} \left(\ptj,\nbtags\right) = \Bigl[N^{\mathrm{CRSF}}_{\ell\ell} \left(\ptj,\nbtags\right)
 - N^{\mathrm{CRDF}}_{\ell\ell} \left(\ptj,\nbtags\right) \, R^{\mathrm{SF}/\mathrm{DF}} \Bigr] \\
  \times R^{\PZ\to\PGn\PAGn/Z\to\ell^{+}\ell^{-}}_{\mathrm{MC}} \left(\ptj,\nbtags\right).
\end{multline}
\end{linenomath}
}
\item For events with $\njets\geq2$, according to:
\ifthenelse{\boolean{cms@external}}{
\begin{linenomath}
\begin{multline}
  \label{eq:zinv2j}
N^{\mathrm{SR}}_{\PZ\to\PGn\PAGn} \left(\Omega,\mttwo\right) = \Bigl[N^{\mathrm{CRSF}}_{\ell\ell} \left(\Omega\right)\\
 - N^{\mathrm{CRDF}}_{\ell\ell} \left(\Omega\right) \, R^{\mathrm{SF}/\mathrm{DF}} \Bigr] \\
  \times R^{\PZ\to\PGn\PAGn/Z\to\ell^{+}\ell^{-}}_{\mathrm{MC}} \left(\Omega\right) \, k_{\PZ\to\PGn\PAGn}\left(\mttwo~|~\Omega\right),
\end{multline}
\end{linenomath}
}{
\begin{linenomath}
\begin{multline}
  \label{eq:zinv2j}
N^{\mathrm{SR}}_{\PZ\to\PGn\PAGn} \left(\Omega,\mttwo\right) = \Bigl[N^{\mathrm{CRSF}}_{\ell\ell} \left(\Omega\right)
 - N^{\mathrm{CRDF}}_{\ell\ell} \left(\Omega\right) \, R^{\mathrm{SF}/\mathrm{DF}} \Bigr] \\
 \times R^{\PZ\to\PGn\PAGn/Z\to\ell^{+}\ell^{-}}_{\mathrm{MC}} \left(\Omega\right) \, k_{\PZ\to\PGn\PAGn}\left(\mttwo~|~\Omega\right),
\end{multline}
\end{linenomath}
}
where $\Omega$ is defined in Eq.~(\ref{eq:omegall}).
\end{itemize}

\begin{figure*}[htb]
  \centering
    \includegraphics[width=0.53\textwidth]{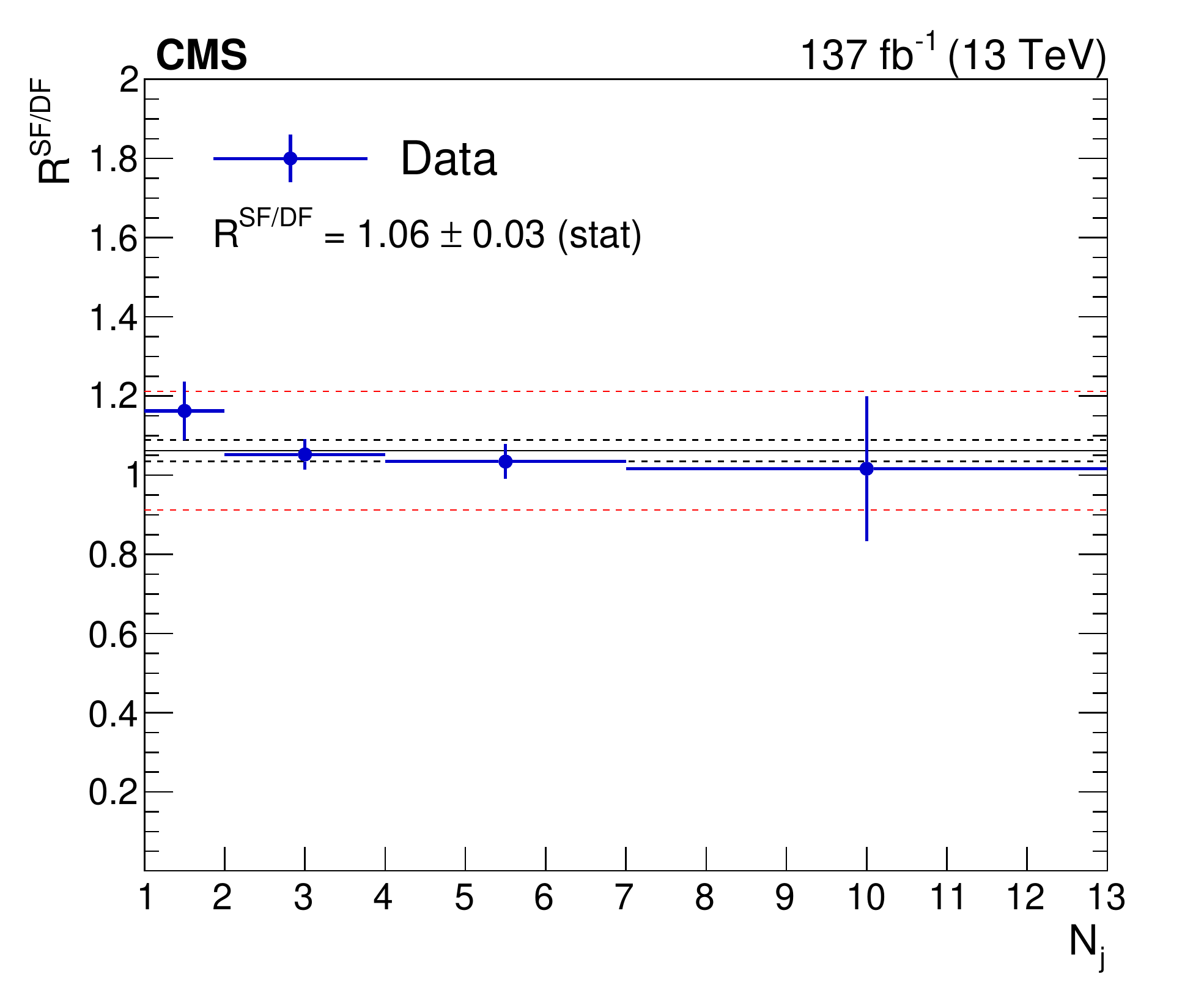}
    \includegraphics[width=0.44\textwidth]{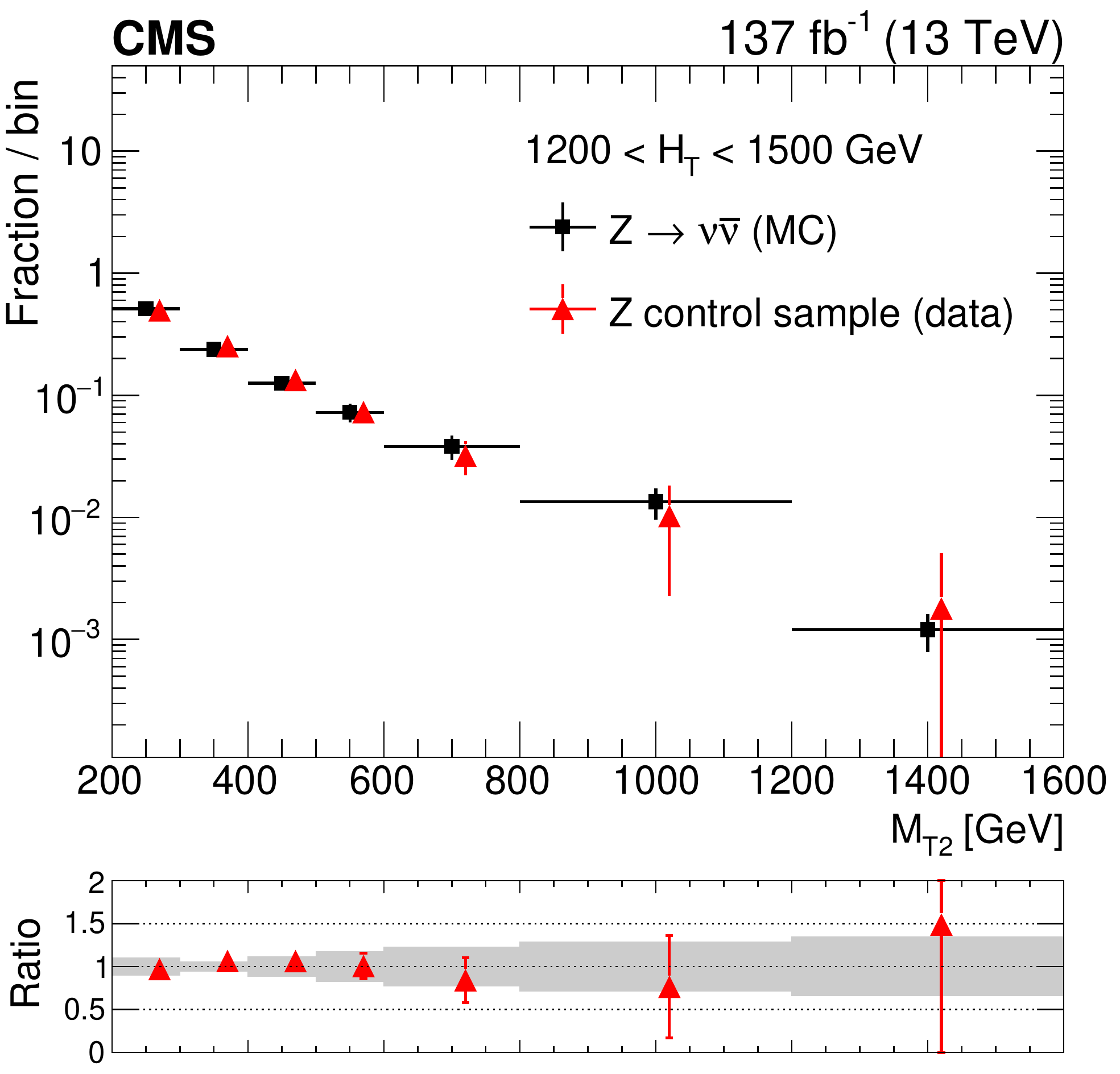}\\
    \caption{(Left) Ratio $R^{\mathrm{SF}/\mathrm{DF}}$ in data as a function of \njets.
      The solid black line enclosed by the red dashed lines corresponds to a value of $1.06\pm0.15$ that is observed to be stable with respect to event kinematic variables,
      while the two dashed black lines denote the statistical uncertainty in the $R^{\mathrm{SF}/\mathrm{DF}}$ value.
      (Right) The shape of the \mttwo distribution in $\PZ\to\PGn\PAGn$ simulation compared to the one obtained from the $\PZ\to\ell^{+}\ell^{-}$ data control sample,
      in a region with $1200<\HT<1500$\GeV and $\njets\geq2$, inclusive in \nbtags. The solid gray band on the ratio plot shows the systematic uncertainty in the \mttwo shape.
      The bins have different widths, denoted by the horizontal bars.
      }
    \label{fig:zinv}
\end{figure*}

{\tolerance=1000
Here $N^{\mathrm{CRSF}}_{\ell\ell}$ and $N^{\mathrm{CRDF}}_{\ell\ell}$ are the number of
SF and DF events in the control region, while
$R^{\PZ\to\PGn\PAGn/\PZ\to\ell^{+}\ell^{-}}_{\mathrm{MC}}$ and
$k_{\PZ\to\PGn\PAGn}$ are defined below.
The factor $R^{\mathrm{SF}/\mathrm{DF}}$ accounts for the difference in acceptance and efficiency
between SF and DF events.
It is determined as the ratio of the number of SF to DF events in a \ttbar enriched control sample,
obtained with the same selection criteria as the $\PZ\to\ell^{+}\ell^{-}$ sample,
but inverting the requirements on the \pt and the invariant mass of the lepton pair.
A measured value of $R^{\mathrm{SF}/\mathrm{DF}}=1.06\pm0.15$ is observed to be stable with respect to event kinematic variables, and is applied in all regions.
Figure~\ref{fig:zinv} (left) shows $R^{\mathrm{SF}/\mathrm{DF}}$ measured as a function of the number of jets.
\par}

{\tolerance=1200
For events with $\njets=1$,
an estimate of the $\PZ\to\PGn\PAGn$ background in each search bin is obtained from the corresponding dilepton control region
via the factor $R^{\PZ\to\PGn\PAGn/\PZ\to\ell^{+}\ell^{-}}_{\mathrm{MC}}$, which accounts for the acceptance and efficiency to select the dilepton pair
and the ratio of branching fractions for the $\PZ\to\ell^{+}\ell^{-}$ and $\PZ\to\PGn\PAGn$ decays.
For events with at least two jets, an estimate of the $\PZ\to\PGn\PAGn$ background is obtained analogously in each (\HT, \njets, \nbtags) region, integrated over \mttwo.
The factor $R^{\PZ\to\PGn\PAGn/\PZ\to\ell^{+}\ell^{-}}_{\mathrm{MC}}$ is obtained from simulation,
including corrections for the differences in the lepton efficiencies between data and simulation.
\par}

For events with $\njets\geq2$, the factor $k_{\PZ\to\PGn\PAGn}$ accounts for the distribution in bins of \mttwo of the estimated background
in each (\HT, \njets, \nbtags) region.
This distribution is constructed using \mttwo shape templates from dilepton data and \znunu simulation
in each (\HT, \njets, \nbtags) region.
The templates obtained from data are used at low values of \mttwo, where the amount of data is sufficient.
On the other hand, at high values of \mttwo we use the templates from simulation.

Studies with simulated samples have demonstrated that the shape
of the \mttwo distribution of the function $k_{\PZ\to\PGn\PAGn}$
is independent of \nbtags for a given \HT and \njets selection, and that
the shape is also independent of \njets for $\HT>1500\GeV$.
The dilepton control sample supports this observation.
Therefore, functions $k_{\PZ\to\PGn\PAGn}$
are obtained for each (\HT, \njets) region, integrated over \nbtags.
For $\HT>1500\GeV$, only one function $k_{\PZ\to\PGn\PAGn}$ is constructed, integrating also over \njets.

The MC modeling of the \mttwo variable is validated in data using control samples enriched in  $\PZ\to\ell^{+}\ell^{-}$ events, in each bin of \HT,
as shown in
the right panel of Fig.~\ref{fig:zinv} for events with $1200<\HT<1500\GeV$.

The largest uncertainty in the estimate of the invisible \PZ\ background in most regions results from the limited size of the dilepton control sample.
The dominant uncertainty of about 5\% in the ratio $R^{Z\to\PGn\PAGn/Z\to\ell^{+}\ell^{-}}_{\mathrm{MC}}$
reflects the uncertainty in the differences between the lepton efficiencies in data and simulation.
The uncertainty in the $k_{\PZ\to\PGn\PAGn}$ factor arises from data statistical uncertainty for bins at low values of \mttwo, where the function $k_{\PZ\to\PGn\PAGn}$ is obtained from data,
while for bins at high values of \mttwo, where the function $k_{\PZ\to\PGn\PAGn}$ is obtained from simulation,
it is due to the uncertainties in the JES and the choice of the $\mu_{\mathrm{R}}$ and $\mu_{\mathrm{F}}$.
These can result in effects as large as 40\%.

The uncertainties in the $\PZ\to\PGn\PAGn$ background prediction are summarized in Table~\ref{tab:invzsyst} together with their typical size ranges across the search bins.
\begin{table}[tbhp]
    \centering
    \topcaption{\label{tab:invzsyst}
    Summary of systematic uncertainties in the $\PZ\to\PGn\PAGn$ background prediction,
    together with their typical size ranges across the search bins.
      }
    \begin{tabular}{ l  c }
      \hline
      Source & Range [\%] \\
      \hline
      Limited size of data control samples & 5--100\\
      Limited size of MC samples & 0--50\\
      Lepton efficiency & 0--5\\
      Jet energy scale & 0--5\\
      Uncertainty in $R^{\mathrm{SF}/\mathrm{DF}}$ & 0--5\\
      \mttwo shape uncertainty (if $k_{\PZ\to\PGn\PAGn}\neq1$) & 0--40\\
      \hline
    \end{tabular}
\end{table}

\subsubsection{Estimation of the multijet background}
\label{sec:bkgds:qcd}

{\tolerance=1000
The background from SM events comprised uniquely of jets produced through the strong interaction
(multijet events)
is estimated from control regions in data
selected using triggers that require \HT to exceed thresholds
ranging from 125 (180) to 900 (1050)\GeV in 2016 (2017--2018) data samples.
In addition, events are required to have
at least two jets with $\pt > 10\GeV$.
\par}

The rebalance and smear (R\&S) method  used to estimate the multijet background
consists of two steps.
First, multijet data events are rebalanced by adjusting the \pt of the jets such that the resulting \Met is approximately zero.
This rebalancing is performed through a likelihood maximization, accounting for
the jet energy resolution~\cite{JER8TeV,JER}.
The output of the rebalancing step is an inclusive sample of  multijet events with approximately zero \Met
that are used as a seed for the second step,
the smearing.
In the smearing step, the \pt of the rebalanced jets is smeared
according to the jet response function,
in order to model the instrumental effects that lead to nonzero \Met.
The smearing step is repeated many times for each rebalanced event.
The output of each smearing step is an independent sample of events, which serves to populate
 the tails of kinematic distributions such as \Met and \mttwo,
and to obtain a more precise estimate of the multijet background than
would be possible using only simulation.

{\tolerance=1200
The method makes use of
jet response templates, \ie,
distributions of the ratio of reconstructed jet \pt to generator-level jet \pt.
The
templates are
derived from simulation in bins of jet \pt and $\eta$, separately for \PQb-tagged and non-\PQb-tagged jets.
Systematic uncertainties are assessed to cover for the modeling of the core and of the tails of the jet response templates.
\par}

Of all jets in the event, a jet qualifies for use in the R\&S procedure if it has $\pt>10\GeV$,
and if it is not identified as a jet from pileup~\cite{jetid13TeV} in the case that $\pt<100\GeV$.
All other jets are left unchanged but are still used in the calculation of \vMet and other jet-related quantities. An event with $n$ qualifying jets is rebalanced by varying the $\pt^\text{reb}$ of each jet, which is an estimate of the true jet \pt, to maximize the likelihood function
\begin{linenomath}
\begin{equation}
\label{eq:rebalance_likelihood}
L = \prod_{i=1}^n \text{P} \left( p_{\text{T},i}^{\text{reco}} | p_{\text{T},i}^{\text{reb}} \right) \, G\left( \frac{p_{\text{T},\text{reb,x}}^\text{miss}}{\sigma_\text{T}^{\text{soft}}}\right) \, G\left( \frac{p_{\text{T},\text{reb,y}}^\text{miss}}{\sigma_\text{T}^{\text{soft}}}\right),
\end{equation}
\end{linenomath}
where
\begin{linenomath}
\begin{equation}
\label{eq:G_def}
G(x) \equiv \re^{-x^2/2},
\end{equation}
\end{linenomath}
and
\begin{linenomath}
\begin{equation}
\label{eq:reb_met_def}
\vec{p}_{\text{T},\text{reb}}^{\text{miss}} \equiv \ptvecmiss - \sum_{i=1}^n \left( \vec{p}_{\text{T},i}^\text{reb} - \vec{p}_{\text{T},i}^\text{reco} \right).
\end{equation}
\end{linenomath}
The term $\text{P} ( p_{\text{T},i}^{\text{reco}} | p_{\text{T},i}^{\text{reb}} )$ in Eq.~(\ref{eq:rebalance_likelihood}) is the probability for a jet with \pt of
$p_{\text{T},i}^{\text{reb}}$ to be assigned a \pt of $p_{\text{T},i}^{\text{reco}}$ after reconstruction.
This probability is taken directly from the jet response templates.
The two $G(x)$ terms in Eq.~(\ref{eq:rebalance_likelihood}) enforce an approximate balancing condition.
The $\vec{p}_{\text{T},\text{reb}}^{\text{miss}}$ terms in Eq.~(\ref{eq:rebalance_likelihood}) represent the \vMet after rebalancing, and are obtained
by simply propagating the changes in jet \pt from rebalancing to \vMet.
For the balancing of the $x$ and $y$ components of the \vMet,
we use $\sigma_\text{T}^{\text{soft}}=20$\GeV, which is
approximately the width of the distributions of the $x$ and $y$ components of \vMet
in minimum bias events. This parameter represents the inherent missing energy due to low-\pt jets, unclustered energy, and jets from pileup that cannot be eliminated by rebalancing.
A systematic uncertainty is assessed to cover for the effects of the variation of $\sigma_\text{T}^\text{soft}$.

The rebalanced events are used as input to the smearing procedure,
where the \pt of each qualifying jet is rescaled by a random factor drawn
from the corresponding jet response template,
and all kinematic quantities are recalculated accordingly.

The background from multijet events is estimated by applying the signal region selection requirements to the above rebalanced and smeared sample,
except events are only used if $p_{\text{T},\text{reb}}^\text{miss}<100\GeV$
to remove potential contamination from electroweak sources.
This additional requirement is found to be fully efficient for multijet events, in simulation.
Hence, no correction is applied to the prediction.

Systematic uncertainties are summarized in Table~\ref{tab:qcdsyst} together with their typical size ranges across the search bins.
\begin{table}[tbhp]
    \centering
    \topcaption{\label{tab:qcdsyst}
    Summary of systematic uncertainties in the multijet background prediction,
    together with their typical size ranges across the search bins.
      }
    \begin{tabular}{ l  c }
    \hline
      Source & Range [\%] \\
      \hline
      Jet energy resolution & 10--20\\
      Tails of jet response in templates & 17--25\\
      $\sigma_\text{T}^{\text{soft}}$ modeling & 1--25\\
      \njets modeling & 1--19\\
      \nbtags modeling & 1--16\\
    \hline
    \end{tabular}
\end{table}

\begin{figure*}[htb!]
    \centering
    \includegraphics[width=0.85\textwidth]{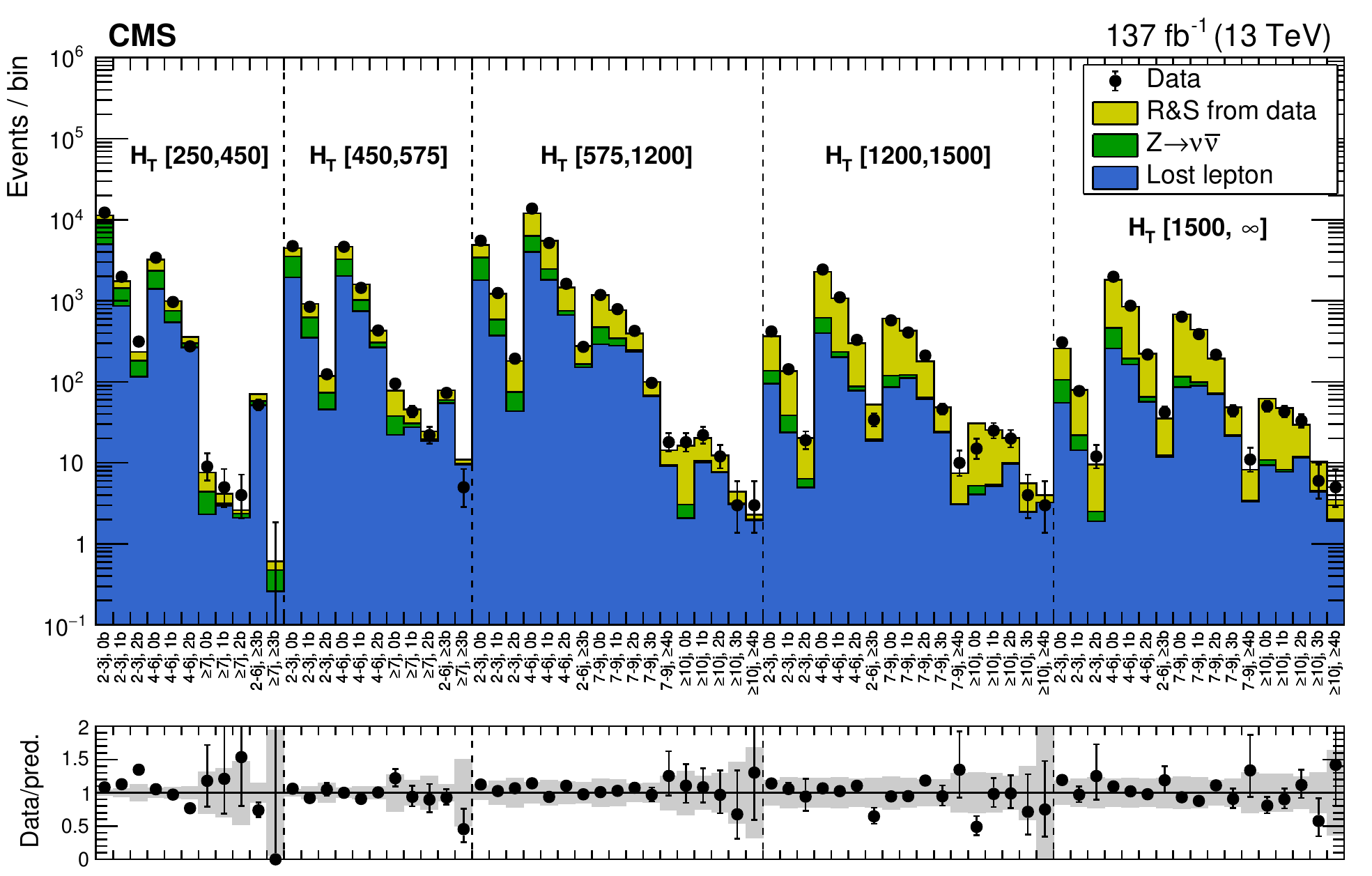}
    \caption{Validation of the R\&S multijet background prediction in control regions in data selected with $\dpmin<0.3$.
    Electroweak backgrounds (LL and $\PZ\to\PGn\PAGn$) are estimated from data.
    In regions where the amount of data is insufficient to estimate the electroweak backgrounds,
    the corresponding yields are taken directly from simulation.
    The bins on the horizontal axis correspond to the (\HT, \njets, \nbtags) topological regions. The gray band on the ratio plot represents the total uncertainty in the prediction.}
    \label{fig:rs_dataCR_InvertDPhi}
\end{figure*}

The resulting background prediction is validated in data using control regions enriched in multijet events.
The results of the validation in a control region selected by inverting
the \dpmin requirement
are shown in Fig.~\ref{fig:rs_dataCR_InvertDPhi}.
The electroweak backgrounds (LL and $\PZ\to\PGn\PAGn$) in this control region are estimated from data
using transfer factors from leptonic control regions as described above.
In regions where the number of events in the data leptonic control regions
are insufficient,
the electroweak background is taken from simulation.
The observation is found to agree with the prediction, within the uncertainties.

\subsection{Search for disappearing tracks}
\label{sec:distrbkgds}

In the search for disappearing tracks, the SM background consists of events with charged hadrons or leptons that interact
in the tracker or are poorly reconstructed, as well as tracks built out of incorrect combinations of hits.
The background is estimated from data,
leveraging
the orthogonal definition of STCs and selected STs (Section~\ref{sec:evtsel:evtcat:distracks}),
as described by Eq.~(\ref{eqn:bkgest}).
\begin{linenomath}
\begin{equation}
\label{eqn:bkgest}
 N_{\mathrm{ST}}^\text{est} = \fshort \, N_{\mathrm{STC}}^\text{obs},
\end{equation}
\end{linenomath}
where $N_{\mathrm{ST}}$ is the number of selected short tracks, $N_{\mathrm{STC}}$ is the number of selected short track candidates, and \fshort is defined as:
\begin{linenomath}
\begin{equation}
\label{eqn:fshort}
\fshort = N_{\mathrm{ST}}^\text{obs} / N_{\mathrm{STC}}^\text{obs}.
\end{equation}
\end{linenomath}

The \fshort ratio is measured directly in data,
in a control region of events selected using
the same triggers
and preselection criteria used for the signal regions,
except the selection on \Met is relaxed to $\Met>30\GeV$ for all \HT values,
and the selection on \mttwo is shifted to $60<\mttwo<100\GeV$.
We exploit the empirical invariance of this ratio with respect to the \HT and \Met selection criteria,
as observed in data control regions,
to reduce the statistical uncertainty in the measurement.
The \fshort ratio is therefore measured in data
separately for each \njets, track \pt, track length category,
and inclusively in \HT.
The \fshort values are measured separately in 2016 and 2017--2018 data,
mainly to account for the upgrade of the CMS tracking detector after 2016.
Since a reliable measurement in data of the  \fshort ratio for long (L) tracks is not achievable
because of the insufficient number of events,
the value measured in data for medium (M) length tracks is used instead,
after applying a correction based on simulation:
\begin{linenomath}
\begin{equation}
\label{eqn:fshortL}
\fshort(\text{L})_\text{data}^\text{est} =  \fshort(\text{M})_\text{data} \, \fshort(\text{L})_{\mathrm{MC}}/\fshort(\text{M})_{\mathrm{MC}}.
\end{equation}
\end{linenomath}
A systematic uncertainty in the measured values of \fshort is assigned
to cover for the empirically motivated assumption of its invariance with respect to \HT and \Met.
Its size is determined
by varying the  \HT and \Met selection requirements
in data events with $60<\mttwo<100\GeV$.
For long tracks, a conservative systematic uncertainty of 100\% is assigned,
as a correction based on simulation is used and there are insufficient data to study
the effect of \HT and \Met variations.

The \fshort ratio is then used to predict the expected background
in events with $\mttwo>100\GeV$, as described in Eq.~(\ref{eqn:bkgest}).

In the presence of BSM physics, the above-defined control regions could be affected by signal contamination.
Although the expected signal contamination is typically negligible,
its potential impact is accounted for in the interpretation of the results, as further described in Section~\ref{sec:interpretation}.

The background prediction
is validated in data in an intermediate \mttwo region ($100<\mttwo<200\GeV$).
No excess event yield is observed.
The event categorization in this validation region is identical to the signal region,
allowing for a bin-by-bin validation of the background prediction.

\begin{figure*}[htbp!]
  \centering
    \includegraphics[width=0.85\textwidth]{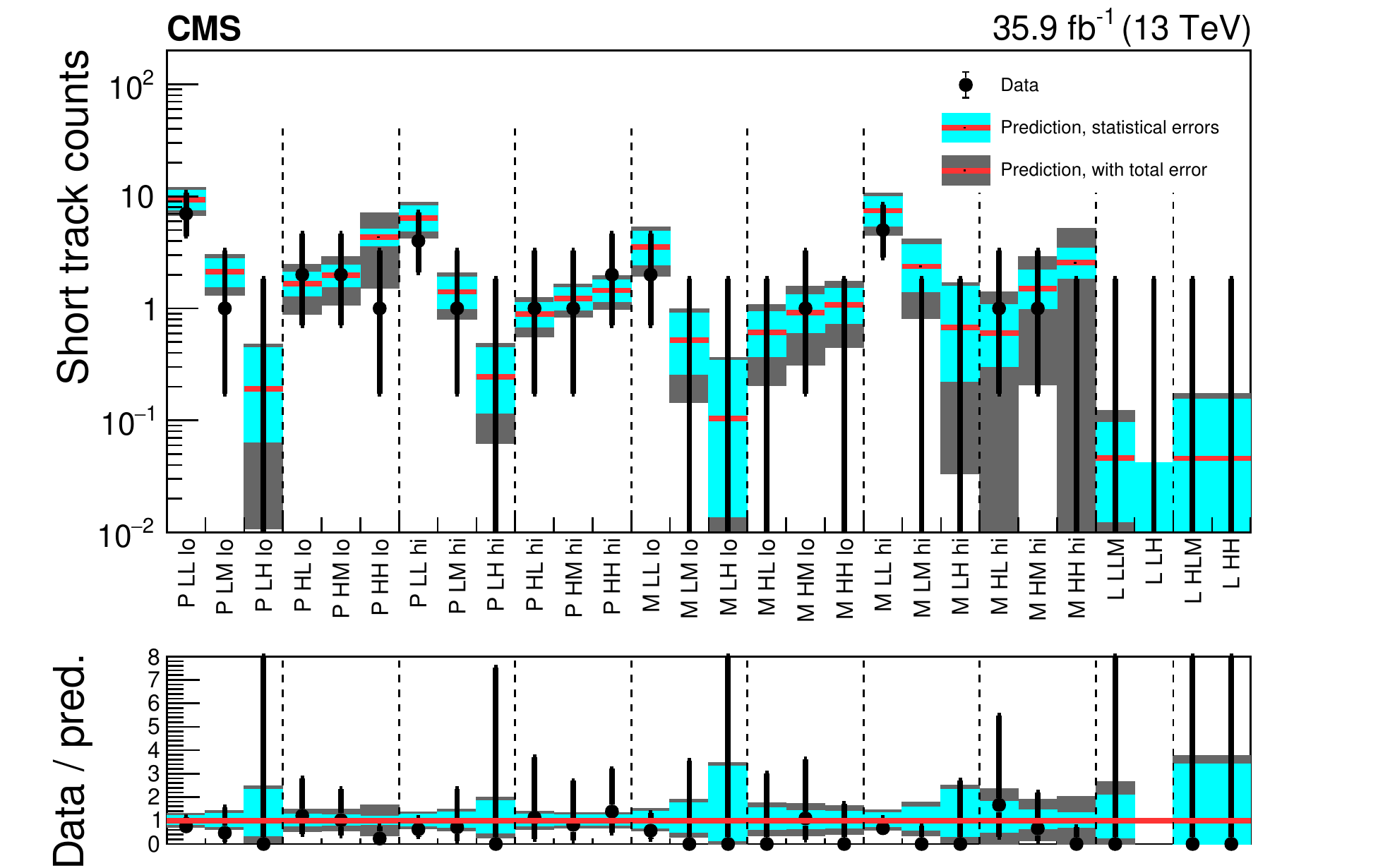}
    \includegraphics[width=0.85\textwidth]{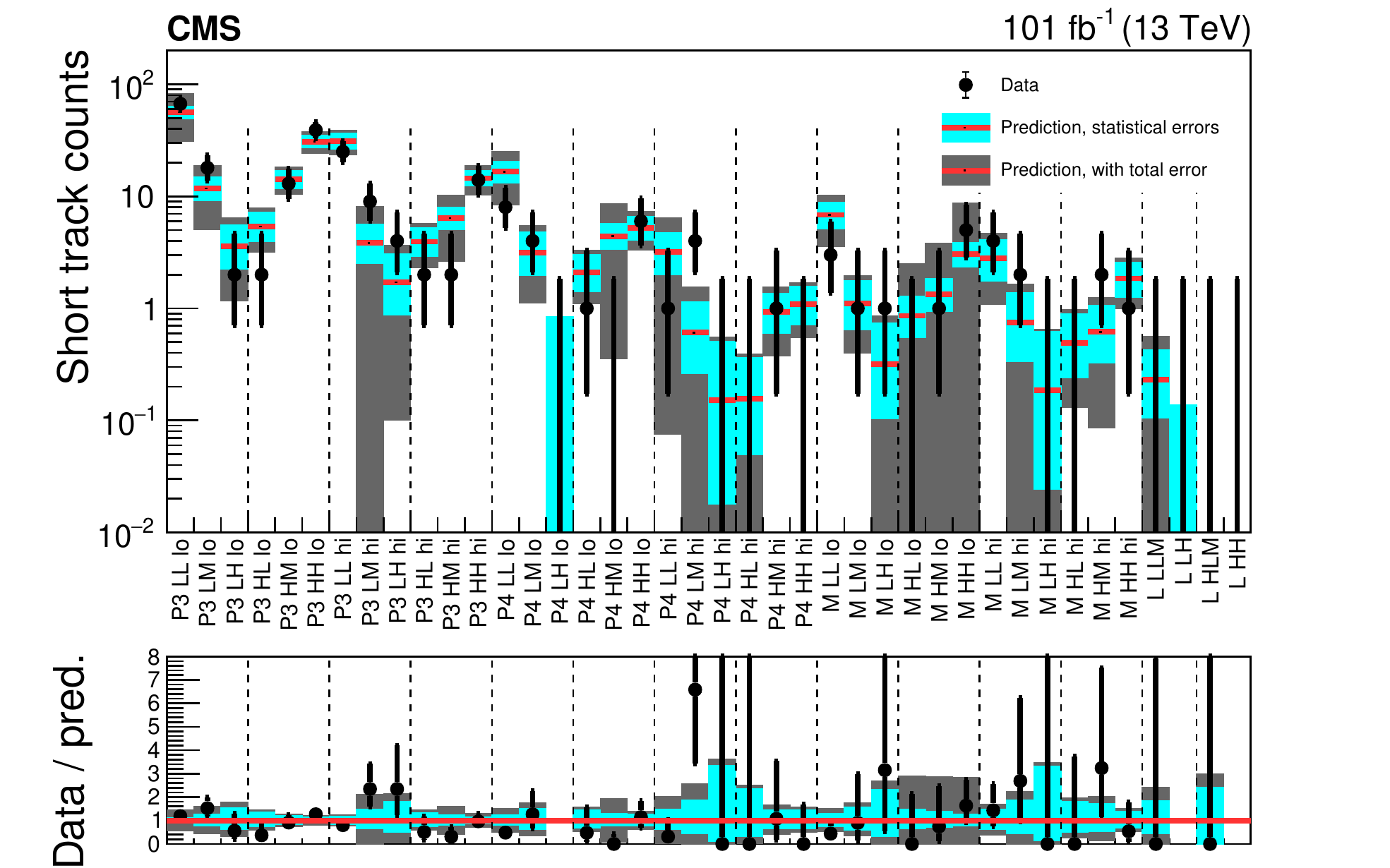}
    \caption{Validation of the background prediction method in (upper) 2016 and (lower) 2017--2018 data with $100<\mttwo<200\GeV$,
      for the disappearing tracks search.
      The red histograms represent the predicted backgrounds, while the black markers are the observed data counts.
      The cyan bands represent the statistical uncertainty in the prediction.
      The gray bands represent the total uncertainty in the prediction.
      The labels on the $x$ axes are explained in Tables~\ref{tab:sr1_distracks}--\ref{tab:sr2_distracks} of Appendix~\ref{app:srs:distr}.
      Regions whose predictions use the same measurement of \fshort are grouped by the vertical dashed lines.
      Bins with no entry in the ratio have zero predicted background.
      \label{fig:vr_data}}
\end{figure*}

Figure~\ref{fig:vr_data} shows the result of
the background prediction validation in 2016 data and in 2017--2018 data.
We find good agreement between the observation and the background prediction in the validation region.
An additional systematic uncertainty is assigned
to cover for discrepancies exceeding statistical uncertainties.
The uncertainties in the background prediction are summarized in Table~\ref{tab:distracksyst} together with their typical size ranges across the search bins.

\begin{table}[htb!]
    \centering
    \topcaption{\label{tab:distracksyst}
    Summary of systematic uncertainties in the disappearing track background prediction, together with their typical size ranges across the search bins.
    The systematic uncertainties arising from the assumption of kinematic invariance of \fshort and from the validation of the background prediction are always taken
    to be at least as large as the statistical uncertainties on the measured values of  \fshort and on the background prediction in the validation region,
    respectively.
      }
      \cmsTableTwo{
    \begin{tabular}{ l  c }
    \hline
      Source & Range [\%] \\
      \hline
      Limited size of data control samples & 1--100\\
      Limited size of data \fshort measurement samples & 5--45\\
      Kinematic invariance of \fshort & 10--80\\
      Validation of background prediction & 25--75\\
    \hline
    \end{tabular}}
\end{table}

\section{Results}
\label{sec:results}

The data yields in the search regions are statistically compatible with the estimated
backgrounds from SM processes.

\subsection{Inclusive \texorpdfstring{\mttwo}{MT2} search}
\label{sec:results:mt2}

\begin{figure*}[!htb]
  \centering
    \includegraphics[width=0.85\textwidth]{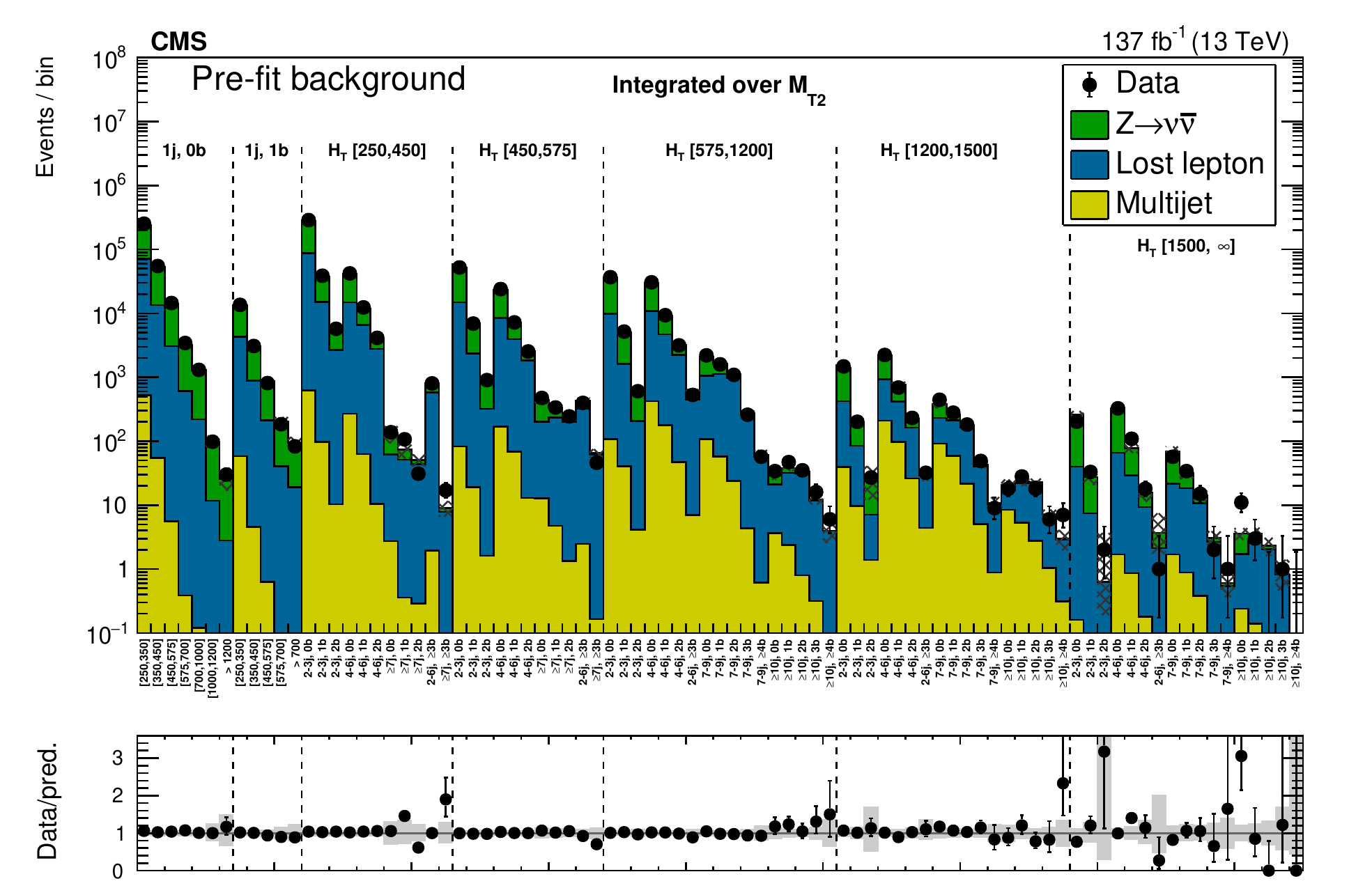}\\
    \includegraphics[width=0.85\textwidth]{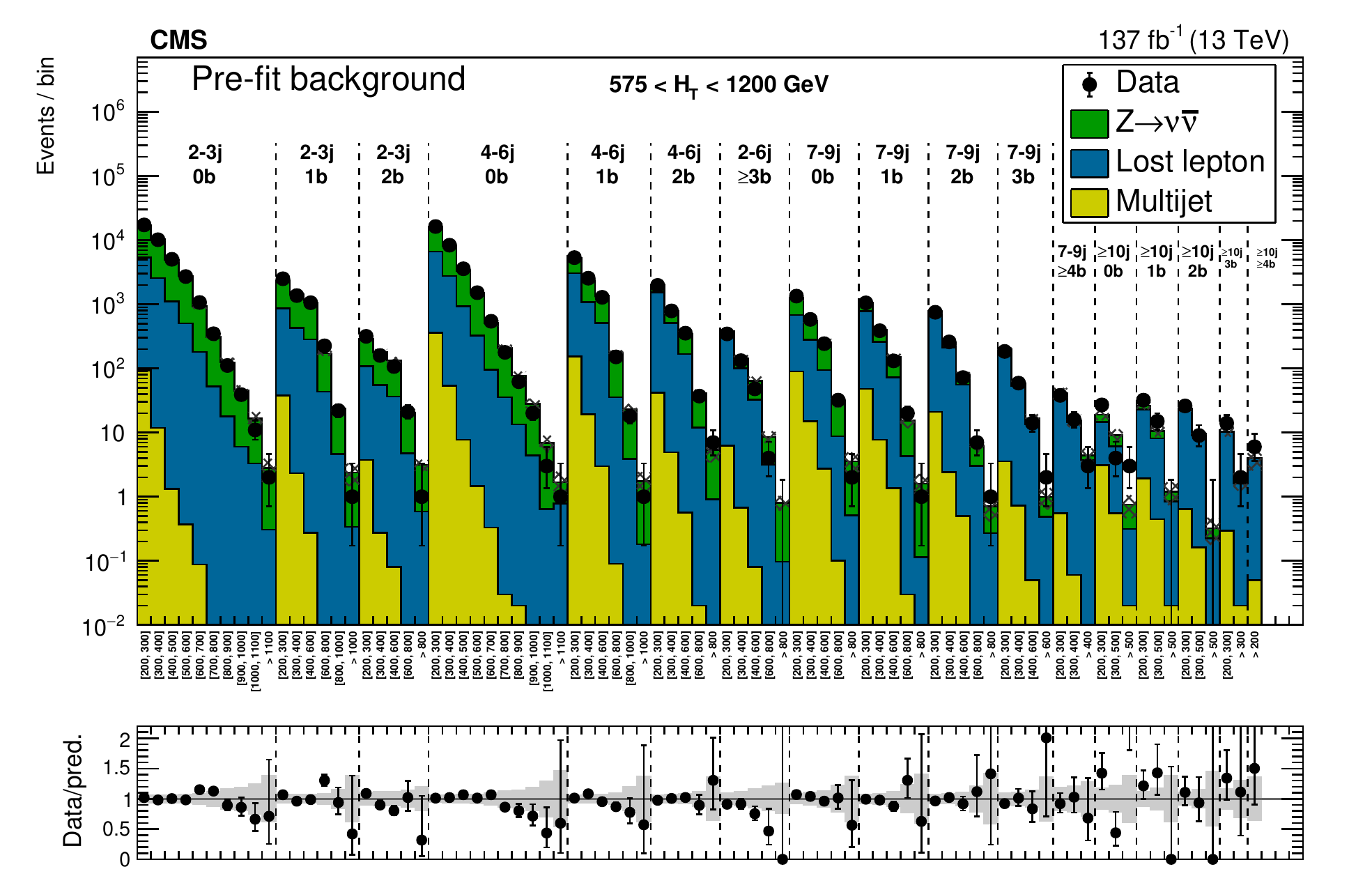}\\
    \caption{
    (Upper) Comparison of the estimated (pre-fit) background and observed data events in each topological region. The hatched bands represent the full uncertainty in the background estimate.
      The monojet regions ($\njets = 1$) are identified by the labels ``1j, 0b'' and ``1j, 1b'', and are binned in jet \pt.
      The multijet regions are shown for each \HT region separately, and are labeled accordingly. The notations j, b are short for \njets, \nbtags.
      (Lower) Same for individual \mttwo search bins in the medium-\HT region. On the $x$ axis, the \mttwo binning is shown in
      units of GeV.}
    \label{fig:results_mt2}
\end{figure*}

A summary of the results of the \mttwo inclusive search is shown in Fig.~\ref{fig:results_mt2}.
Each bin in Fig.~\ref{fig:results_mt2} (upper) corresponds to a single (\HT, \njets, \nbtags) topological region
integrated over \mttwo.
Figure~\ref{fig:results_mt2} (lower)  breaks down the background estimates and observed data yields
into \mttwo bins for the region $575 < \HT <1200\GeV$: each bin corresponds to a single \mttwo bin,
and vertical lines identify (\HT, \njets, \nbtags) topological regions.
Distributions for the other \HT regions can be found in Figs.~\ref{fig:otherResults1}--\ref{fig:otherResults2} in Appendix~\ref{app:results:mt2}.
Background predictions and observed yields in all search regions are also summarized in Tables~\ref{tab:yieldsmonojet}--\ref{tab:yieldsUHh} in Appendix~\ref{app:srs:mt2}.
The background estimates and corresponding uncertainties
rely exclusively
on the inputs from control samples and simulation described in
Section~\ref{sec:mt2bkgds}, prior to the fit to the data detailed in Section~\ref{sec:interpretation},
and are referred to
in the rest of the text as pre-fit background results.

To allow simpler reinterpretation, we also provide results for super signal regions, which cover subsets of the full
analysis with simpler inclusive selection criteria and that can be used to obtain approximate interpretations of this search.
The definitions of these regions are given in Table~\ref{tab:ssr_def_yields},
with the predicted and observed number of events
and the 95\% confidence level (\CL) upper limit on the number of signal events contributing to each region.
Limits are set using a modified frequentist approach, employing the \CLs
criterion and relying on asymptotic approximations
to calculate the distribution of the profile likelihood test-statistic
used~\cite{Junk:1999kv,Read:2002hq,Cowan:2010js,ATLAS:2011tau}.

\begin{table*}[!htbp]
 \centering
   \topcaption{\label{tab:ssr_def_yields} Definitions of super signal
     regions, along with predictions, observed data, and the observed
     95\% \CL upper limits on the number of signal events contributing to
     each region ($N_{95}^{\mathrm{max}}$).  The limits are shown as a range
     corresponding to an assumed uncertainty in the signal acceptance of 0 or 15\% ($N_{95}^{\mathrm{max},0}$--$N_{95}^{\mathrm{max},15}$).
     A dash in the selection criteria means that no requirement is applied.
     All selection criteria as in the full analysis are applied. For regions with $\njets=1$, $\HT\equiv\ptj$.
     The mono-$\phi$ super signal region corresponds to the subset of analysis bins identified in Refs.~\cite{monojetexcess1,monojetexcess2}
     as showing a significant excess in data based on the results of Ref.~\cite{MT2_2016}.}
 \renewcommand{\arraystretch}{1.3}
 \cmsTableAlt{
 \begin{tabular}{l c c c c c c c}                                                                                                                                                                            \hline
      Region & \njets & \nbtags & \HT [\GeVns{}] & \mttwo [\GeVns{}] & Prediction & Data & $N_{95}^{\mathrm{max},0}$--$N_{95}^{\mathrm{max},15}$ \\
      \hline
2j loose & $\geq$2 & -- & $>$1200 & $>$1200 & $37\pm14$ & 41 & 26.0--27.2 \\
2j tight & $\geq$2 & -- & $>$1500 & $>$1400 & $10.7^{+4.2}_{-4.1}$ & 13 & 11.7--12.3 \\ [\cmsTabSkip]
4j loose & $\geq$4 & -- & $>$1200 & $>$1000 & $54\pm13$ & 72 & 41.5--43.8 \\
4j tight & $\geq$4 & -- & $>$1500 & $>$1400 & $6.4\pm2.5$ & 10 & 10.9--11.4 \\ [\cmsTabSkip]
7j loose & $\geq$7 & -- & $>$1200 & $>$600 & $63^{+13}_{-12}$ & 72 & 33.4--35.0 \\
7j tight & $\geq$7 & -- & $>$1500 & $>$800 & $14.9^{+4.3}_{-4.2}$ & 14 & 10.1--10.4 \\ [\cmsTabSkip]
10j loose & $\geq$10 & -- & $>$1200 & $>$400 & $17.3\pm4.0$ & 25 & 18.6--19.5 \\
10j tight & $\geq$10 & -- & $>$1500 & $>$600 & $3.6^{+1.2}_{-1.1}$ & 5 & 6.8--7.1 \\ [\cmsTabSkip]
2b loose & $\geq$2 & $\geq$2 & $>$1200 & $>$600 & $32.0\pm4.5$ & 33 & 15.3--15.9 \\
2b tight & $\geq$2 & $\geq$2 & $>$1500 & $>$600 & $12.0^{+2.8}_{-2.7}$ & 12 & 9.1--9.4 \\ [\cmsTabSkip]
3b loose & $\geq$2 & $\geq$3 & $>$1200 & $>$400 & $17.6\pm4.0$ & 16 & 10.0--10.3 \\
3b tight & $\geq$2 & $\geq$3 & $>$1500 & $>$400 & $7.5\pm2.1$ & 5 & 5.3--5.5 \\ [\cmsTabSkip]
4b loose & $\geq$2 & $\geq$4 & $>$1200 & $>$400 & $2.1\pm0.7$ & 2 & 4.2--4.4 \\
4b tight & $\geq$2 & $\geq$4 & $>$1500 & $>$400 & $0.8^{+0.4}_{-0.3}$ & 1 & 3.5--3.6 \\ [\cmsTabSkip]
7j 3b loose & $\geq$7 & $\geq$3 & $>$1200 & $>$400 & $10.9^{+3.0}_{-2.9}$ & 8 & 8.7--8.9 \\
7j 3b tight & $\geq$7 & $\geq$3 & $>$1500 & $>$400 & $4.6^{+2.0}_{-1.9}$ & 4 & 5.5--5.7 \\ [\cmsTabSkip]
7j 4b loose & $\geq$7 & $\geq$4 & $>$1200 & $>$400 & $1.7\pm0.7$ & 2 & 4.3--4.5 \\
7j 4b tight & $\geq$7 & $\geq$4 & $>$1500 & $>$400 & $0.7\pm0.4$ & 1 & 3.6--3.7 \\ [\cmsTabSkip]
10j 4b loose & $\geq$10 & $\geq$4 & $>$1200 & $>$400 & $0.6^{+0.5}_{-0.4}$ & 1 & 3.6--3.7 \\
10j 4b tight & $\geq$10 & $\geq$4 & $>$1500 & $>$400 & $0.1^{+0.5}_{-0.1}$ & 0 & 2.0--2.1 \\ [\cmsTabSkip]
\multirow{2}{*}{Mono-$\phi$} & \multirow{2}{*}{1--3} & \multirow{2}{*}{0} & \multirow{2}{*}{250--450} & 200--300 & \multirow{2}{*}{$(5.2\pm0.3) \times10^{5}$} &
\multirow{2}{*}{$5.5\times10^{5}$} & \multirow{2}{*}{(0.6--0.8)$\times10^{5}$} \\
            &     &   &         & (if $\njets\geq2$) &               &         &  \\
    \hline

\end{tabular}
}
\end{table*}

\subsection{Search for disappearing tracks}
\label{sec:results:distr}

\begin{figure*}[!htb]
  \centering
    \includegraphics[width=0.85\textwidth]{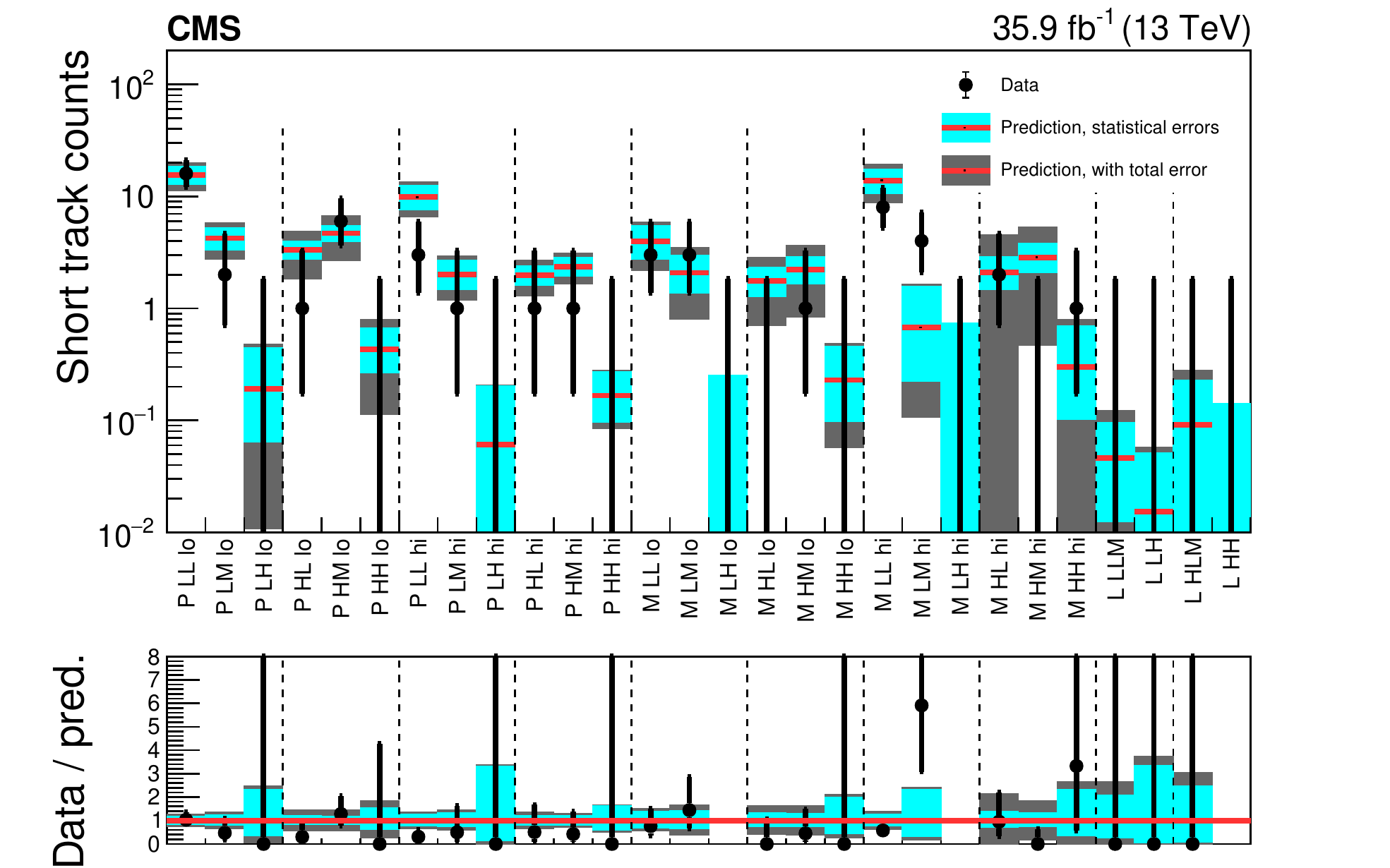}\\
    \includegraphics[width=0.85\textwidth]{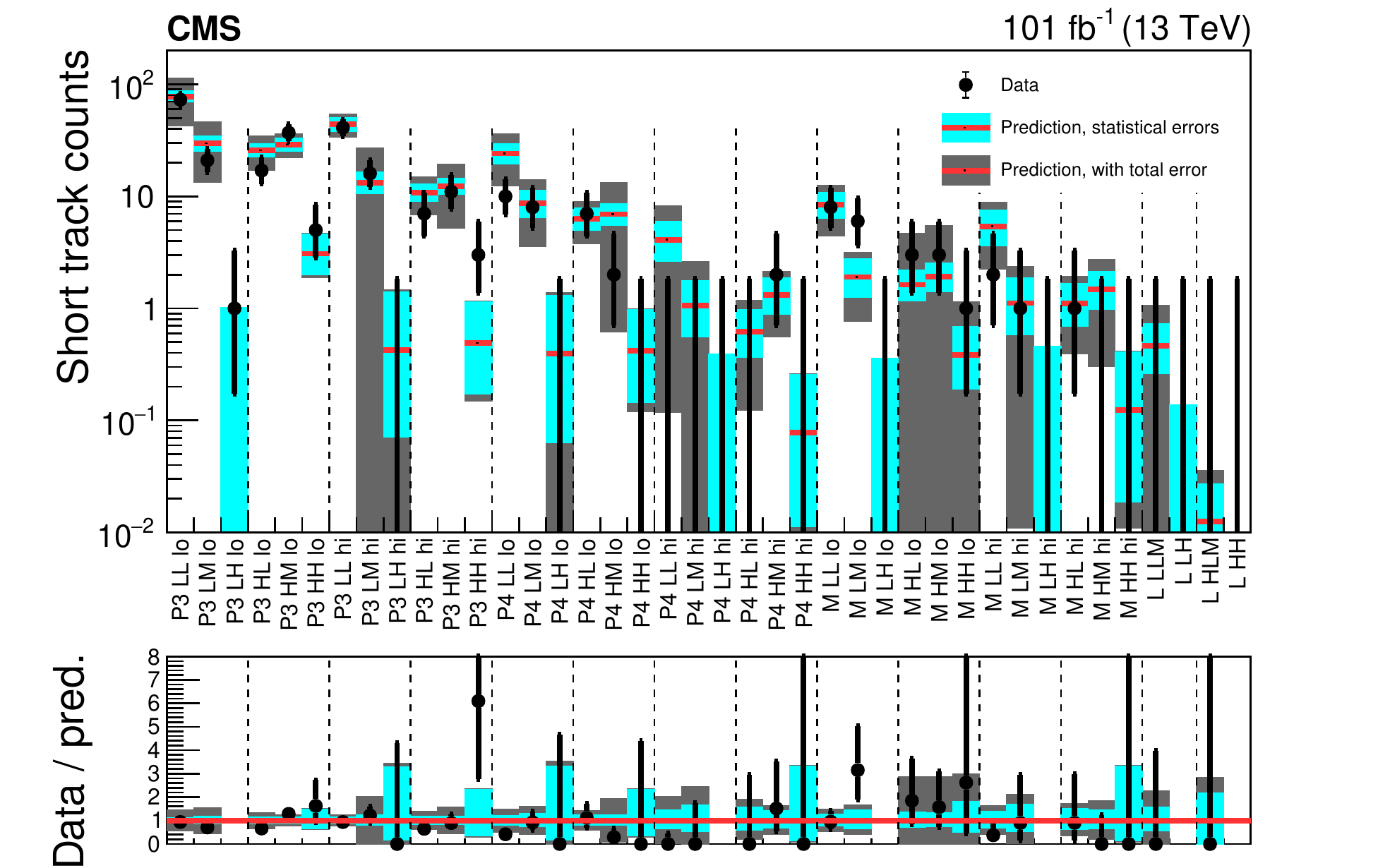}\\
    \caption{
            Comparison of the estimated (pre-fit) background and observed data events in (upper) each of the 2016 search regions, and in (lower) each of the 2017--2018 search regions, in the search for disappearing tracks.
      The red histogram represents the predicted background, while the black markers are the observed data counts.
      The cyan band represents the statistical uncertainty in the prediction.
      The gray band represents the total uncertainty.
      The labels on the $x$ axes are explained in Tables~\ref{tab:sr1_distracks}--\ref{tab:sr2_distracks} of Appendix~\ref{app:srs:distr}.
      Regions whose predictions use the same measurement of \fshort are grouped by the vertical dashed lines.
      Bins with no entry in the ratio have zero pre-fit predicted background.
      }
    \label{fig:results_distracks}
\end{figure*}

The results of the search for disappearing tracks are shown in Fig.~\ref{fig:results_distracks}.
Just as in the case of the inclusive search, the background estimates and the uncertainties
rely exclusively
on the inputs from control samples and simulation
(Section~\ref{sec:distrbkgds}), prior to the fit to the data described in Section~\ref{sec:interpretation}.
We refer to them in the rest of the text as pre-fit background results.
Background predictions and observed yields in all search regions are also summarized in Tables~\ref{tab:sr1_distracks}--\ref{tab:sr2_distracks} in Appendix~\ref{app:srs:distr}.

\section{Interpretation of the results}
\label{sec:interpretation}

The measurements are interpreted in the context of models of new physics.
Maximum likelihood fits to the data in the signal regions are carried out under either
background-only or background+signal hypotheses.
The uncertainties in the modeling of the backgrounds, summarized in Section~\ref{sec:bkgds}, are inputs to the fitting procedure.
The likelihoods are constructed as the product of Poisson probability density functions, one for each signal region, with
additional log-normal
constraint terms that account for the uncertainties in the background estimates and, if considered, in the signal yields.

The background+signal fits are used to set 95\% \CL upper limits on the cross sections
for the signal models under consideration.  These limits are then used, in conjunction
with the theoretical cross section calculations, to exclude ranges of masses for
the BSM particles of the signal models.
Before the fits are performed, the signal yields are corrected to account for the expected signal contamination
of the data control regions used to estimate the SM background.

For the interpretation of the results, simplified BSM physics models~\cite{sms1,sms2,sms3,sms4,sms5} are used.
Simplified models are defined by sets of hypothetical particles and sequences of their production and decay.
The theoretical parameters are thus reduced to a small number of masses and cross sections,
providing an effective tool to characterize potential signals of BSM physics.

\begin{figure*}[htbp]
  \centering
    \includegraphics[width=0.3\textwidth]{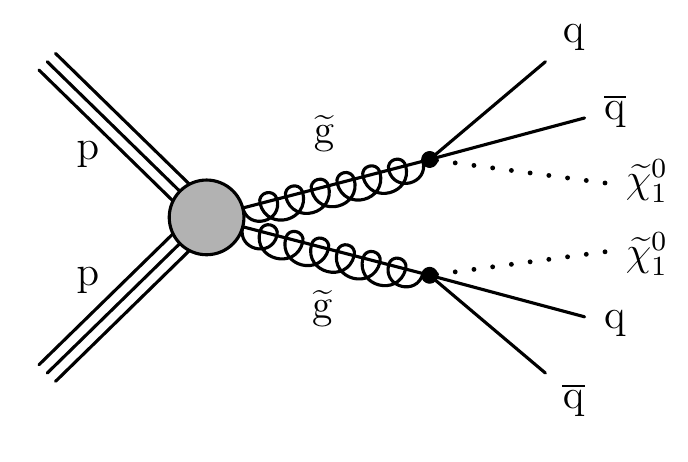}
    \includegraphics[width=0.3\textwidth]{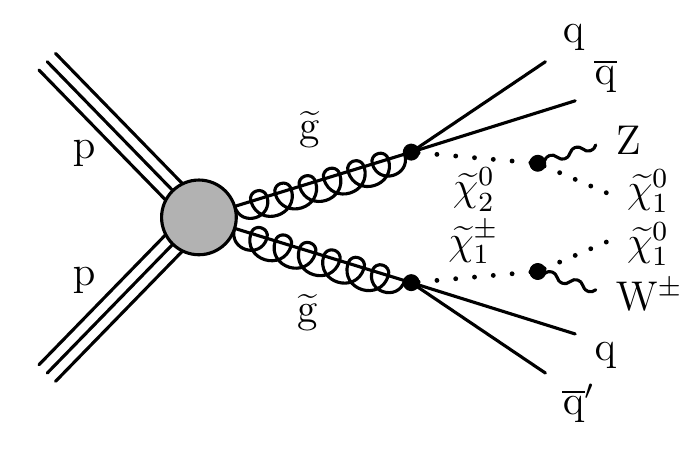}
    \includegraphics[width=0.3\textwidth]{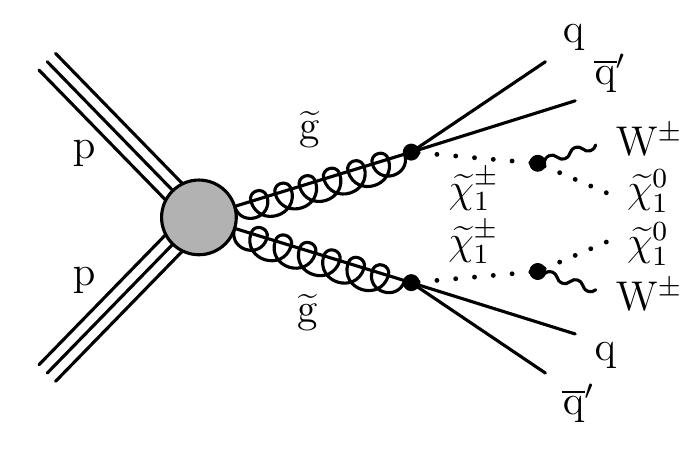}\\
    \includegraphics[width=0.3\textwidth]{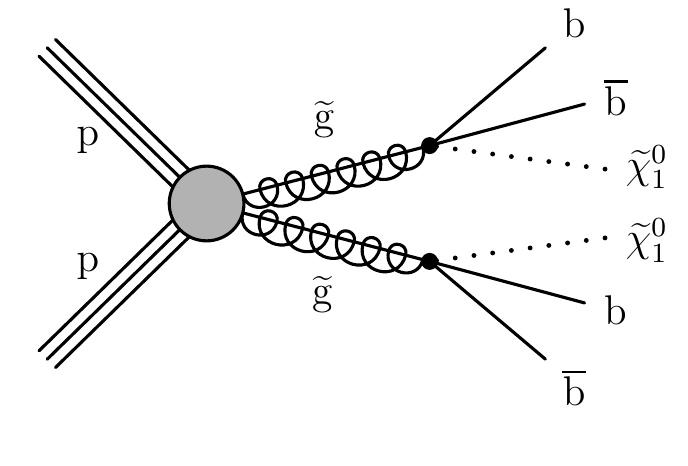}
    \includegraphics[width=0.3\textwidth]{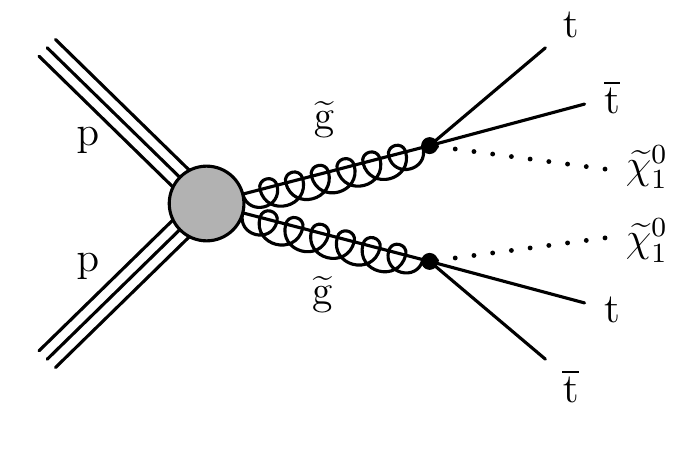} \\
    \includegraphics[width=0.3\textwidth]{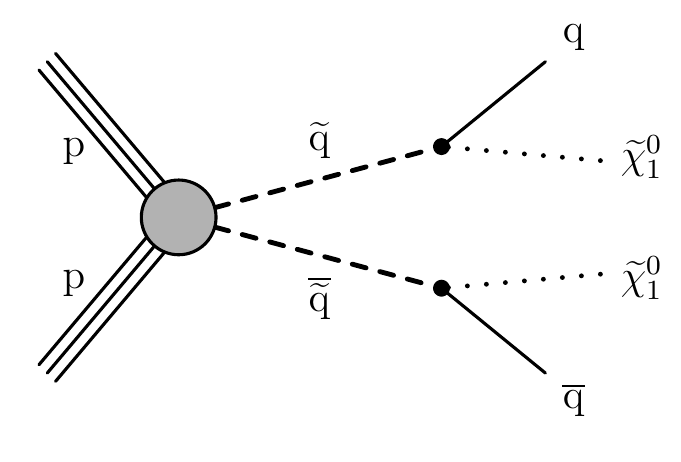}
    \includegraphics[width=0.3\textwidth]{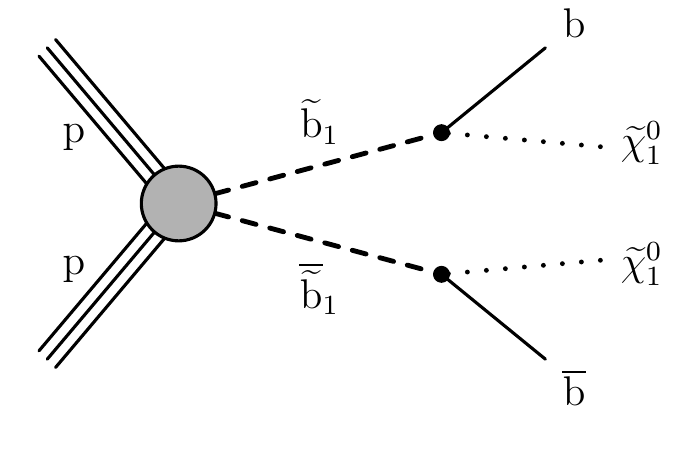}
    \includegraphics[width=0.3\textwidth]{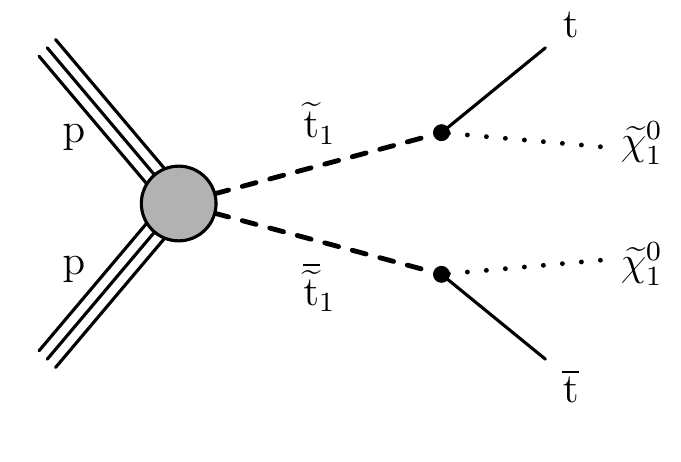} \\
    \includegraphics[width=0.3\textwidth]{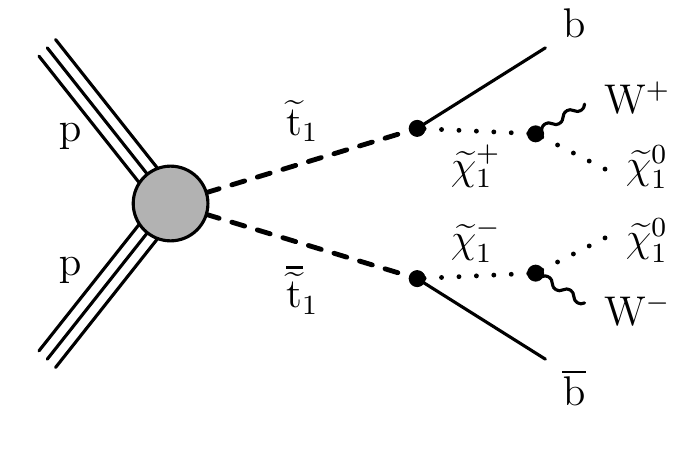}
    \includegraphics[width=0.3\textwidth]{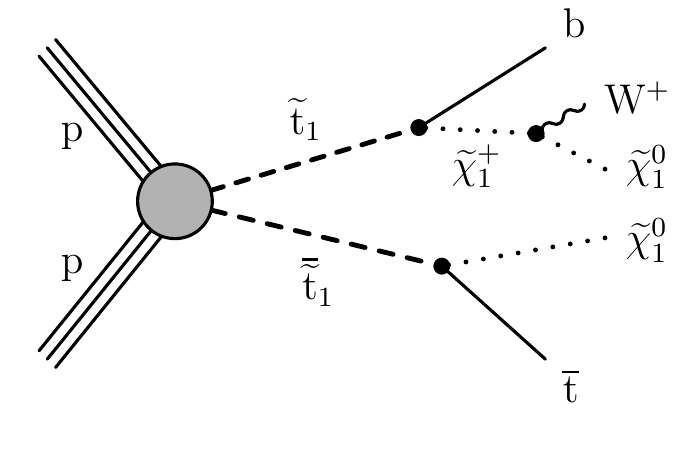}
    \includegraphics[width=0.3\textwidth]{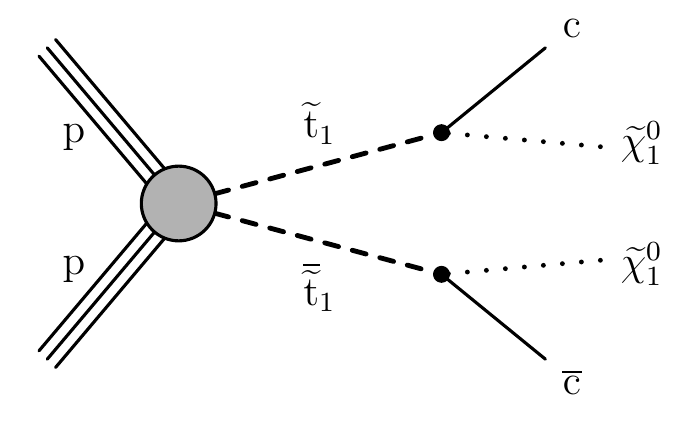} \\
    \caption{(Upper) Diagrams for three scenarios of direct gluino pair production where each gluino undergoes a three-body decay to light-flavor (\cPqu, \cPqd, \cPqs, \cPqc) quarks,
    with different decay modes.
    For mixed-decay scenarios, we assume equal branching fraction for each decay mode.
    (Upper middle) Diagrams for the direct gluino pair production where gluinos decay to bottom and top quarks.
      (Lower middle) Diagrams for the direct pair production of light-flavor, bottom, and top squark pairs.
      (Lower) Diagrams for three alternate scenarios of direct top squark pair production with different decay modes.
      For mixed-decay scenarios, we assume equal branching fraction for each decay mode.}
    \label{fig:SMS_feynDiagrams}
\end{figure*}

The results of the inclusive \mttwo search are used to constrain each of the simplified models of SUSY shown in Fig.~\ref{fig:SMS_feynDiagrams}.
For each scenario of gluino (squark) pair production, the simplified models assume that
all SUSY particles other than
those shown in the corresponding diagram
are too heavy to be produced directly, and that the gluino (squark) decays promptly.
The models assume that each gluino (squark) decays with a 100\% branching fraction into the decay products depicted in Fig.~\ref{fig:SMS_feynDiagrams}.
For models where the decays of the two gluinos or squarks in the same diagram differ, a 1/3 (1/2)
branching fraction for each of the three (two) decay modes is assumed.
In particular, for the diagram of gluino pair production where the decays of the two gluinos differ, each gluino can decay via a \chizz, \charginoplus,
or \charginominus.
For scenarios with top squarks decaying into top quarks, the polarization of the
top quark can be model dependent and a function of the top squark and neutralino mixing matrices.
To maintain independence of any particular model realization, events are generated with unpolarized
top quarks.
Signal cross sections are calculated at approximately NNLO+NNLL (next-to-next-to-leading-logarithm) order in \alpS~\cite{bib-nlo-nll-01,bib-nnll-05,bib-nlo-nll-02,bib-nlo-nll-03,bib-nlo-nll-04,bib-nnll-06,bib-nlo-nll-05,bib-nnll-02,bib-nnll-03,bib-nnll-04,bib-nnll-07,bib-nnll}.
For direct light-flavor squark pair production we assume either one single squark, or eight degenerate squarks ($\PSQL+\PSQR$, with $\PSQ=\PSQu,~\PSQd,~\PSQs,~\PSQc$).
For direct bottom and top squark pair production, we assume one single squark.

\begin{figure*}[htbp]
  \centering
    \includegraphics[width=0.35\textwidth]{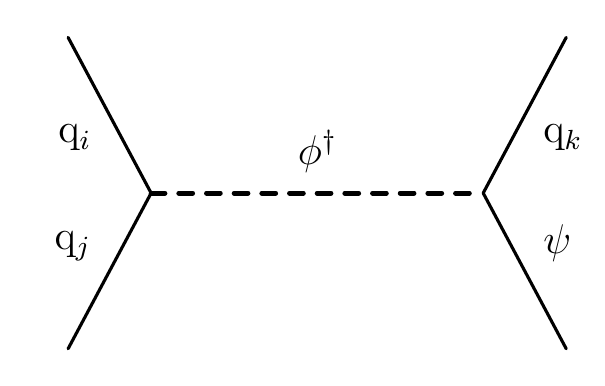}
    \caption{Diagram for the mono-$\phi$ model, where a colored scalar $\phi$ is resonantly produced, and it decays to an invisible massive Dirac fermion $\psi$ and an SM quark.}
    \label{fig:SMS_monophi}
\end{figure*}

The mono-$\phi$ model depicted in Fig.~\ref{fig:SMS_monophi},
that was recently proposed~\cite{monojetexcess1,monojetexcess2} based on
a reinterpretation of the results of Refs.~\cite{MT2_2016,RA2b_2016,CMSmonojet,ATLASat2016:monojet,ATLASat2016:hadronic},
is also probed
by the inclusive \mttwo search.
In this case, the cross section for the signal is only calculated at LO order in \alpS.

\begin{figure*}[htbp]
  \centering
    \includegraphics[width=0.3\textwidth]{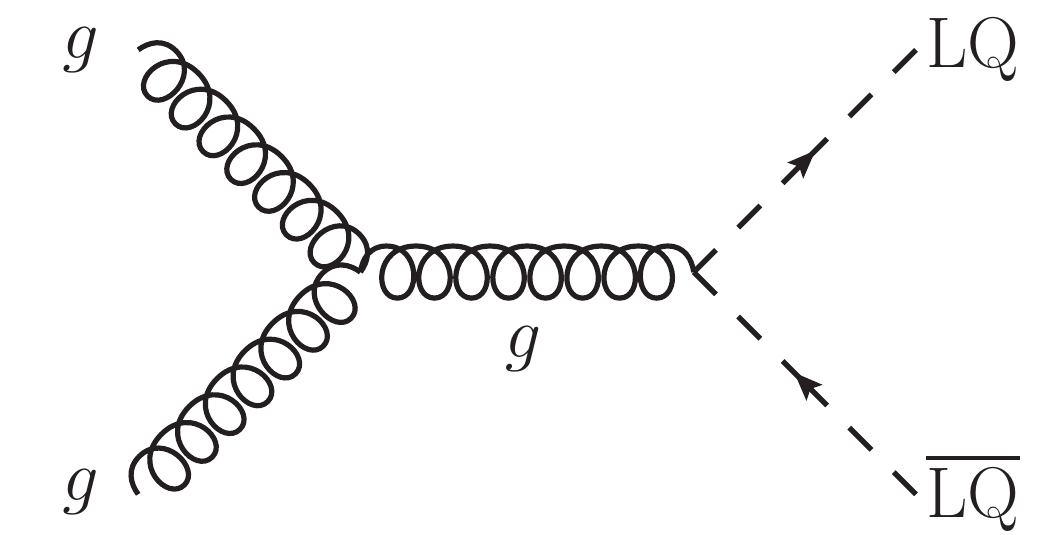}
    \includegraphics[width=0.3\textwidth]{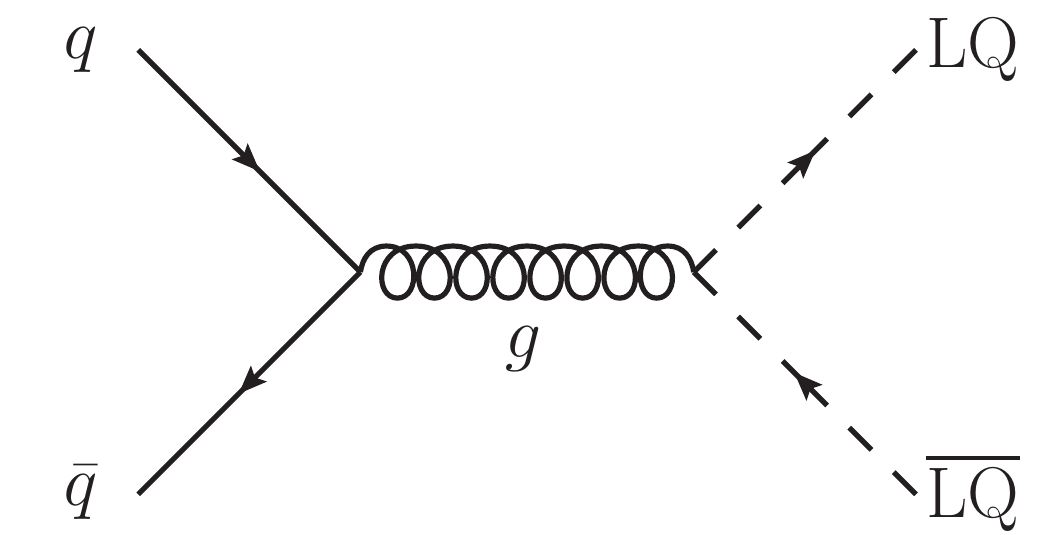}\\
    \vspace{3mm}
    \includegraphics[width=0.3\textwidth]{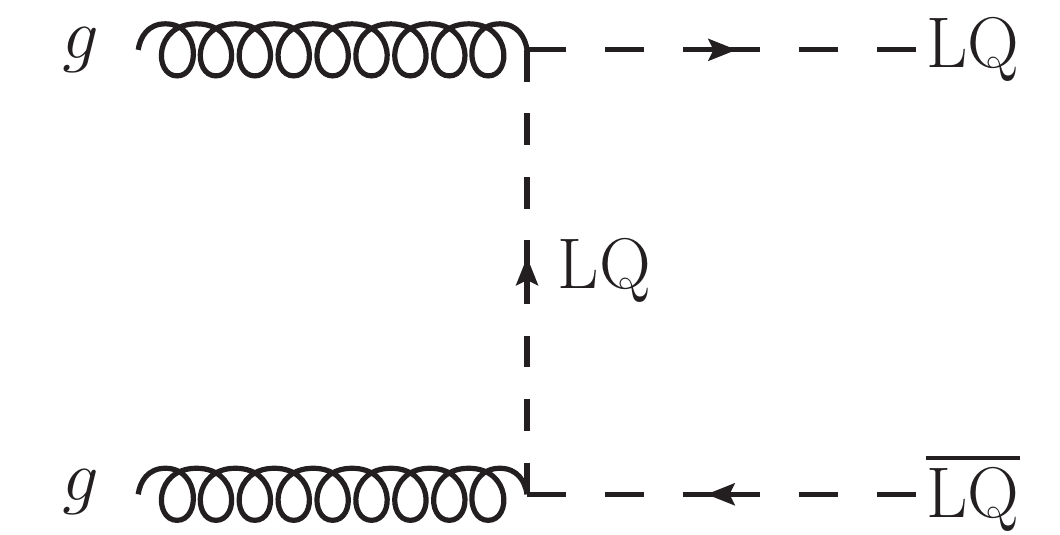}
    \includegraphics[width=0.3\textwidth]{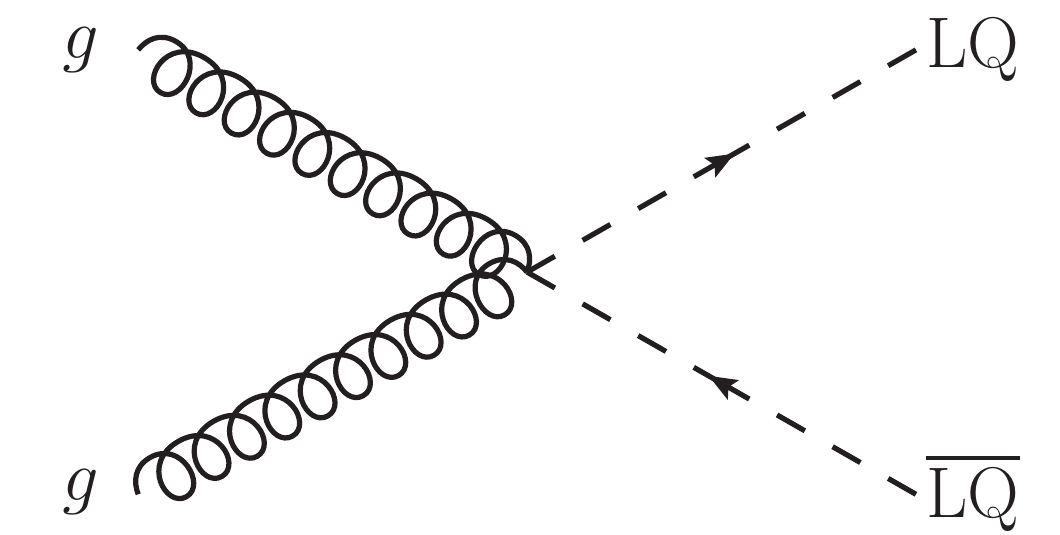}
    \vspace{3mm}
    \caption{Diagrams for LQ pair production.}
    \label{fig:SMS_LQ}
\end{figure*}

Another interpretation of
the inclusive \mttwo results places cross section limits on LQ pair production (depicted in Fig.~\ref{fig:SMS_LQ})
as a function of the LQ mass, similarly to Ref.~\cite{MT2_2016LQ}.
We consider production of either $\mathrm{LQ_{S}}$ or $\mathrm{LQ_{V}}$.
In each case, we assume that only one LQ state is within mass reach of the LHC,
and that the LQ decays with 100\% branching fraction to a neutrino and a single type of quark: a light-flavor quark (\cPq $=$ \cPqu, \cPqd, \cPqs, or \cPqc),
a bottom quark, or a top quark.
The cross sections for $\mathrm{LQ_{S}}$ ($\mathrm{LQ_{V}}$) pair production are computed to NLO (LO) order in \alpS following Ref.~\cite{lqtoolbox}.
The $\mathrm{LQ_{S}}$ pair production cross section depends only on the LQ mass. For $\mathrm{LQ_{V}}$, additional constraints are imposed by unitarity at high energy scales,
leading to model dependent solutions and thus production cross sections. In the model of Ref.~\cite{lqtoolbox}, developed to explain the flavor physics anomalies,
the additional relevant parameter for the $\mathrm{LQ_{V}}$ pair production cross section is $\kappa$, a dimensionless coupling that is 1 in the Yang--Mills case
and 0 in the minimal coupling case. We consider both values. For $\kappa=1$, the cross section for the $\mathrm{LQ_{V}}$ pair production is a factor 5--20 times larger
than that of $\mathrm{LQ_{S}}$,
depending on the LQ mass. In the $\mathrm{LQ_{V}}$ model, other free parameters are $g_{\PQtL}$ and $g_{\PQbL}$,
the couplings of the $\mathrm{LQ_{V}}$ to $\PQt\PGn$ and $\PQb\PGt$ pairs, respectively.
However, $g_{\PQtL}$ and $g_{\PQbL}$ do not affect the cross section or the kinematics for the $\mathrm{LQ_{V}}$ pair production, and we assume $g_{\PQtL} = g_{\PQbL} = 0.1$,
as predicted to explain the flavor physics anomalies.

\begin{figure*}[htbp]
  \centering
    \includegraphics[width=0.3\textwidth]{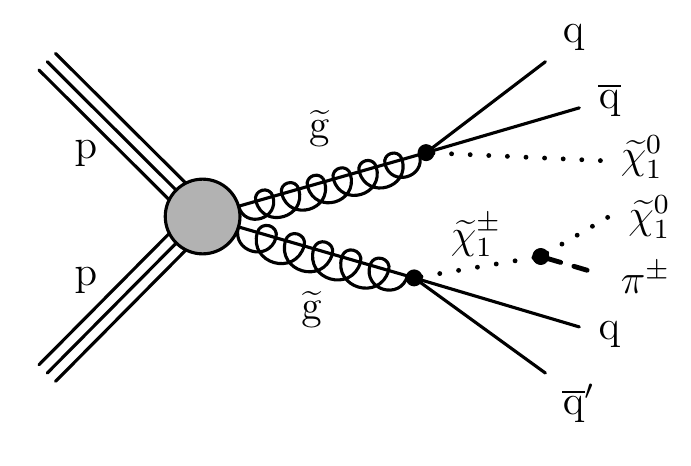}
    \includegraphics[width=0.3\textwidth]{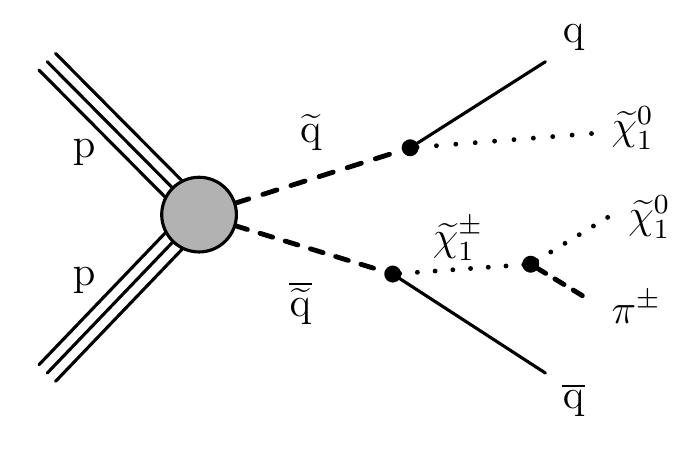}
    \includegraphics[width=0.3\textwidth]{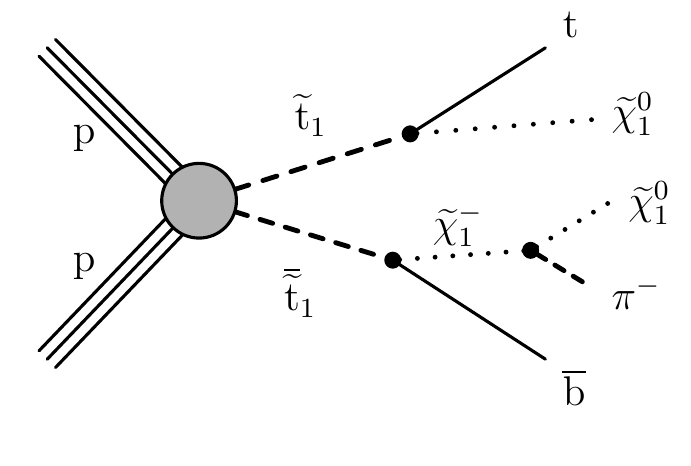}
    \caption{Diagrams for direct (left) gluino, (middle) light-flavor (\cPqu, \cPqd, \cPqs, \cPqc) squark, and (right) top squark pair production,
    where the directly produced gluinos and squarks can decay via a long-lived \chargino.
    For gluinos, we assume a 1/3
    decay branching fraction to each \lsp, \charginoplus, and \charginominus, and each gluino decays to light-flavor quarks.
    For squarks, we assume a 1/2
    branching fraction for decays to \lsp and to the \chargino allowed by charge conservation.
    The mass of the \chargino is larger than the mass of the \lsp by hundreds of \MeV.
    The \chargino decays to a \lsp via a pion, which is too soft to be detected.}
    \label{fig:SMS_chargino}
\end{figure*}

The results of the search for disappearing tracks are used to constrain simplified models of SUSY
where gluinos
and squarks
are produced in pairs, and each one
decays either directly to the lightest neutralino (\lsp), or first to a long-lived chargino (\chargino) as shown in Fig.~\ref{fig:SMS_chargino}.
All possible decays are assumed to occur with equal probability.
Thus, the gluino branching fraction is 1/3
each for the decay to \lsp, \charginoplus, and \charginominus,
and the squark branching fraction is 1/2
to \lsp and 1/2
to the \chargino of opposite charge.
The \chargino and \lsp are assumed to be wino-like, and their masses
to differ by a few hundred~MeV~\cite{AMSB_1,AMSB_2}.
Thus, the phase space for the decay of the
\chargino
to a \lsp and a charged pion is small.
As a consequence, the \chargino has lifetime of the order of a few nanoseconds, and the momentum of the pion originating from its decay does not exceed a few hundred~MeV.
Hence, the final state shows negligible dependence on small variations of the mass difference between \chargino and \lsp.
Lifetimes of the \chargino are probed in the range $c\tau_{0}(\chargino) =$~1--2000\cm.

Uncertainties in the signal yield for the simplified models considered are listed in Table~\ref{tab:sig_systs}.
The sources of uncertainty and the methods used to evaluate their
effect on the interpretation are the same as those discussed in Refs.~\cite{MT2_2016,MT2at13TeV}.
For each data sample corresponding to the different periods of data taking (2016, 2017, and 2018),
uncertainties in the luminosity measurement~\cite{lumi2016,lumi2017,lumi2018}, ISR modeling, fast simulation \Met distributions, and \PQb tagging
and lepton efficiencies are treated as correlated across search bins.
Uncertainties in fast simulation \Met distributions, \PQb tagging, and lepton efficiencies are treated as correlated also across data samples.
The remaining uncertainties are taken as uncorrelated.
In the search for disappearing tracks, all other tagging and lepton efficiencies are neglected.
Other uncertainties associated with the modeling of disappearing tracks are treated as correlated across search bins.
Specifically, an uncertainty in the signal yield is assigned, equal to one half of the track selection inefficiency: 25 (17.5)\% for P (M and L) tracks in 2016,
and 10\% for tracks of all lengths in 2017--2018.
Additionally, a 6\% uncertainty in the 2017--2018 signal yield is assigned to account for inaccuracies in the fast simulation modeling of the signal acceptance.

\begin{table}[htb]
\topcaption{
Systematic uncertainties in the signal yields for the simplified models of BSM physics.
The large statistical uncertainties in the simulated signal sample come from a small number of bins with low acceptance,
which are typically not among the most sensitive bins contributing to a given model benchmark point.
    \label{tab:sig_systs}}
\centering
\begin{tabular}{lc}
    \hline
Source & Range [\%] \\
\hline
Integrated luminosity                     & 2.3--2.5     \\
Limited size of MC samples                & 1--100  \\
\PQb tagging efficiency, heavy flavors       & 0--40   \\
\PQb tagging efficiency, light flavors       & 0--20   \\
Lepton efficiency                         & 0--20   \\
Jet energy scale                          & 5       \\
Fast simulation \Met\ modeling            & 0--5     \\
ISR modeling                              & 0--30   \\
$\mu_{\mathrm{R}}$ and $\mu_{\mathrm{F}}$  & 5       \\
    \hline
\end{tabular}
\end{table}

\subsection{Inclusive \texorpdfstring{\mttwo}{MT2} search}
\label{sec:interpretation:mt2}

\begin{figure*}[htbp]
  \centering
    \includegraphics[width=0.48\textwidth]{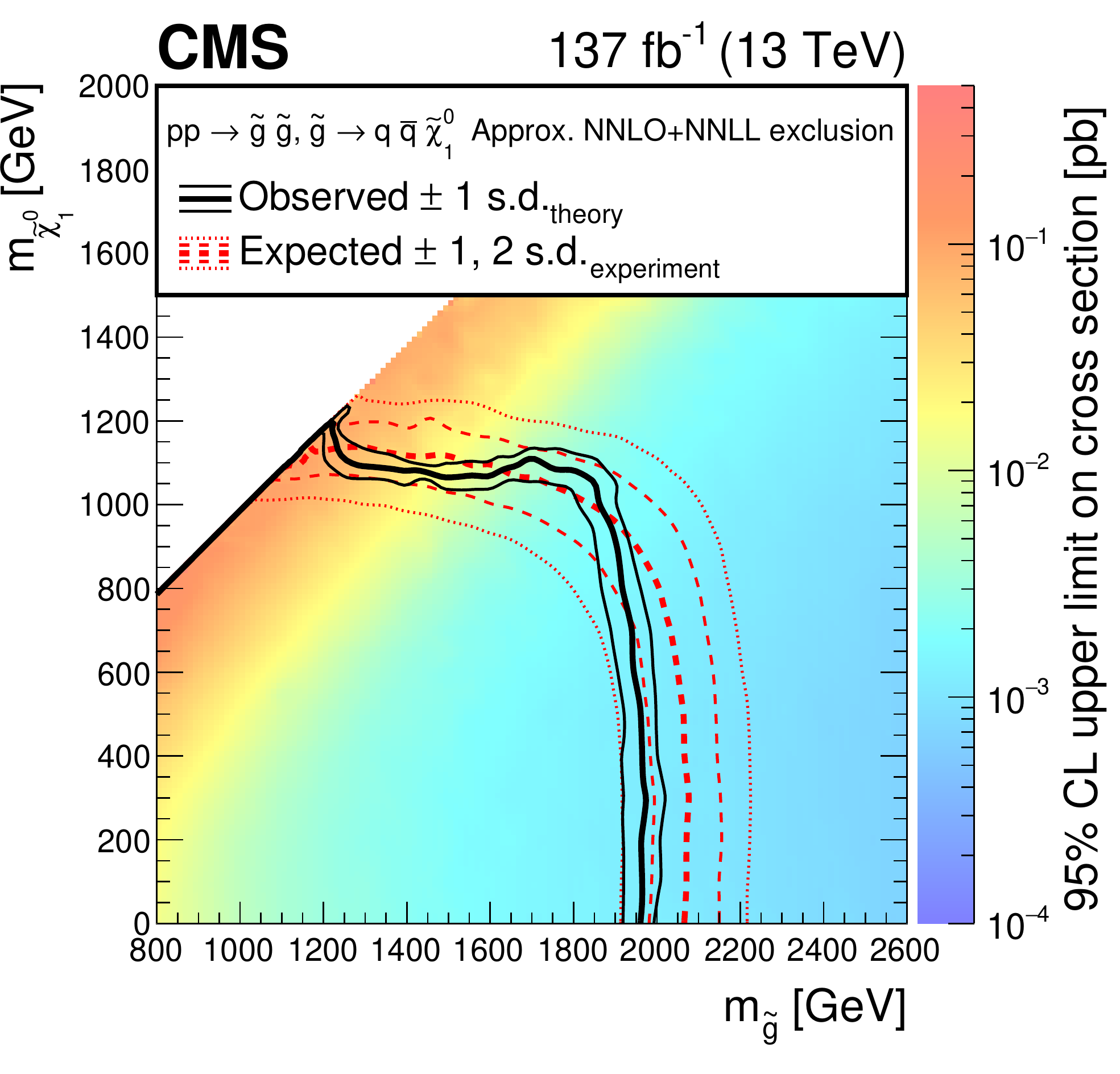}\\
    \includegraphics[width=0.48\textwidth]{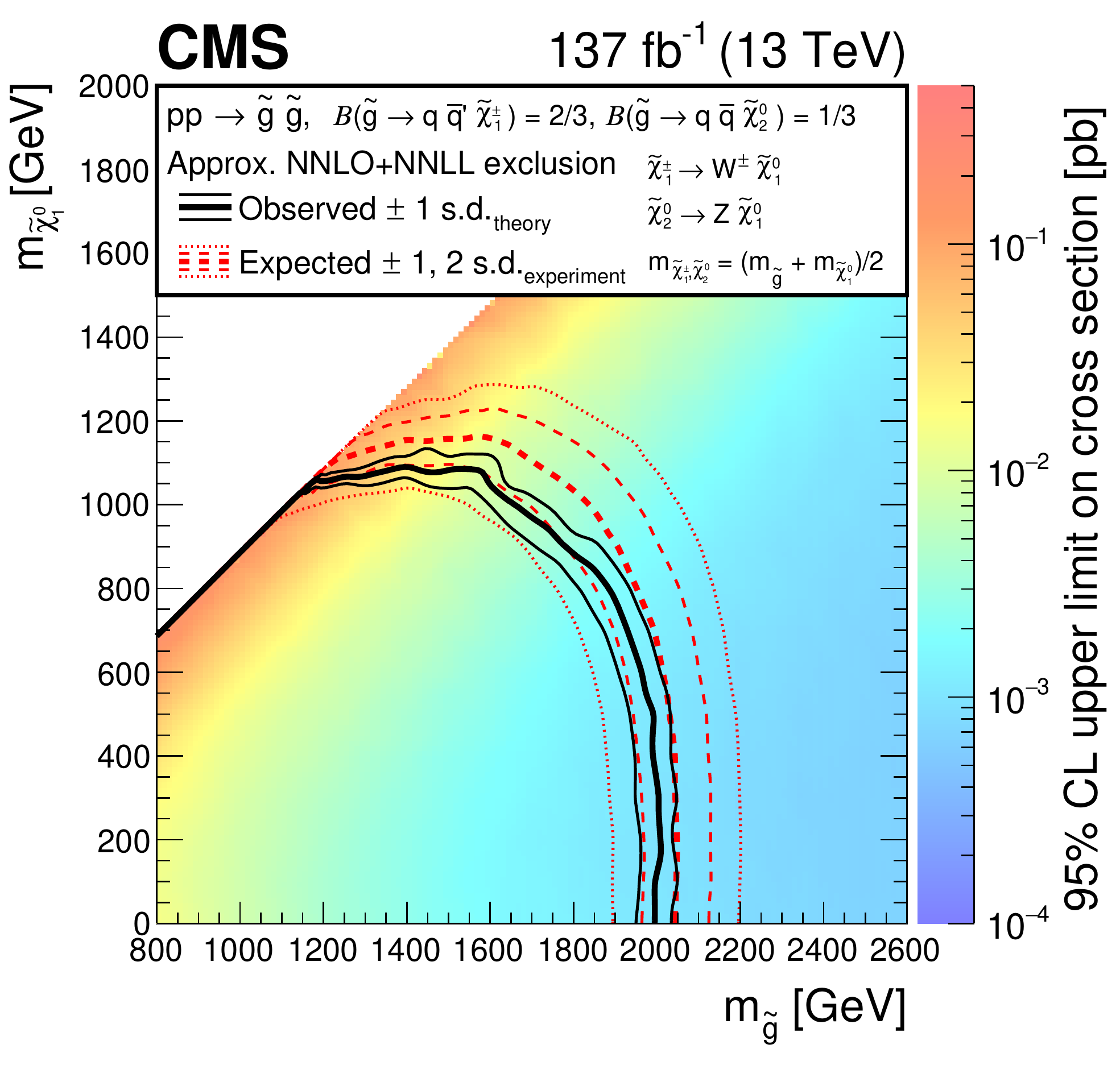}
    \includegraphics[width=0.48\textwidth]{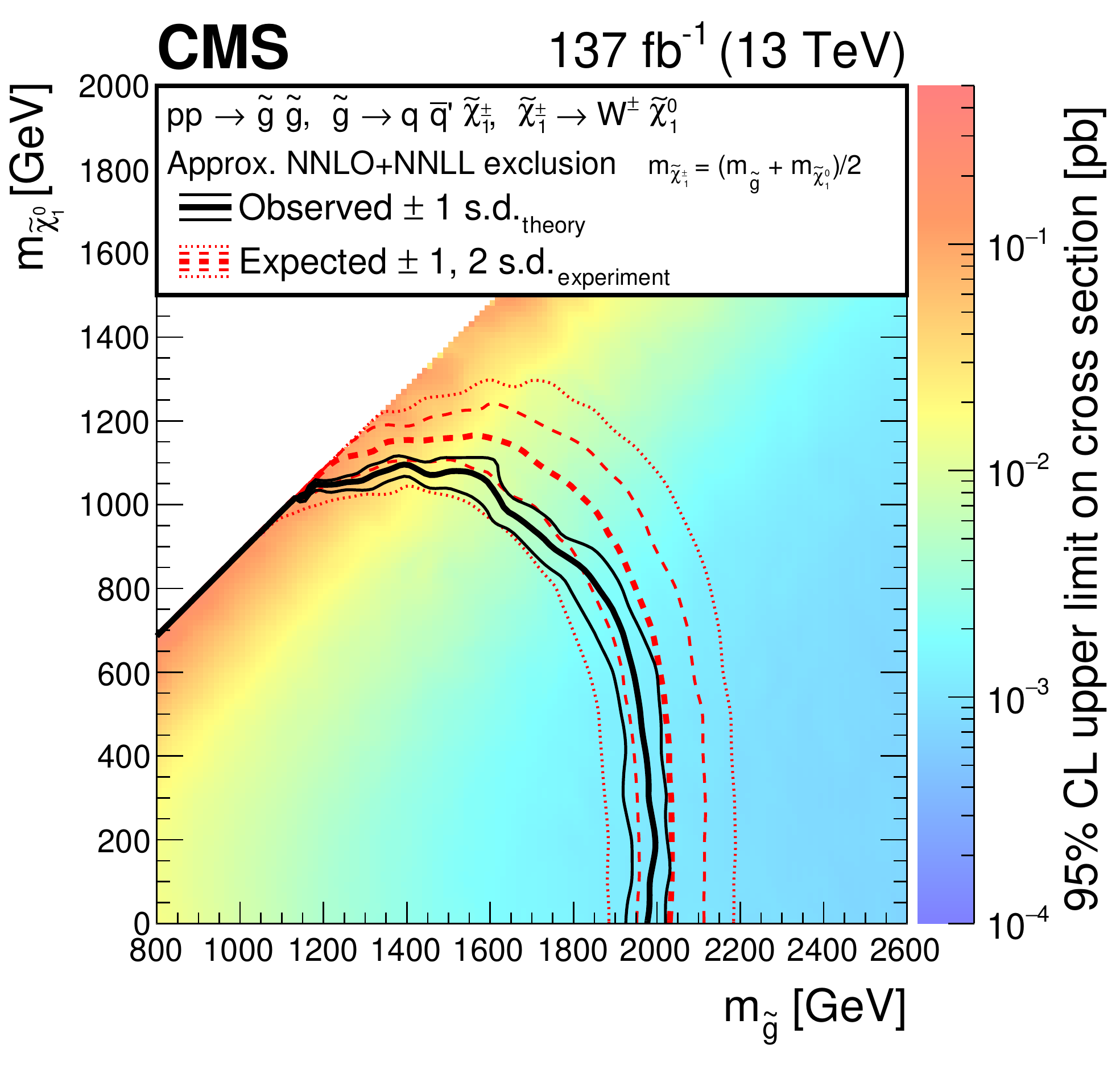}
    \caption{
    Exclusion limits at  95\% \CL for direct gluino pair production,
    where (upper) $\gluino \to \qqbar\lsp$, (lower left) $\gluino \to \qqbar\chizz$ and $\chizz \to \PZ\lsp$, or $\gluino \to \qqbarpr\chargino$ and $\chargino \to \PWpm\lsp$, and
    (lower right) $\gluino \to \qqbarpr\chargino$ and $\chargino \to \PWpm\lsp$ (with \cPq $=$ \cPqu, \cPqd, \cPqs, or \cPqc).
    For the scenarios where the gluinos decay via an intermediate \chizz or \chargino, \chizz and \chargino are assumed to be mass-degenerate, with
     $m_{\chargino,~\chizz}=0.5(m_{\gluino}+m_{\lsp})$.
      The area enclosed by the thick black curve represents the observed exclusion region,
      while the dashed red lines indicate the expected limits and
      their $\pm$1 and $\pm$2 standard deviation (s.d.) ranges.
      The thin black lines show the effect of the theoretical
      uncertainties in the signal cross section.
      Signal cross sections are calculated at approximately NNLO+NNLL order in \alpS~\cite{bib-nlo-nll-01,bib-nnll-05,bib-nlo-nll-02,bib-nlo-nll-03,bib-nlo-nll-04,bib-nnll-06,bib-nlo-nll-05,bib-nnll-02,bib-nnll-03,bib-nnll-04,bib-nnll-07,bib-nnll},
      assuming 1/3
      branching fraction ($\mathcal{B}$) for each decay mode in the mixed-decay scenarios, or unity branching fraction for the indicated decay.}
    \label{fig:t5x}
\end{figure*}

\begin{figure*}[htbp]
  \centering
    \includegraphics[width=0.48\textwidth]{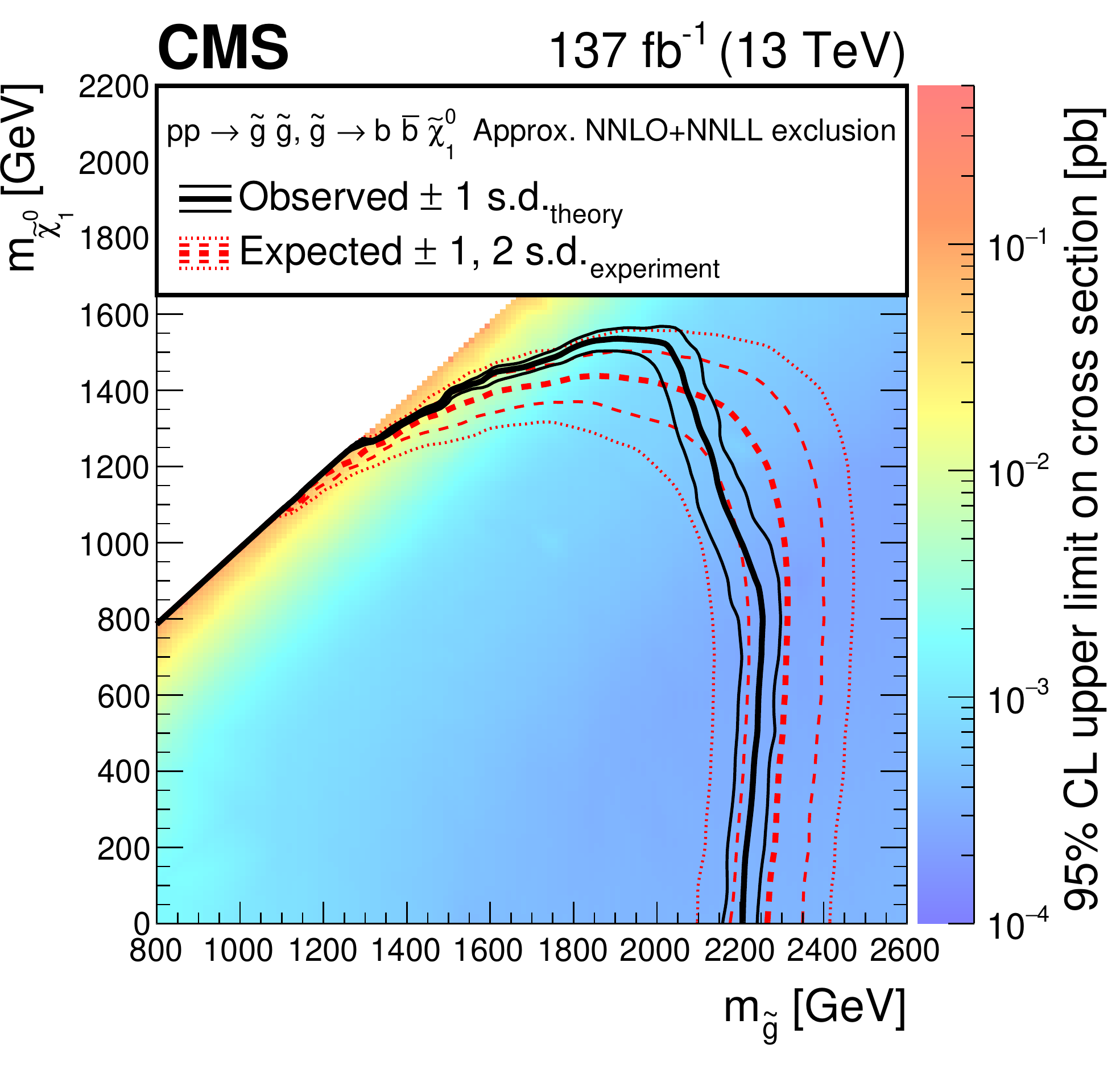}
    \includegraphics[width=0.48\textwidth]{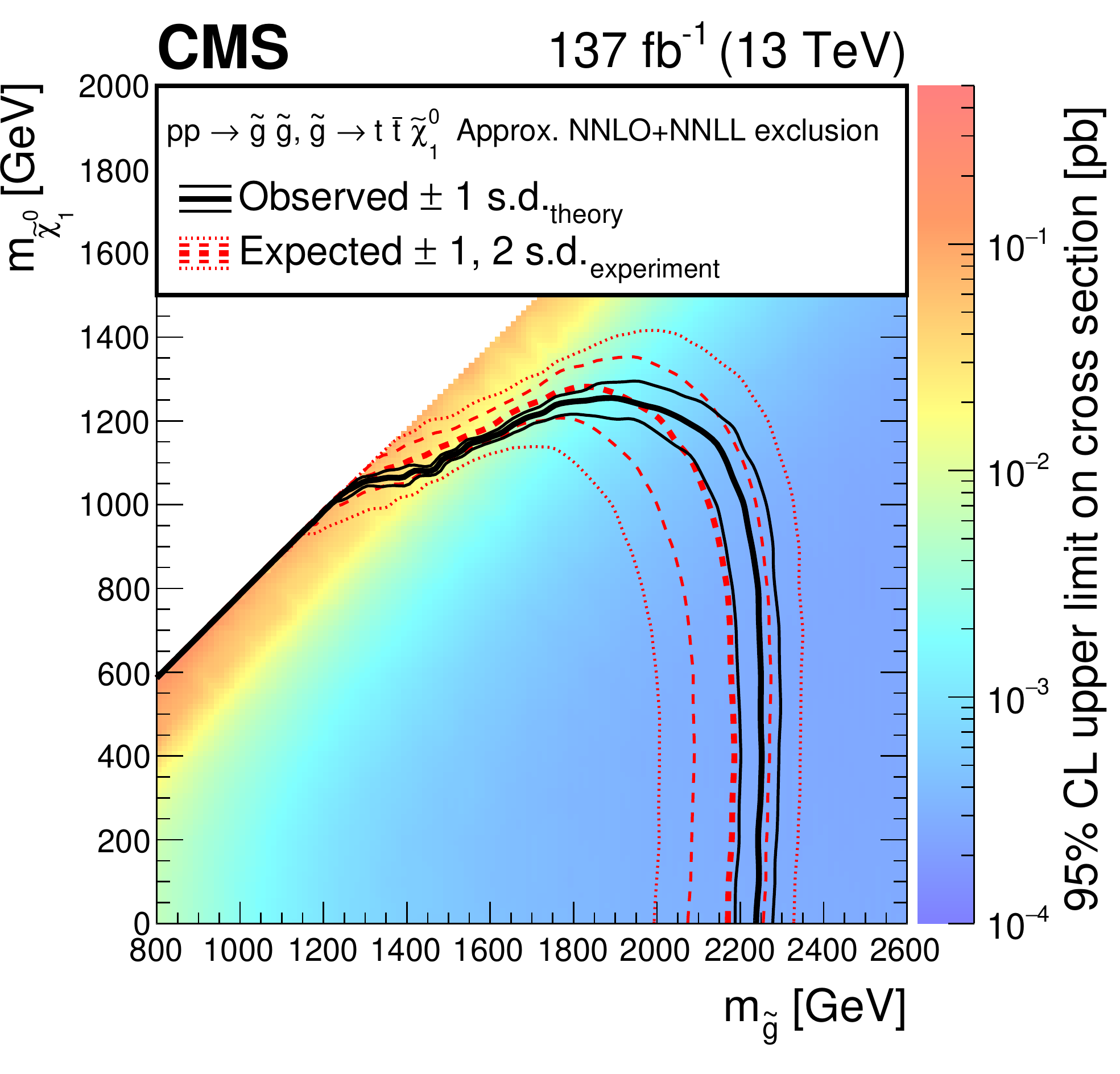}
    \caption{
    Exclusion limits at  95\% \CL for
     direct gluino pair production where the gluinos decay to (left) bottom quarks and (right) top quarks.
      The area enclosed by the thick black curve represents the observed exclusion region,
      while the dashed red lines indicate the expected limits and
      their $\pm$1 and $\pm$2
      standard deviation (s.d.) ranges.
      The thin black lines show the effect of the theoretical
      uncertainties in the signal cross section.
      Signal cross sections are calculated at approximately NNLO+NNLL order in \alpS~\cite{bib-nlo-nll-01,bib-nnll-05,bib-nlo-nll-02,bib-nlo-nll-03,bib-nlo-nll-04,bib-nnll-06,bib-nlo-nll-05,bib-nnll-02,bib-nnll-03,bib-nnll-04,bib-nnll-07,bib-nnll},
      assuming unity branching fraction for the indicated decay.}
    \label{fig:t1x}
\end{figure*}

\begin{figure*}[htbp!]
  \centering
    \includegraphics[width=0.48\textwidth]{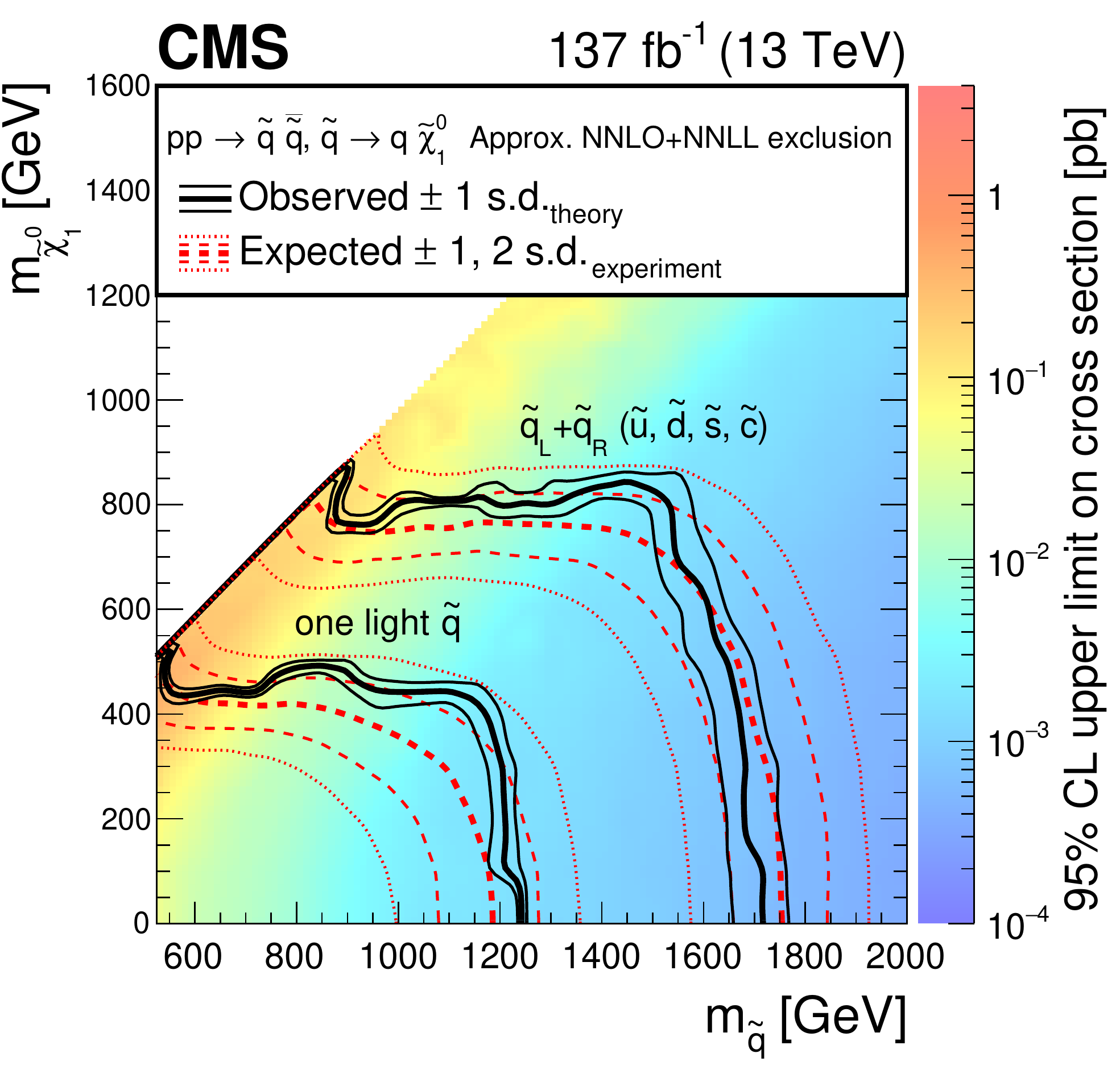}
    \includegraphics[width=0.48\textwidth]{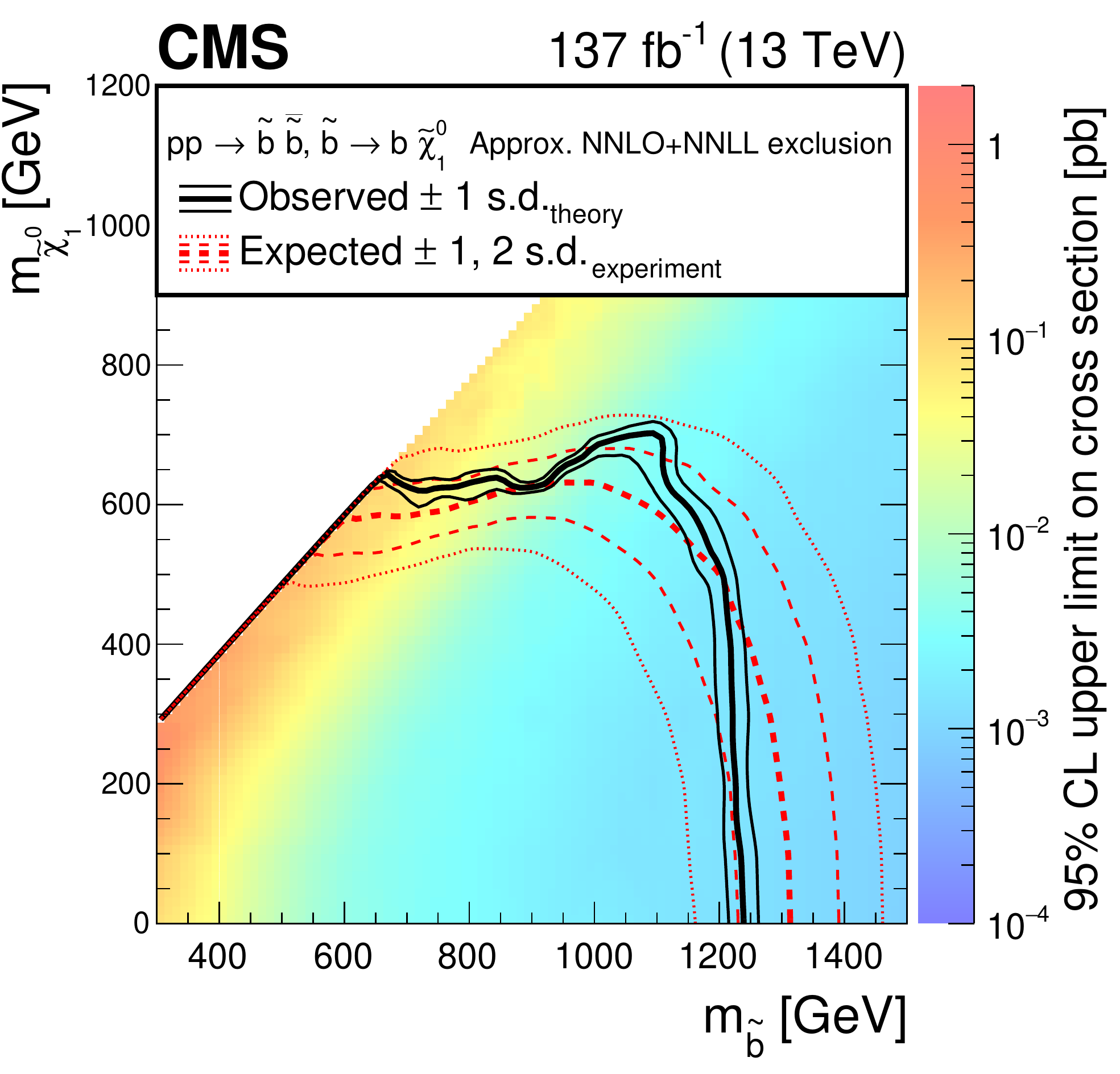}
    \includegraphics[width=0.48\textwidth]{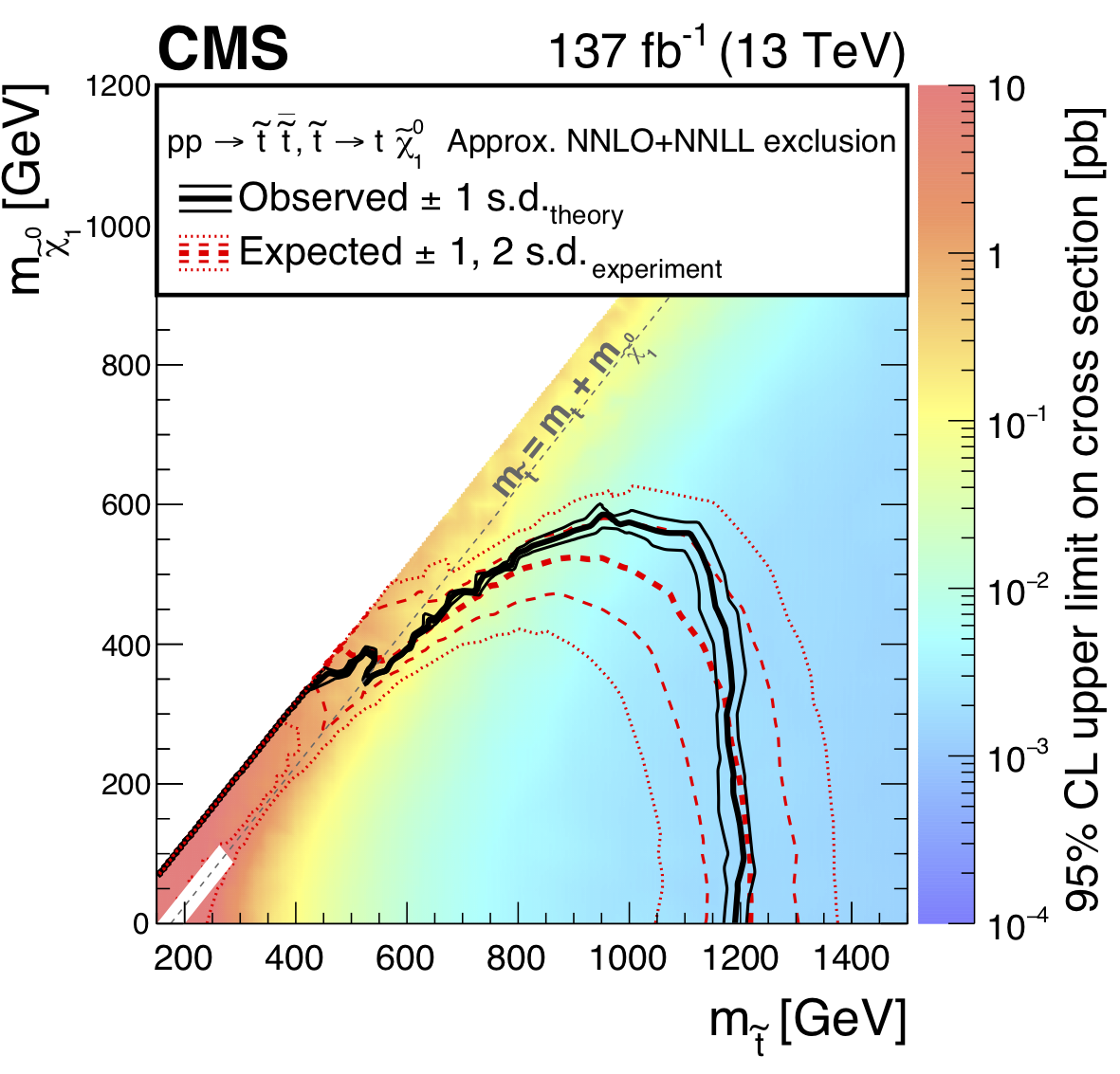}
    \caption{
    Exclusion limit at 95\% \CL for (upper left) light-flavor squark pair production, (upper right) bottom squark pair production,
    and (lower) top squark pair production.
      The area enclosed by the thick black curve represents the observed exclusion region,
      while the dashed red lines indicate the expected limits and
      their $\pm$1 and $\pm$2
      standard deviation (s.d.) ranges.
      The thin black lines show the effect of the theoretical
      uncertainties in the signal cross section.
      The white diagonal band in the top squark pair production exclusion limit corresponds to the region
      $\abs{m_{\PSQt}-m_{\PQt}-m_{\lsp}}< 25\GeV$ and small $m_{\lsp}$. Here the efficiency of the selection
      is a strong function of $m_{\PSQt}-m_{\lsp}$, and as a result the precise
      determination of the cross section upper limit is uncertain
      because of the finite granularity of the available
      MC samples in this region of the ($m_{\PSQt}, m_{\lsp}$)  plane. In the same exclusion limit, the dashed black diagonal line corresponds to $m_{\PSQt}=m_{\PQt}+m_{\lsp}$.
      Signal cross sections are calculated at approximately NNLO+NNLL order in \alpS~\cite{bib-nlo-nll-01,bib-nnll-05,bib-nlo-nll-02,bib-nlo-nll-03,bib-nlo-nll-04,bib-nnll-06,bib-nlo-nll-05,bib-nnll-02,bib-nnll-03,bib-nnll-04,bib-nnll-07,bib-nnll},
      assuming unity branching fraction for the indicated decay.}
    \label{fig:t2x}
\end{figure*}

\begin{figure*}[htbp!]
 \centering
   \includegraphics[width=0.49\textwidth]{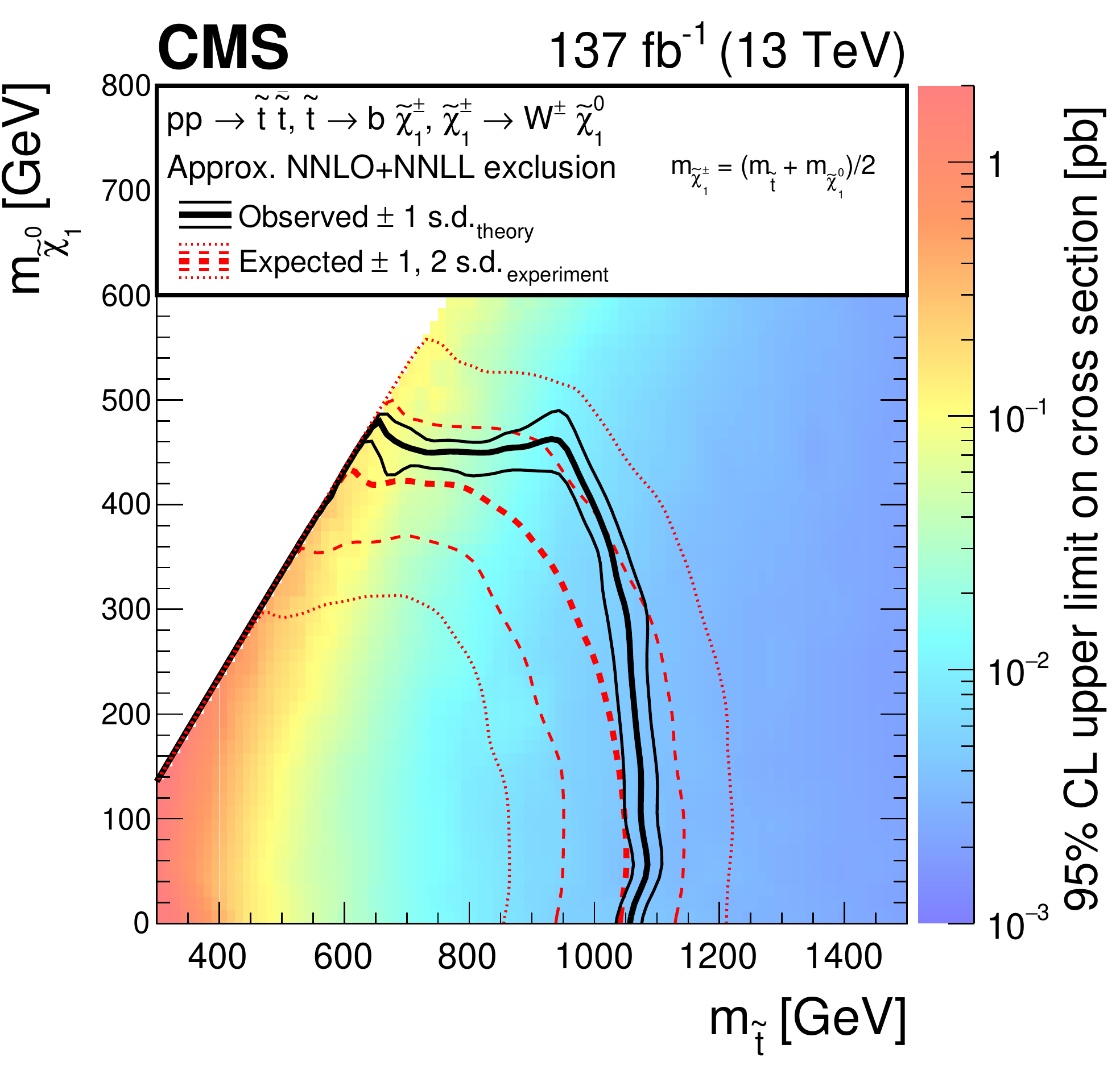}
   \includegraphics[width=0.49\textwidth]{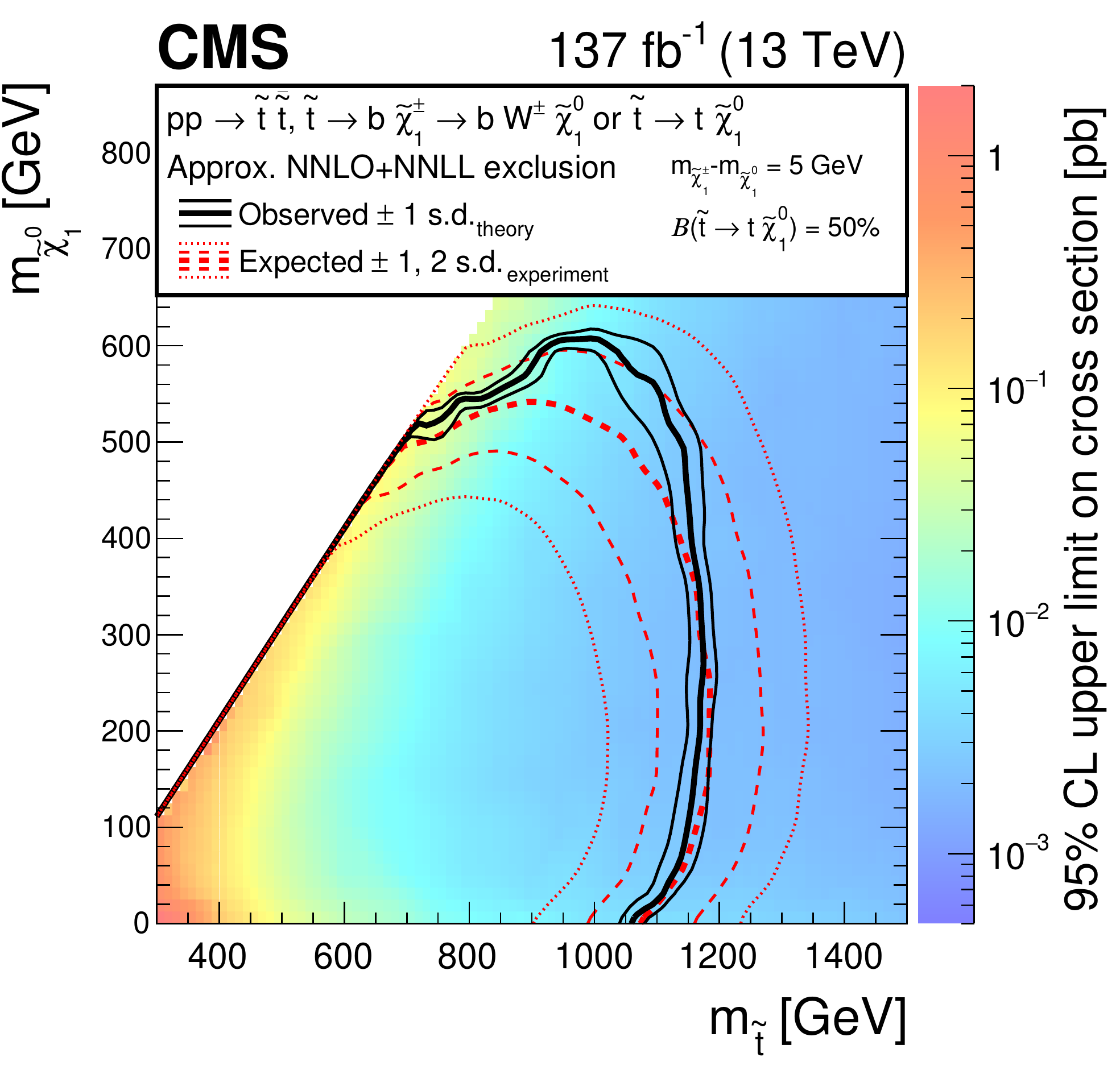}
   \includegraphics[width=0.49\textwidth]{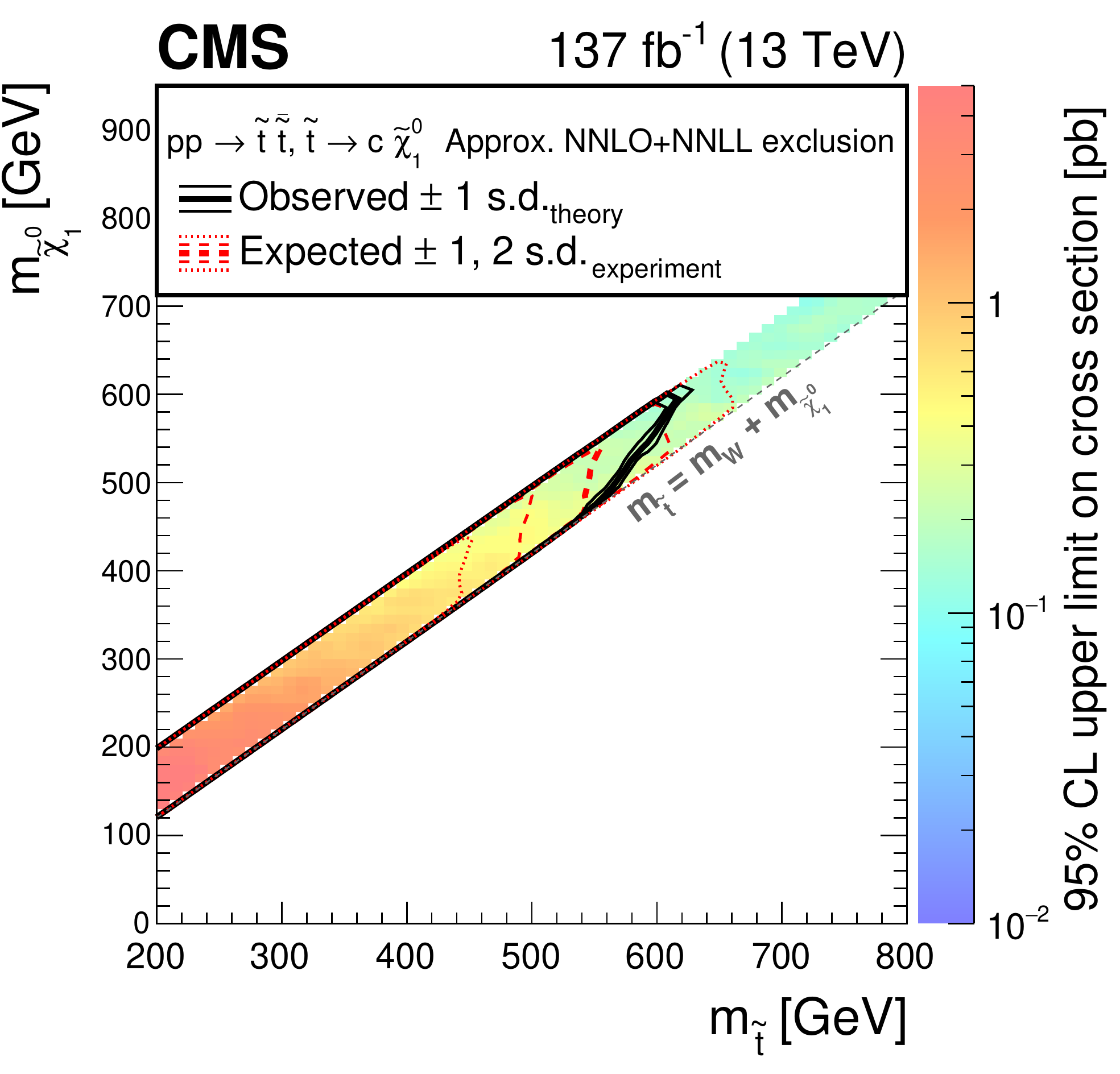}
   \caption{Exclusion limit at 95\% \CL for top squark pair production for different decay modes of the top squark.
     (Upper left) For the scenario where $\Pp\Pp\to\PSQt\astop\to
     \PQb\PAQb\PSGcpmDo\PSGcmpDo$, $\PSGcpmDo\to\PW^{\pm} \PSGczDo$,
     the mass of the chargino is chosen to be half way in between the masses of the top squark and the neutralino.
     (Upper right) A mixed-decay scenario, $\Pp\Pp\to\PSQt\astop$ with equal branching fractions for the top squark decays $\PSQt\to\PQt\PSGczDo$
     and $\PSQt\to\PQb\PSGcpDo$, $\PSGcpDo\to\PW^{*+}\PSGczDo$,
     is also considered, with the chargino mass chosen such that
     $\Delta m\left(\PSGcpmDo,\PSGczDo\right) = 5\GeV$.
     (Lower) Finally, we also consider
     a compressed spectrum scenario where
     $\Pp\Pp\to\PSQt\astop\to
     \PQc\PAQc\PSGczDo\PSGczDo$. In this scenario, mass ranges are considered where the $\PSQt\to\PQc\PSGczDo$ branching fraction can be significant.
     The area enclosed by the thick black curve represents the observed exclusion region,
     while the dashed red lines indicate the expected limits and
     their $\pm$1  and $\pm$2
     standard deviation (s.d.) ranges.
     The thin black lines show the effect of the theoretical
     uncertainties in the signal cross section.
     Signal cross sections are calculated at approximately NNLO+NNLL order
     in \alpS~\cite{bib-nlo-nll-01,bib-nnll-05,bib-nlo-nll-02,bib-nlo-nll-03,bib-nlo-nll-04,bib-nnll-06,bib-nlo-nll-05,bib-nnll-02,bib-nnll-03,bib-nnll-04,bib-nnll-07,bib-nnll},
     assuming 50\% branching fraction ($\mathcal{B}$) for each decay mode in the mixed-decay scenarios, or unity branching fraction for the indicated decay.}
   \label{fig:stop_other}
\end{figure*}

Figure~\ref{fig:t5x} shows the exclusion limits at 95\% \CL for
direct gluino pair production where the gluinos decay to light-flavor quarks under three different decay scenarios.
Exclusion limits for
direct gluino pair production where the gluinos decay to bottom and top quarks are
shown in Fig.~\ref{fig:t1x}, and those for the direct production of
squark pairs are shown in Fig.~\ref{fig:t2x}.
Three alternate decay scenarios
are also considered for the direct pair production of top squarks, and their
exclusion limits
are shown in Fig.~\ref{fig:stop_other}.

Table~\ref{tab:lim} summarizes the limits on the masses of SUSY particles excluded for the
simplified model scenarios considered.
These results extend the constraints on gluino and squark masses by about 100--350\GeV and on the \lsp mass by 100--250\GeV with respect to the limits in Ref.~\cite{MT2_2016}.
\begin{table*}[htb]
  \topcaption{Summary of the observed 95\% \CL exclusion limits on the masses of SUSY particles for different simplified model scenarios.
    The highest limits on the mass of the directly produced particles and on the mass of the \lsp are quoted.
    \label{tab:lim}}
\centering
\begin{tabular}{lrr}
    \hline
Simplified & Highest limit on directly produced  & Highest limit on \\
model & SUSY particle mass [\GeVns{}] & \lsp mass [\GeVns{}] \\
\hline
Direct gluino pair production: & & \\
$\gluino\to \qqbar\lsp$ & 1970 & 1200 \\
$\gluino \to \qqbar\PZ\lsp$ or $\gluino \to \qqbarpr\PWpm\lsp$ & 2020 & 1090 \\
$\gluino\to \bbbar\lsp$ & 2250 & 1525 \\
$\gluino\to \ttbar\lsp$ & 2250 & 1250 \\
[\cmsTabSkip]
Direct squark pair production: & & \\
Eight degenerate light squarks & 1710 & 870 \\
Single light squark & 1250 & 525 \\
Bottom squark & 1240 & 700 \\
Top squark & 1200 & 580 \\
    \hline
\end{tabular}
\end{table*}

\begin{figure}[htbp]
 \centering
   \includegraphics[width=0.49\textwidth]{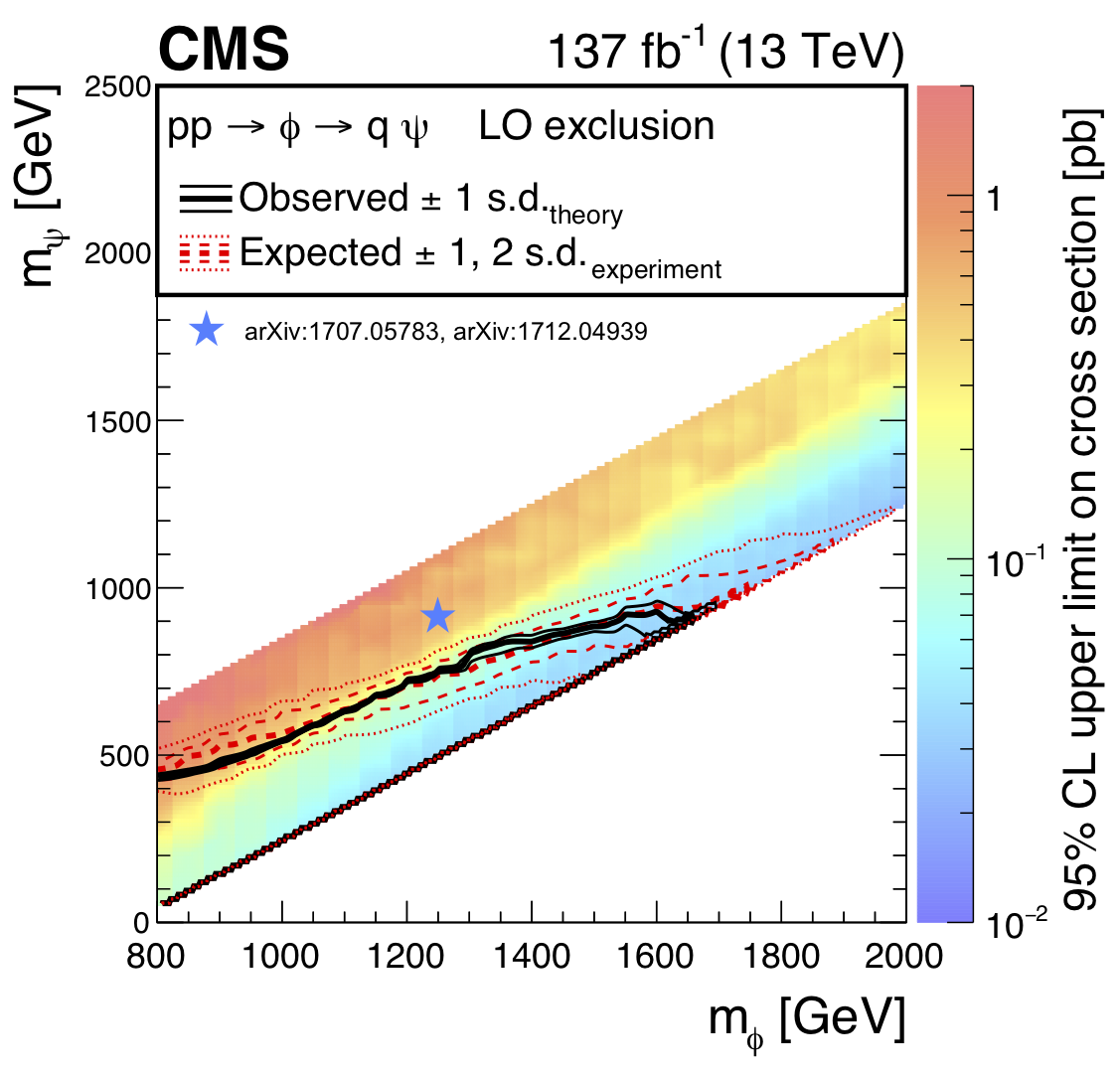}
   \caption{Exclusion limit at 95\% \CL for the mono-$\phi$ model.
     We consider the mass range where such a model could be interesting based on a
     reinterpretation of previous analyses~\cite{monojetexcess1,monojetexcess2}.
     The area enclosed by the thick black curve represents the observed exclusion region,
     while the dashed red lines indicate the expected limits and
     their $\pm$1 and $\pm$2 standard deviation (s.d.) ranges.
     The thin black lines show the effect of the theoretical
     uncertainties in the signal cross section.
     The blue star at $\left(m_{\phi},~m_{\psi}\right)=\left(1250,~900\right)\GeV$ indicates the best fit mass point reported in Refs.~\cite{monojetexcess1,monojetexcess2}.
     Signal cross sections are calculated at LO order
     in \alpS.}
   \label{fig:monophi}
\end{figure}

Figure~\ref{fig:monophi} shows the exclusion limits
for the mono-$\phi$ model~\cite{monojetexcess1,monojetexcess2}.
Based on the LO cross section calculation, we obtain mass limits as large as 1660 and 925\GeV on $m_{\phi}$ and on $m_{\psi}$, respectively.
In this model, the analysis of Refs.~\cite{monojetexcess1,monojetexcess2} reports best fit parameters $\left(m_{\phi},~m_{\psi}\right)=\left(1250,~900\right)\GeV$ and
product of the cross section and branching fraction of about 0.3\unit{pb}.
For this mass point, we find a modest (1.1 standard deviations) excess, and we set an upper limit on the product of the cross section and branching fraction
of about 0.6 (0.4 expected)\unit{pb},
equal to 4.7 (3.2) times the assumed LO theoretical cross section.

\begin{figure*}[htbp]
 \centering
   \includegraphics[width=0.49\textwidth]{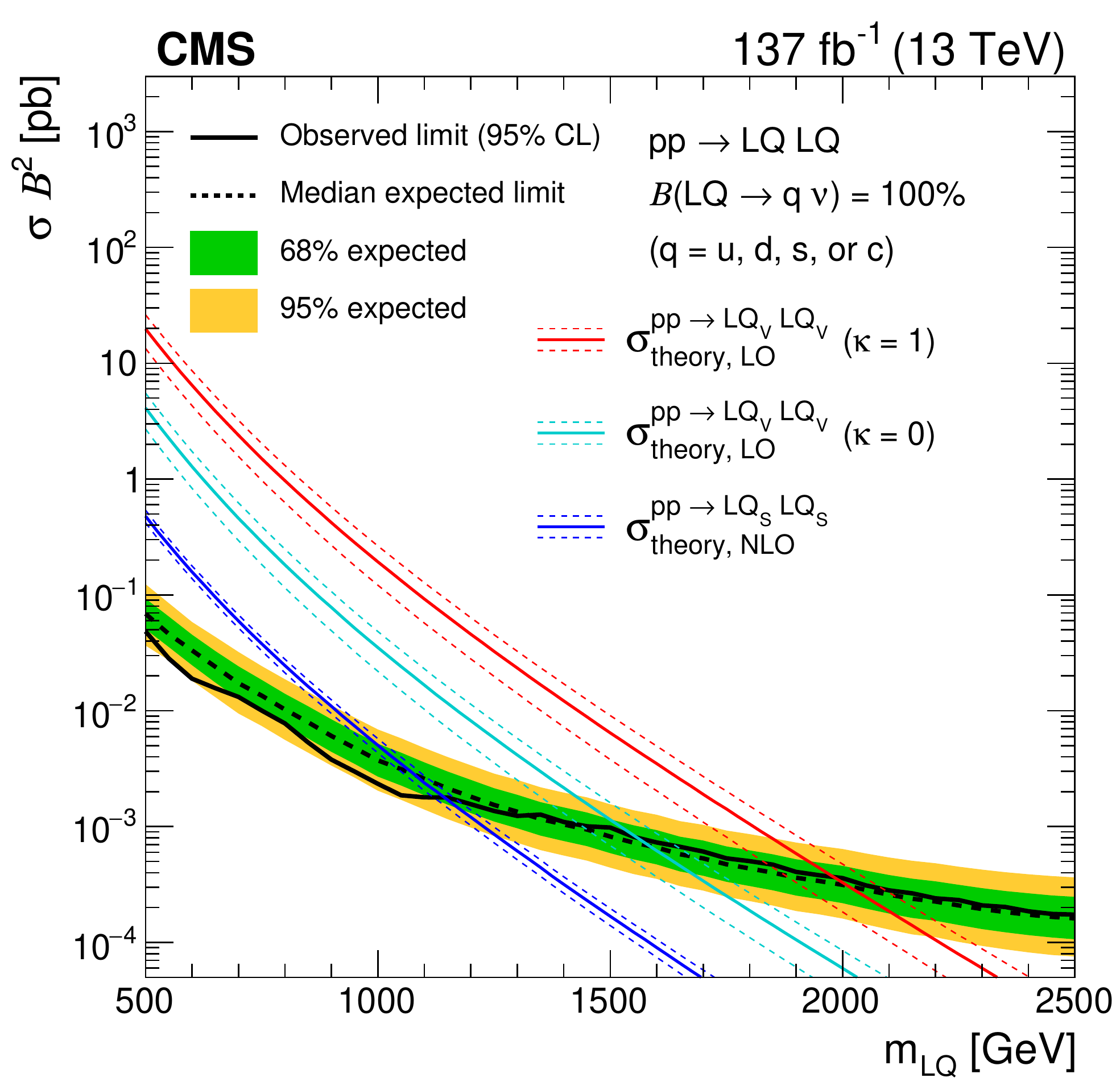}
   \includegraphics[width=0.49\textwidth]{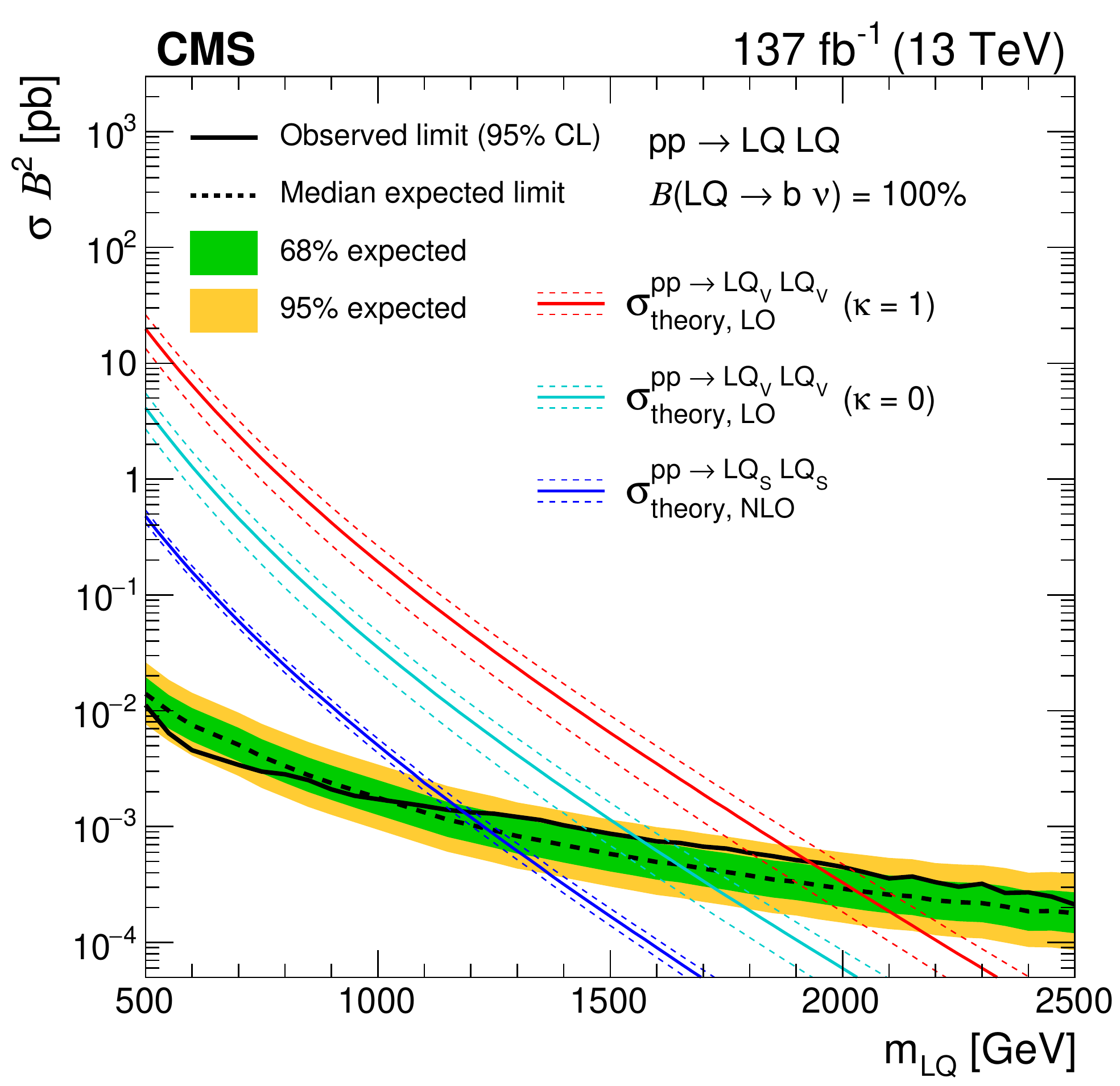}
   \includegraphics[width=0.49\textwidth]{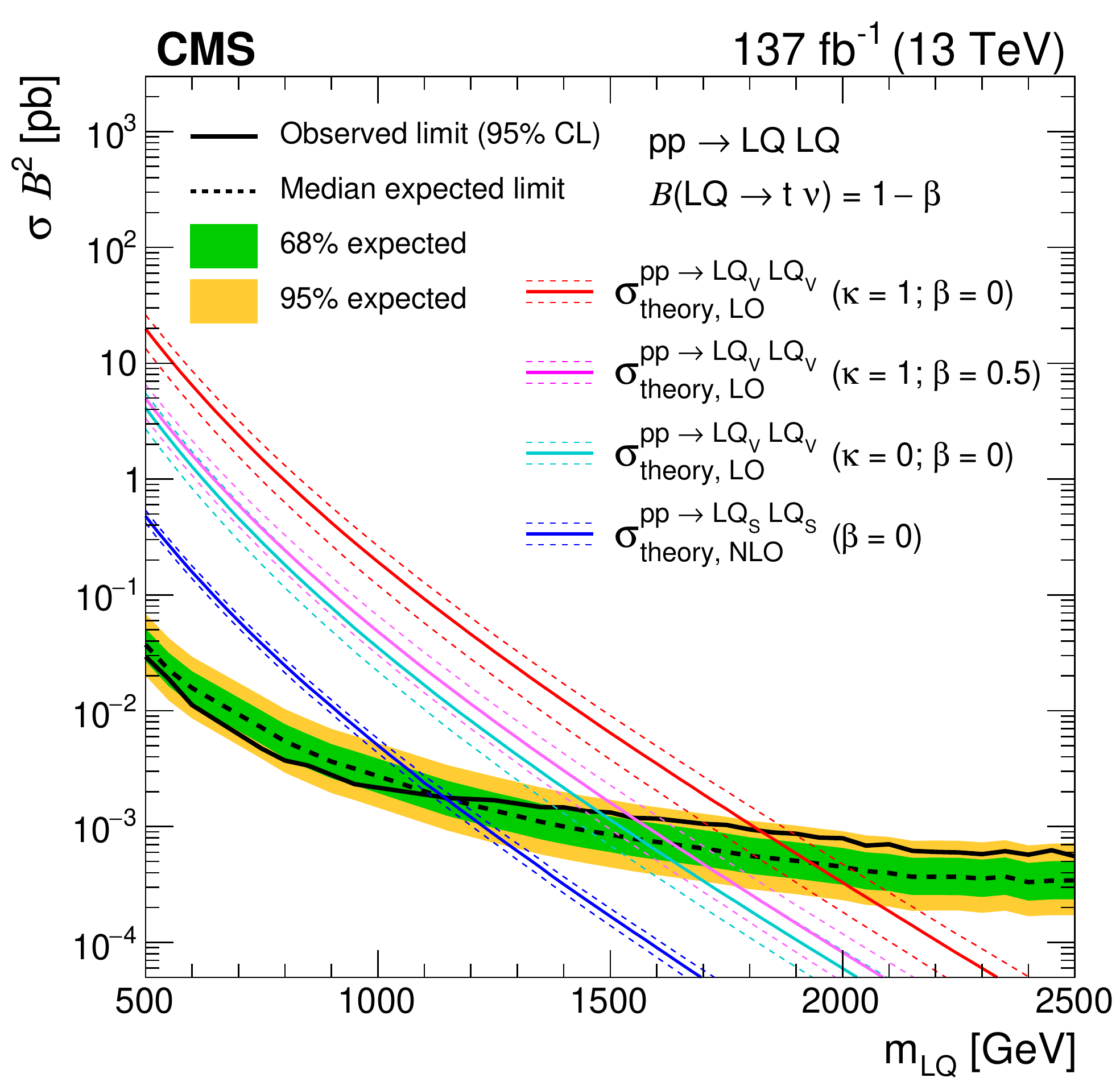}
   \caption{The 95\% \CL upper limits on the production cross sections as a function of LQ mass
for LQ pair production decaying with 100\% branching fraction ($\mathcal{B}$) to a neutrino and (upper left) a light quark
(one of \PQu, \PQd, \PQs, or \PQc),
(upper right) a bottom quark, or (lower) a top quark.
The solid (dashed) black line represents the observed (median expected) exclusion.
The inner green (outer yellow) band indicates the region containing 68 (95)\%
of the distribution of limits expected under the background-only hypothesis.
The dark blue lines show the theoretical cross section for \lqs pair production with its uncertainty.
The red (light blue) lines show the same for \lqv\ pair production assuming $\kappa = 1$ (0).
(Lower) Also shown in magenta is the product of the theoretical cross section and the square of the branching fraction ($\sigma \mathcal{B}^{2}$),
for vector LQ pair production assuming $\kappa = 1$ and a 50\% branching fraction to $\PQt\Pgngt$, with the remaining 50\% to $\PQb\PGt$.
Signal cross sections are calculated at NLO (LO) in \alpS for scalar (vector) LQ pair production.}
   \label{fig:lq}
\end{figure*}

{\tolerance=1200
The LQ limits from the \mttwo search are
shown in Fig.~\ref{fig:lq},
where only one LQ state is assumed to be within reach of the LHC, and where each LQ is assumed to decay to a neutrino and a single type of quark.
\par}
In Refs.~\cite{lqmodel,lqtoolbox}, a model is proposed as a coherent explanation of the flavor physics anomalies.
It is based on an $\mathrm{LQ_{V}}$ that can decay to
$\PQt\PGn$ and to $\PQb\PGt$ final states, each with 50\% branching fraction.
In our analysis, events are selected with a charged-lepton veto,
including hadronically decaying \PGt leptons.
Hence, only the 25\% of events where both LQs decay to $\PQt\PGn$ are considered to set constraints on this model,
and the theoretical prediction for this branching fraction
is shown as a separate curve in Fig.~\ref{fig:lq}~(lower).

Table~\ref{tab:lim_lq} summarizes the limits on the masses of the LQs excluded for the considered scenarios.
These results extend the constraints on LQ masses by up to about 200\GeV with respect to the limits of Ref.~\cite{MT2_2016LQ},
providing the most stringent constraint to date in models of LQ pair production.
\begin{table*}[htb]
  \topcaption{Summary of the observed 95\% \CL exclusion limits on the masses of LQs for the considered scenarios. The columns show scalar or vector LQ with the choice of $\kappa$,
while the rows show the LQ decay channel. For mixed-decay scenarios, the assumed branching fractions ($\mathcal{B}$) are indicated.
    \label{tab:lim_lq}}
\centering
\begin{tabular}{l c c c}
    \hline
 & $\mathrm{LQ_{S}}$  & $\mathrm{LQ_{V}},~\kappa=1$ & $\mathrm{LQ_{V}},~\kappa=0$ \\
 & mass [\GeVns{}] & mass [\GeVns{}] & mass [\GeVns{}] \\
\hline
$\mathrm{LQ}\to \PQq\PGn$ (\cPq $=$ \cPqu, \cPqd, \cPqs, or \cPqc) & 1140 & 1980 & 1560\\
$\mathrm{LQ}\to \PQb\PGn$ & 1185 & 1925 & 1560 \\
$\mathrm{LQ}\to \PQt\PGn$ & 1140 & 1825 & 1475\\
$\mathrm{LQ}\to\left\{
\begin{tabular}{l}
$\PQt\PGn~\left(\mathcal{B}=50\%\right)$\\
$\PQb\PGt~\left(\mathcal{B}=50\%\right)$
\end{tabular}\right.$
& --- & 1550 & 1225 \\
    \hline
\end{tabular}
\end{table*}

The 95\% \CL upper limits on signal cross sections obtained using the
most sensitive super signal regions of Table~\ref{tab:ssr_def_yields}
are typically less stringent by a factor of $\sim$1.5--3 compared to
those obtained in the fully binned analysis.
This difference in performance arises from the larger signal acceptance of the full analysis, as well as from the more favorable signal-to-background ratio achieved in
its individual bins, compared to the super signal regions.

\subsection{Search for disappearing tracks}
\label{sec:interpretation:distracks}

\begin{figure*}[htbp]
  \centering
    \includegraphics[width=0.48\textwidth]{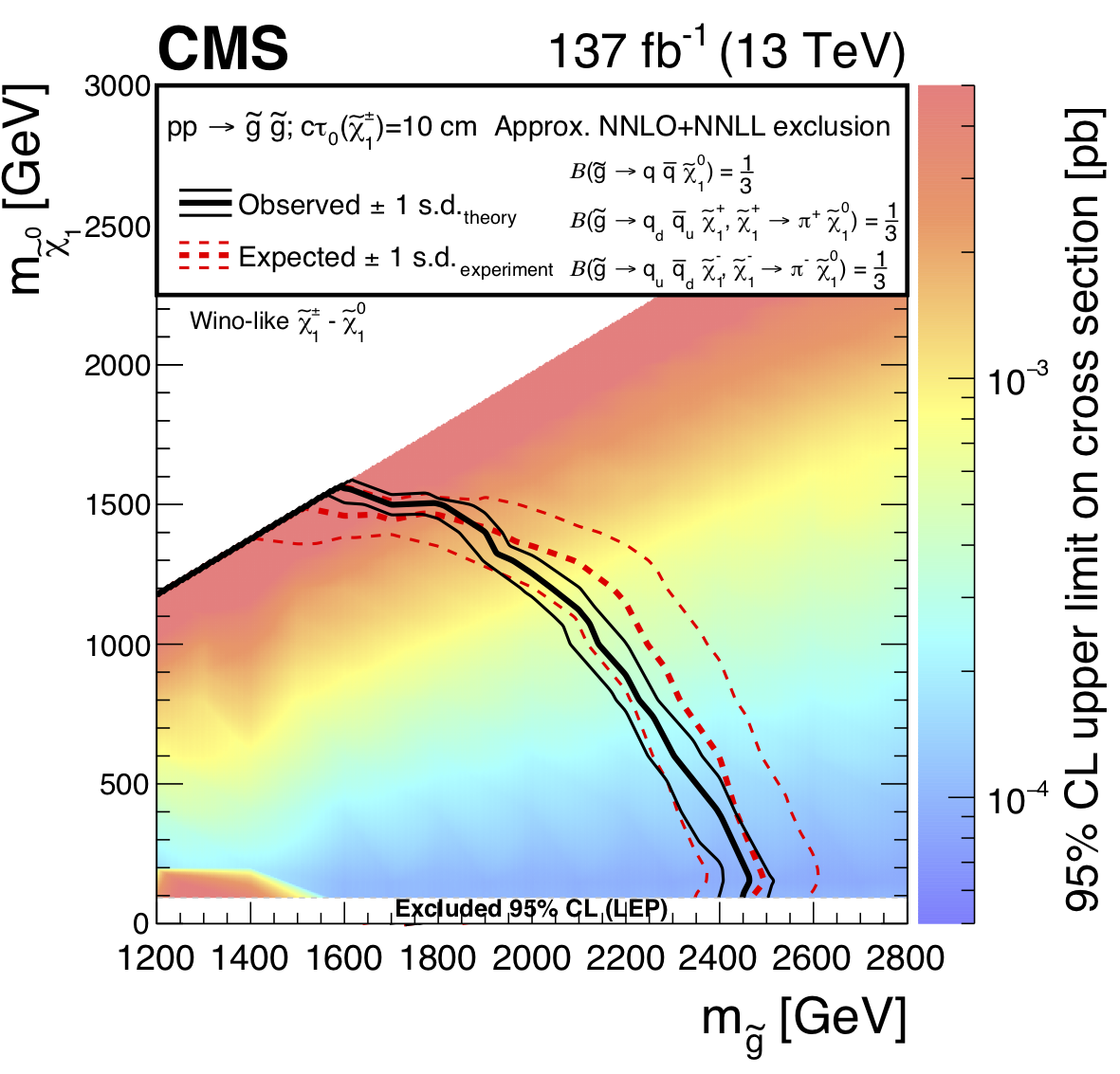}
    \includegraphics[width=0.48\textwidth]{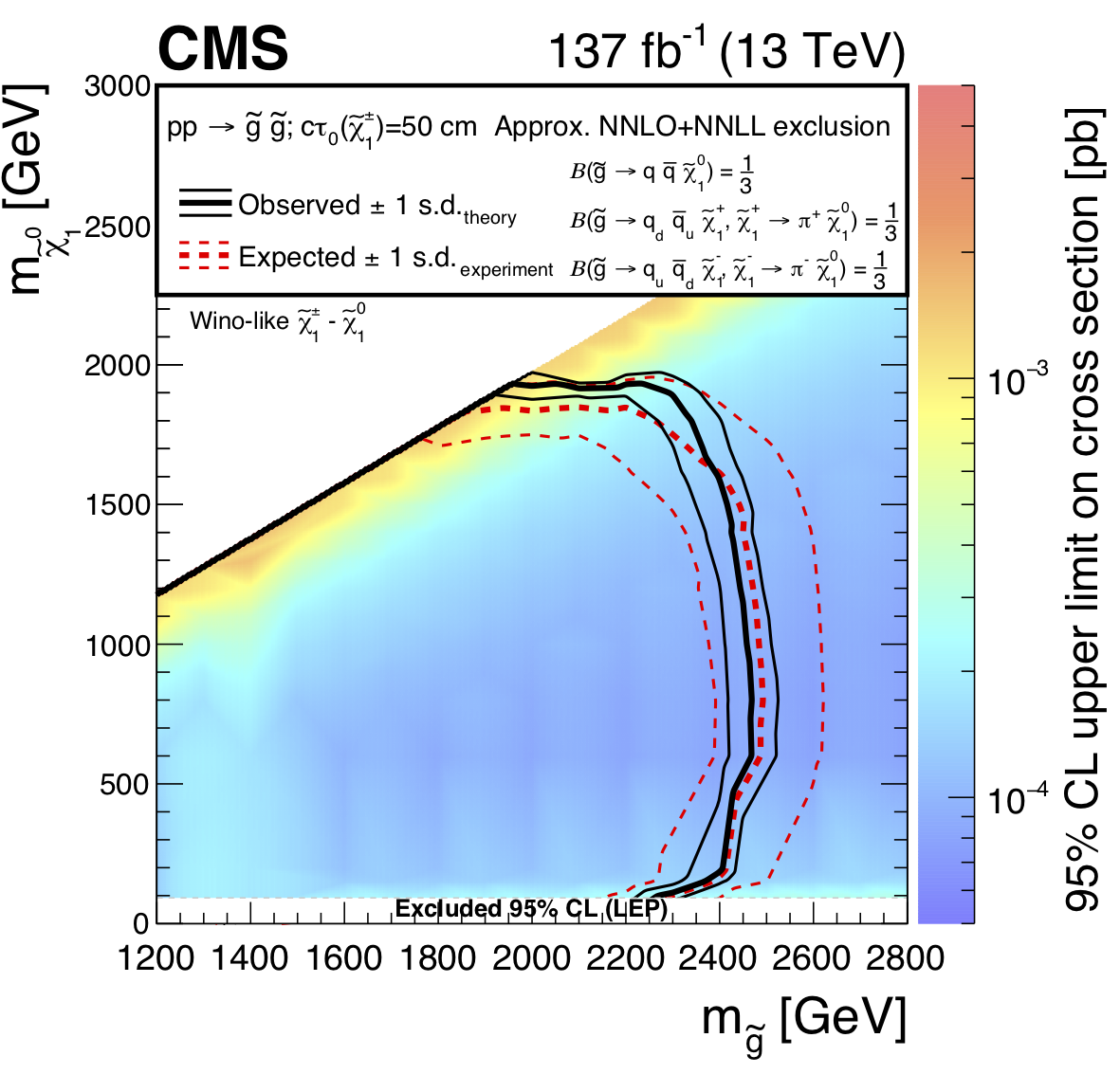}
    \includegraphics[width=0.48\textwidth]{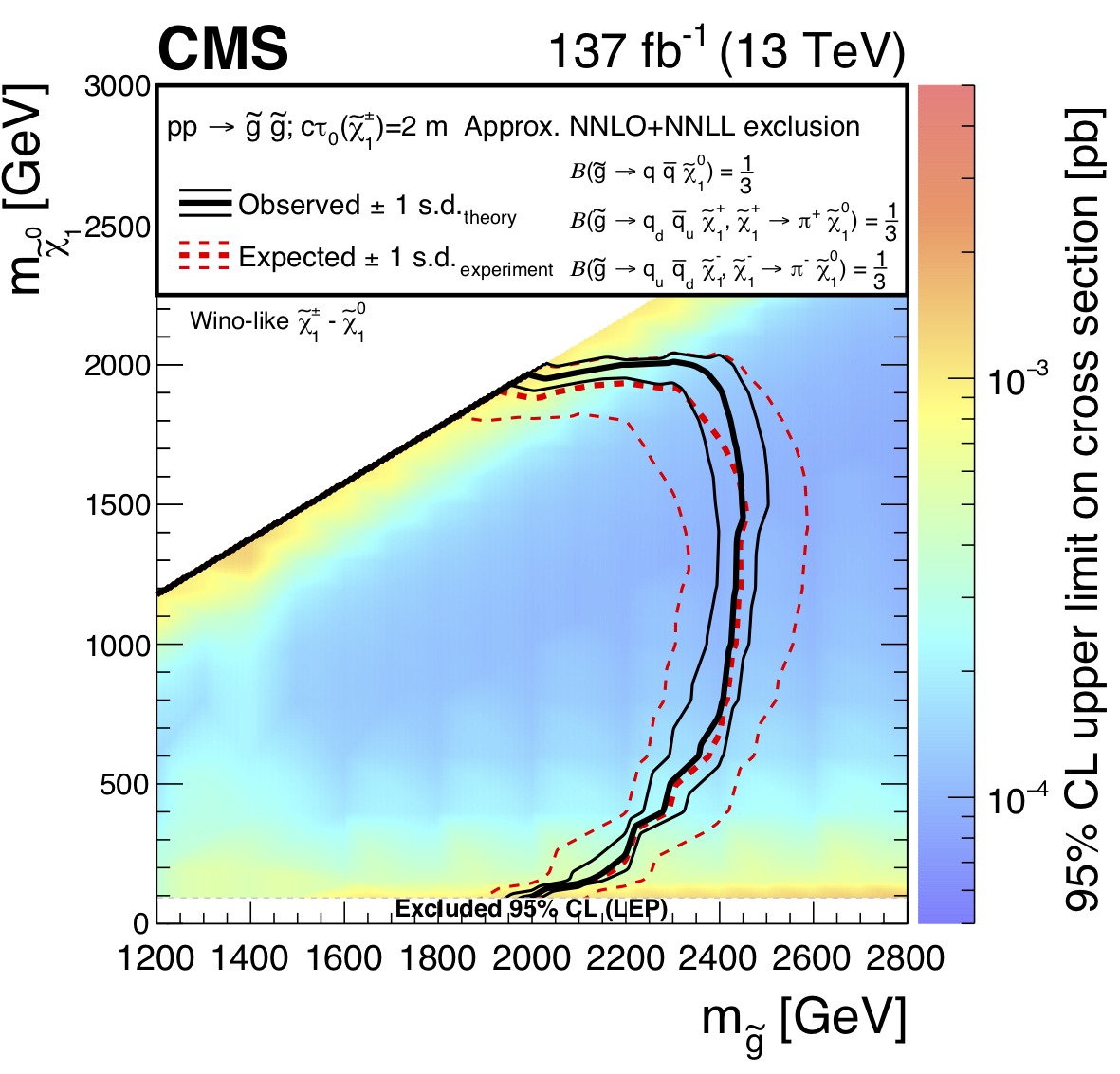}
    \caption{
    Exclusion limits at  95\% \CL for direct gluino pair production where the gluinos decay to light-flavor (\cPqu, \cPqd, \cPqs, \cPqc) quarks, with $c\tau_{0}(\chargino) =$ (upper left) 10\cm,
    (upper right) 50\cm, and (lower) 200\cm.
      The area enclosed by the thick black curve represents the observed exclusion region,
      while the dashed red lines indicate the expected limits and
      their $\pm$1 standard deviation (s.d.) ranges.
      The thin black lines show the effect of the theoretical
      uncertainties in the signal cross section.
      The white band for masses of the \lsp below 91.9\GeV represents the region of the mass plane excluded at the CERN LEP~\cite{lep_chargino}.
      Signal cross sections are calculated at approximately NNLO+NNLL order in \alpS~\cite{bib-nlo-nll-01,bib-nnll-05,bib-nlo-nll-02,bib-nlo-nll-03,bib-nlo-nll-04,bib-nnll-06,bib-nlo-nll-05,bib-nnll-02,bib-nnll-03,bib-nnll-04,bib-nnll-07,bib-nnll},
      assuming decay branching fractions ($\mathcal{B}$) as indicated in the figure.}
    \label{fig:t1st_qqqq}
\end{figure*}

\begin{figure*}[htbp]
  \centering
    \includegraphics[width=0.48\textwidth]{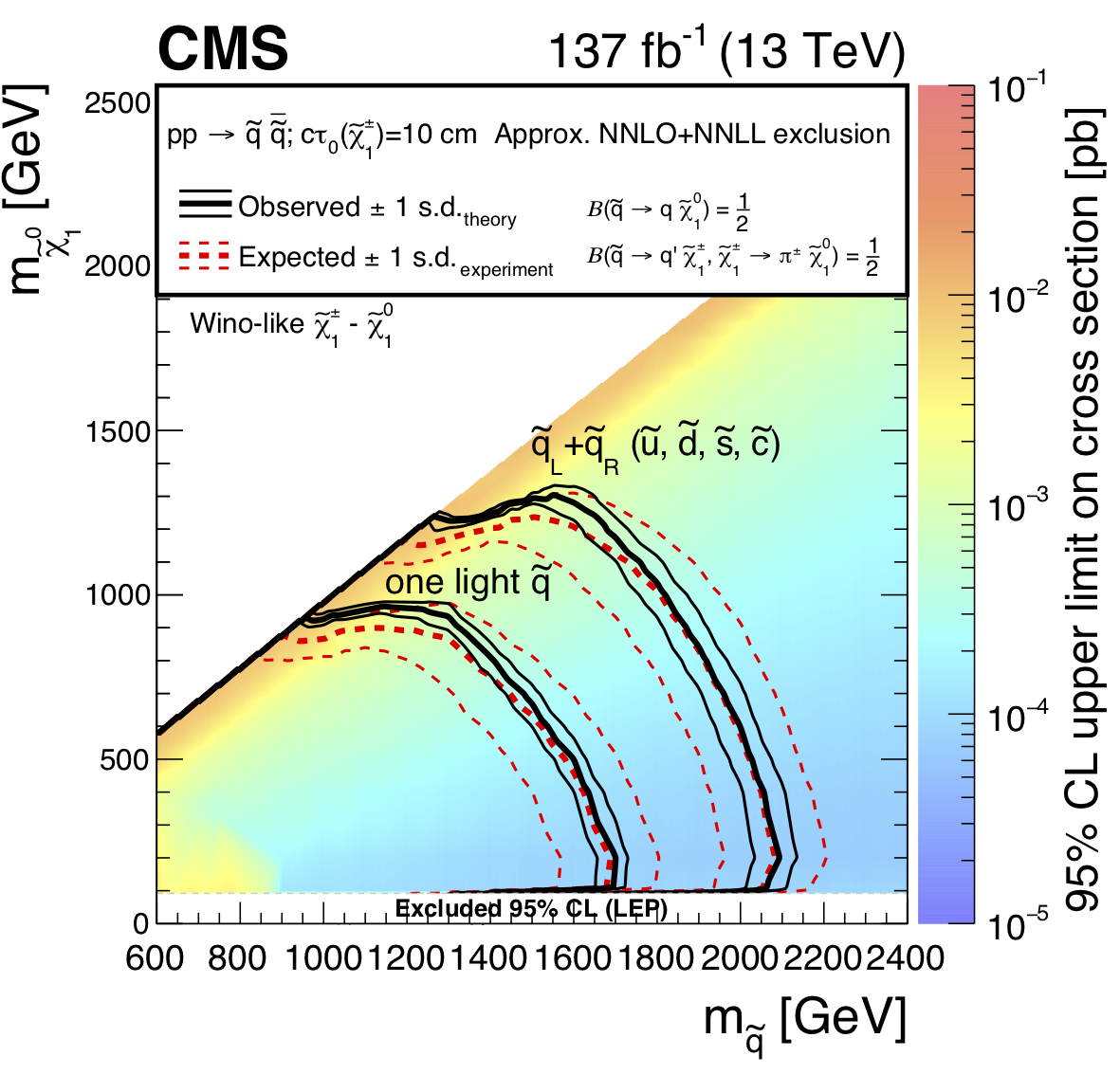}
    \includegraphics[width=0.48\textwidth]{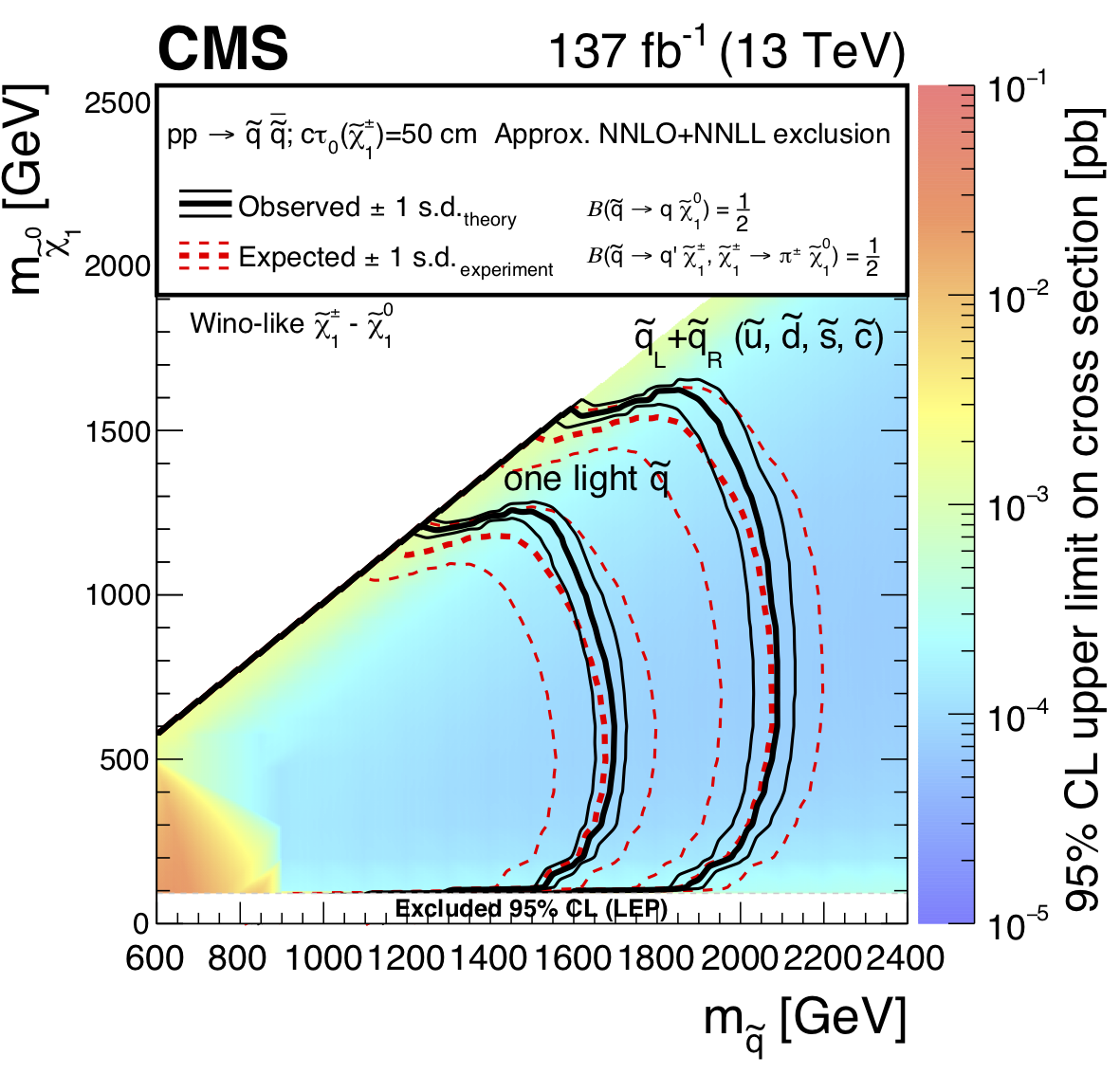}
    \includegraphics[width=0.48\textwidth]{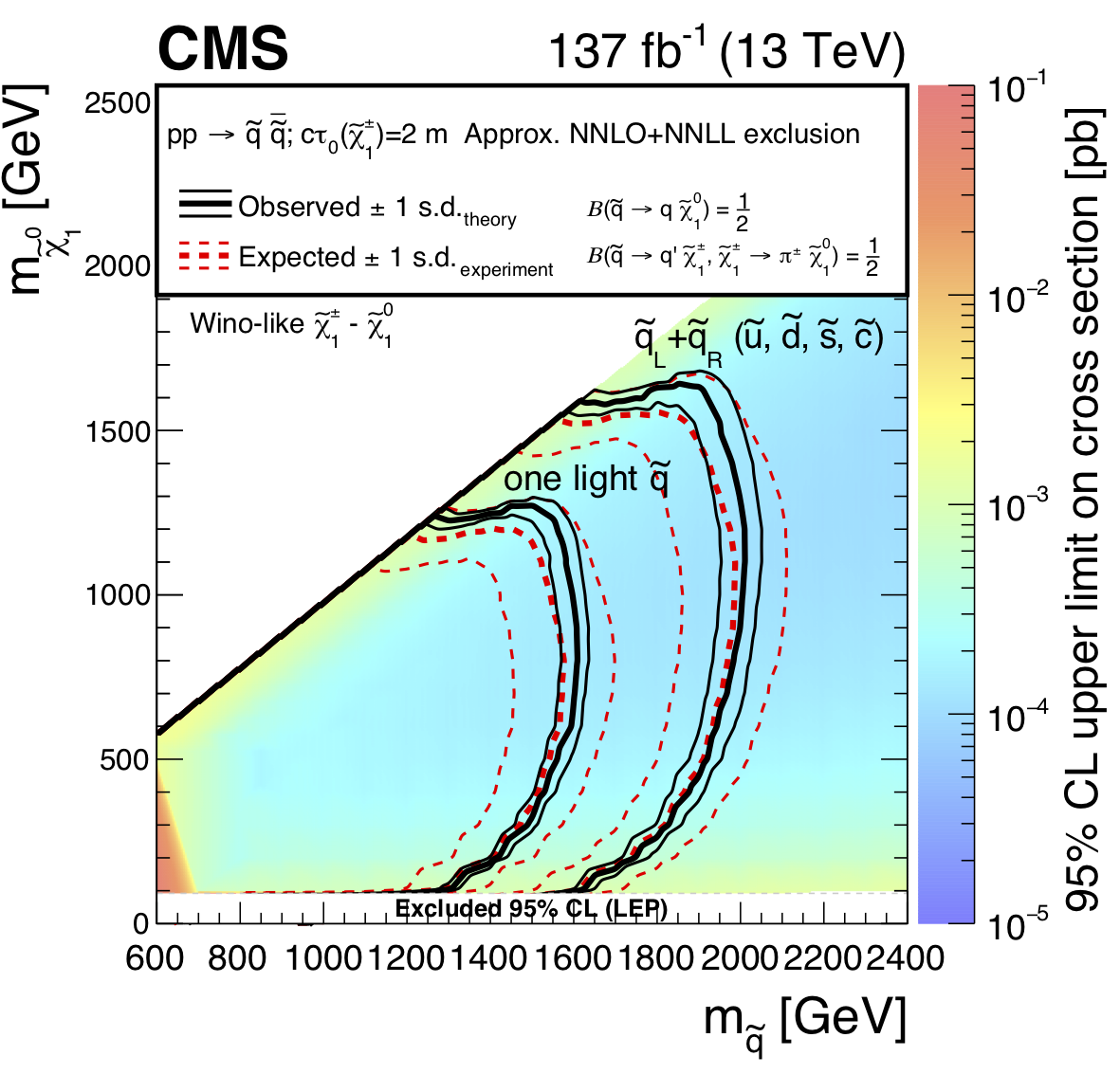}
    \caption{
    Exclusion limits at 95\% \CL for light squark pair production with $c\tau_{0}(\chargino) =$ (upper left) 10\cm,
    (upper right) 50\cm, and (lower) 200\cm.
      The area enclosed by the thick black curve represents the observed exclusion region,
      while the dashed red lines indicate the expected limits and
      their $\pm$1 standard deviation (s.d.) ranges.
      The thin black lines show the effect of the theoretical
      uncertainties in the signal cross section.
      The white band for masses of the \lsp below 91.9\GeV represents the region of the mass plane excluded at the CERN LEP~\cite{lep_chargino}.
      Signal cross sections are calculated at approximately NNLO+NNLL order in \alpS~\cite{bib-nlo-nll-01,bib-nnll-05,bib-nlo-nll-02,bib-nlo-nll-03,bib-nlo-nll-04,bib-nnll-06,bib-nlo-nll-05,bib-nnll-02,bib-nnll-03,bib-nnll-04,bib-nnll-07,bib-nnll},
      assuming decay branching fractions ($\mathcal{B}$) as indicated in the figure.}
    \label{fig:t2st_qq}
\end{figure*}

\begin{figure*}[htbp]
  \centering
    \includegraphics[width=0.48\textwidth]{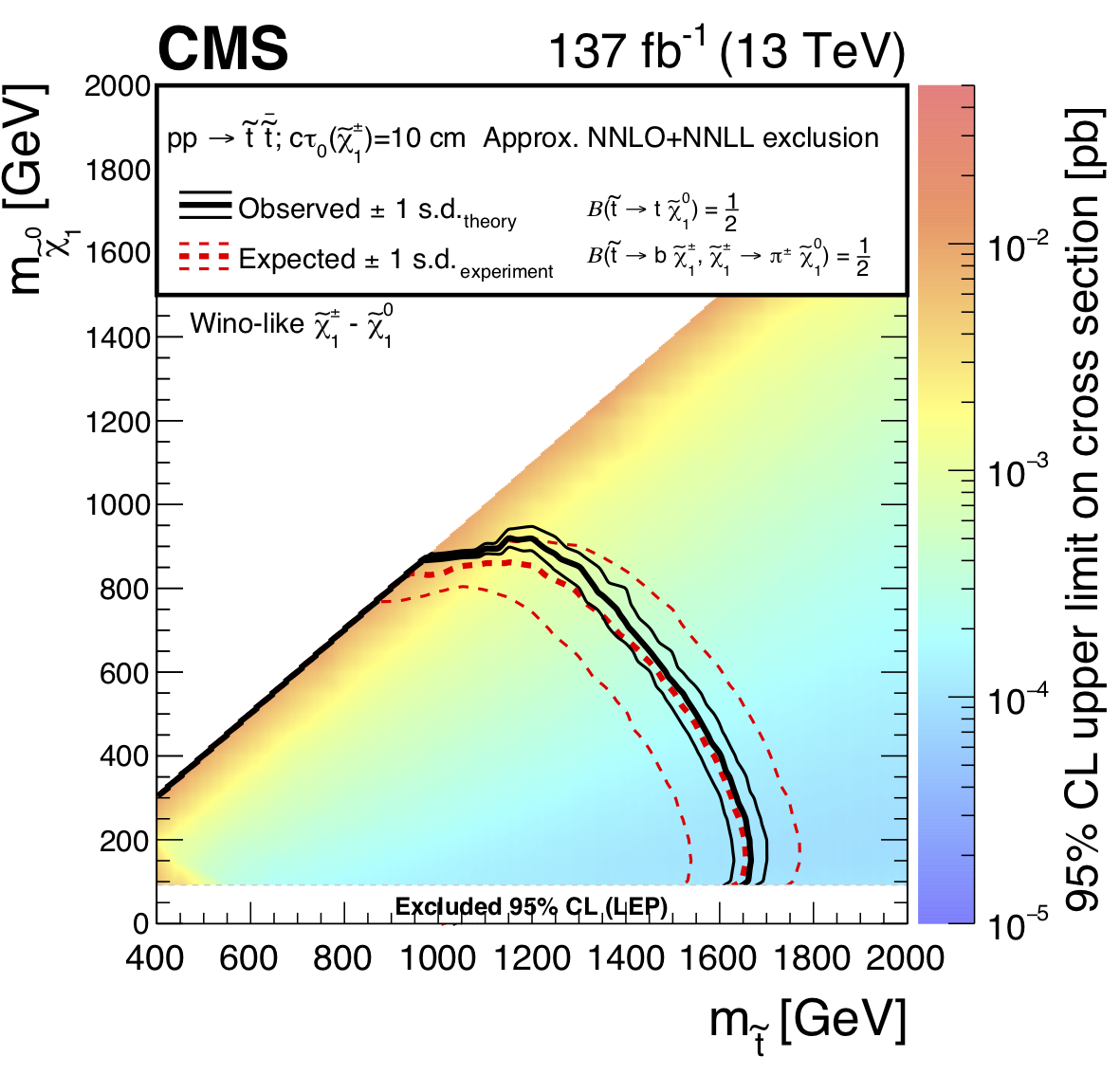}
    \includegraphics[width=0.48\textwidth]{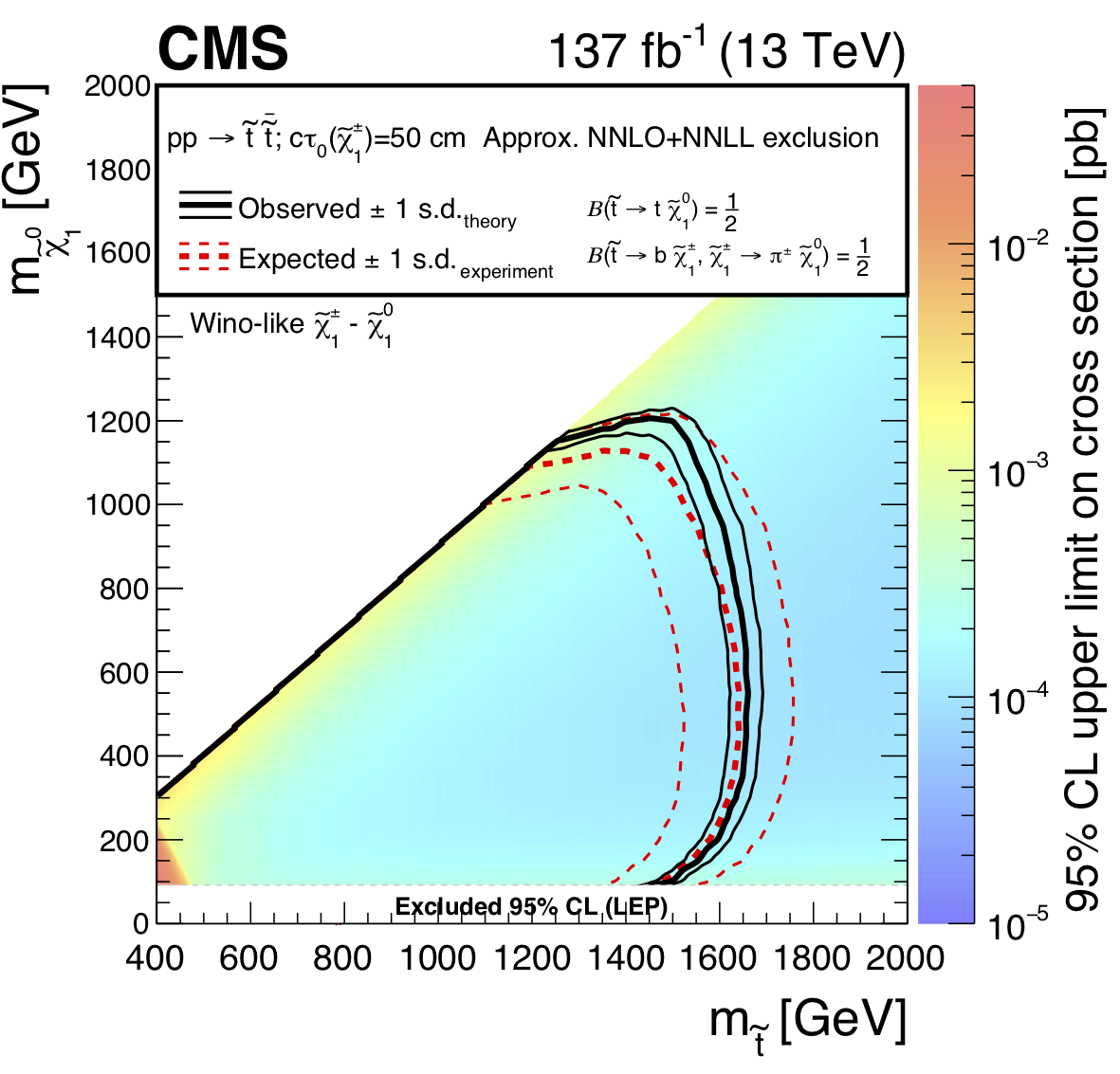}
    \includegraphics[width=0.48\textwidth]{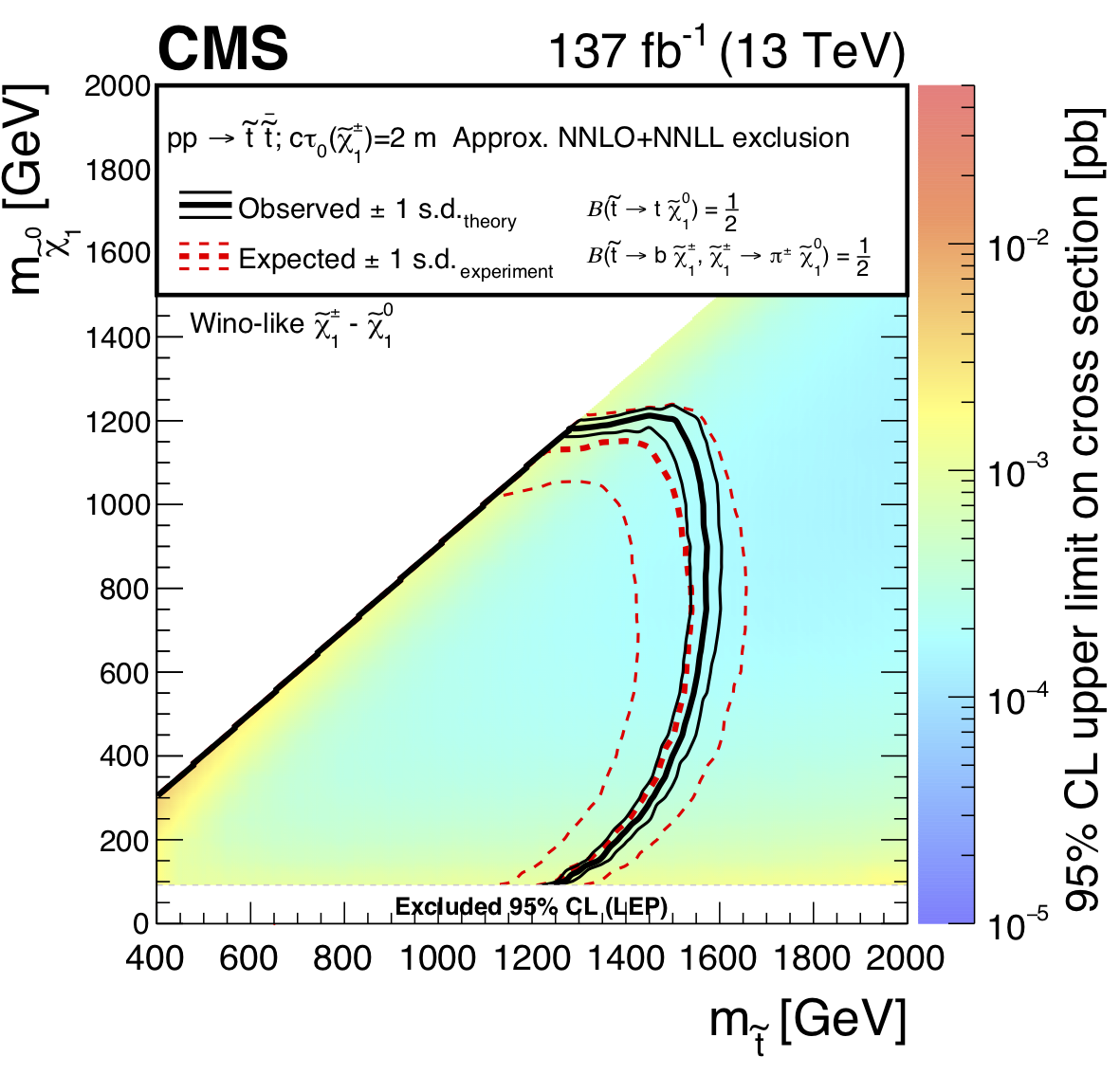}
    \caption{
    Exclusion limits at 95\% \CL for top squark pair production with $c\tau_{0}(\chargino) =$ (upper left) 10\cm,
    (upper right) 50\cm, and (lower) 200\cm.
      The area enclosed by the thick black curve represents the observed exclusion region,
      while the dashed red lines indicate the expected limits and
      their $\pm$1 standard deviation (s.d.) ranges.
      The thin black lines show the effect of the theoretical
      uncertainties in the signal cross section.
      The white band for masses of the \lsp below 91.9\GeV represents the region of the mass plane excluded at the CERN LEP~\cite{lep_chargino}.
      Signal cross sections are calculated at approximately NNLO+NNLL order in \alpS~\cite{bib-nlo-nll-01,bib-nnll-05,bib-nlo-nll-02,bib-nlo-nll-03,bib-nlo-nll-04,bib-nnll-06,bib-nlo-nll-05,bib-nnll-02,bib-nnll-03,bib-nnll-04,bib-nnll-07,bib-nnll},
      assuming decay branching fractions ($\mathcal{B}$) as indicated in the figure.}
    \label{fig:t2st_tt}
\end{figure*}

Figure~\ref{fig:t1st_qqqq} shows the exclusion limits at 95\% \CL for
direct gluino pair production where the gluinos decay to light-flavor (\cPqu, \cPqd, \cPqs, \cPqc) quarks,
with $c\tau_{0}(\chargino) =$ 10, 50, and 200\cm.
Exclusion limits for the direct production of
light-flavor and top squark pairs are shown in
Figs.~\ref{fig:t2st_qq}~and~\ref{fig:t2st_tt}, respectively,
also for $c\tau_{0}(\chargino) =$ 10, 50, and 200\cm.

Exclusion limits from the disappearing track search tend to be strongest in longer $c\tau_{0}(\chargino)$ models, when $m_{\lsp}$ is near the mass of the gluino or squark,
and in shorter $c\tau_{0}(\chargino)$ models, when a large mass splitting generates a large boost for the \chargino, and in models characterized by large jet multiplicities.
Models with these properties tend to populate the background depleted disappearing track regions with high \njets and longer tracks.
In the massless \chargino and \lsp limit, the \chargino receives a large Lorentz boost.
Therefore, it tends not to decay inside the tracking detector, with a consequent reduction in the signal acceptance and in the analysis sensitivity.

When a \chargino decays within the volume of the tracking detector, it is not counted as a PF candidate
and, being almost mass degenerate with the \lsp, its decay products provide negligible visible energy in the detector.
To a good approximation, as confirmed in simulation,
the limits presented in Section~\ref{sec:interpretation:mt2} from the inclusive \mttwo search should apply also to these models with an intermediate \chargino.

For SUSY models with long-lived \chargino, the search for disappearing tracks significantly extends the sensitivity of the inclusive \mttwo search.
Table~\ref{tab:limST} summarizes the limits on the masses of the SUSY particles excluded for the
simplified model scenarios considered.
\begin{table*}[htb]
  \topcaption{Summary of the observed 95\% \CL exclusion limits on the masses of SUSY particles for different simplified model scenarios,
    where the produced particles  decay with equal probability to \charginoplus, \charginominus, and \lsp, and the \chargino are long lived.
    The highest limits on the mass of the directly produced particles and on the mass of the \lsp are quoted.
    \label{tab:limST}}
\centering
\begin{tabular}{lrr}
    \hline
Simplified & Highest limit on directly produced  & Highest limit on \\
model & SUSY particle mass [\GeVns{}] & \lsp mass [\GeVns{}] \\
\hline
Direct gluino pair production: & & \\
$\gluino \to \qqbar\lsp$ or $\gluino \to \qqbarpr\chargino$ & 2460 & 2000 \\
[\cmsTabSkip]
Direct squark pair production: & & \\
Eight degenerate light squarks & 2090 & 1650 \\
Single light squark & 1700 & 1275 \\
Top squark & 1660 & 1210 \\
    \hline
\end{tabular}
\end{table*}

\begin{figure*}[htbp]
  \centering
    \includegraphics[width=0.48\textwidth]{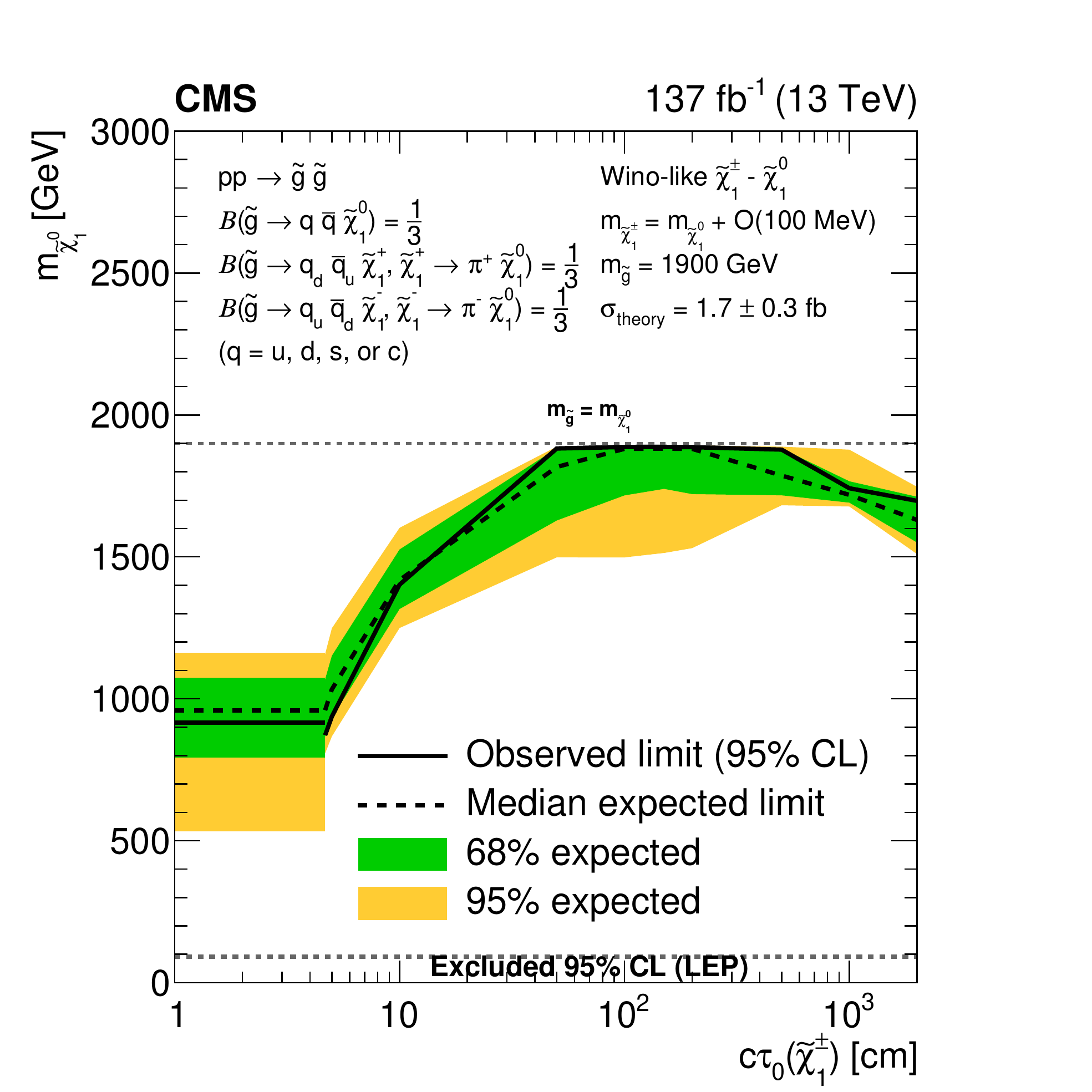}\\
    \includegraphics[width=0.48\textwidth]{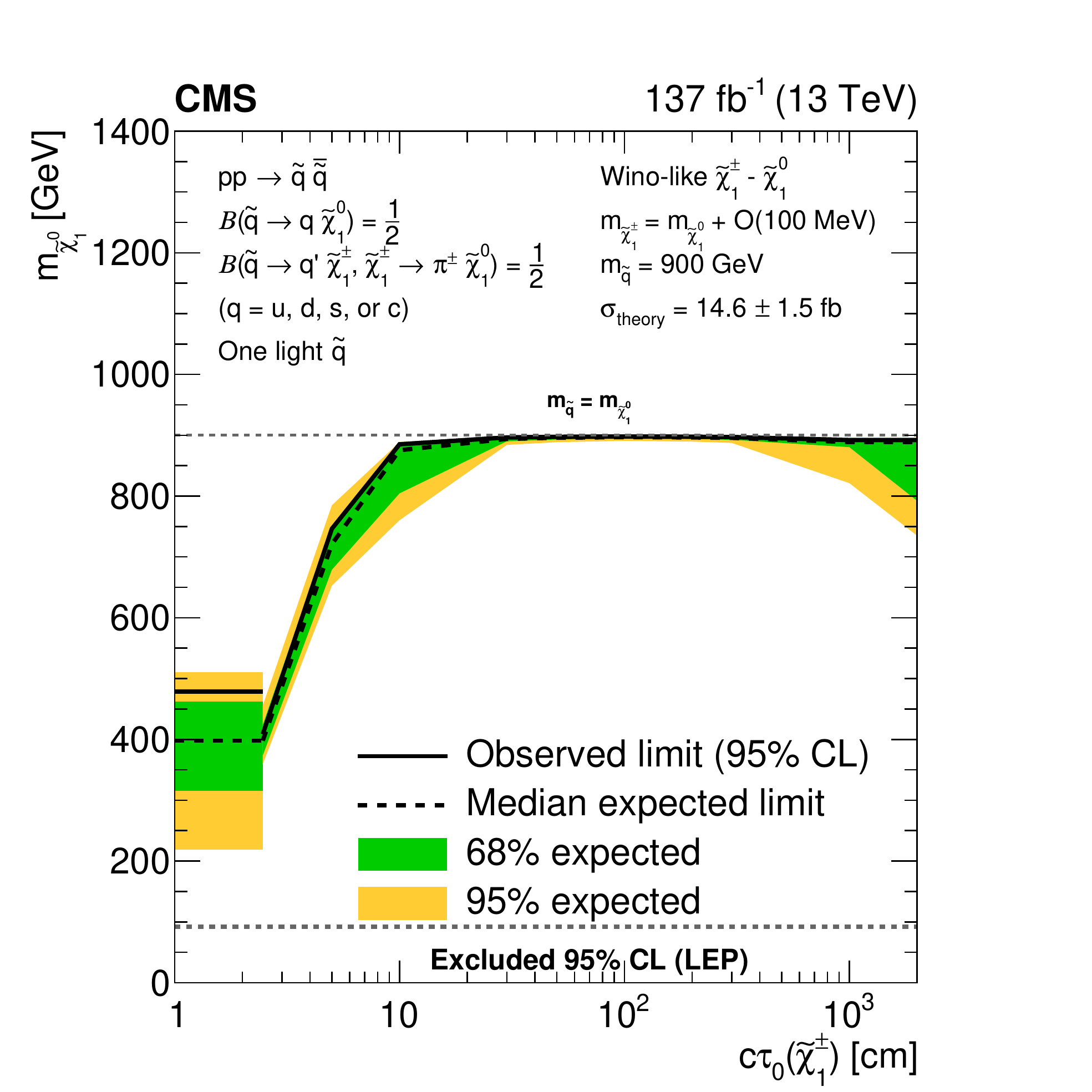}
    \includegraphics[width=0.48\textwidth]{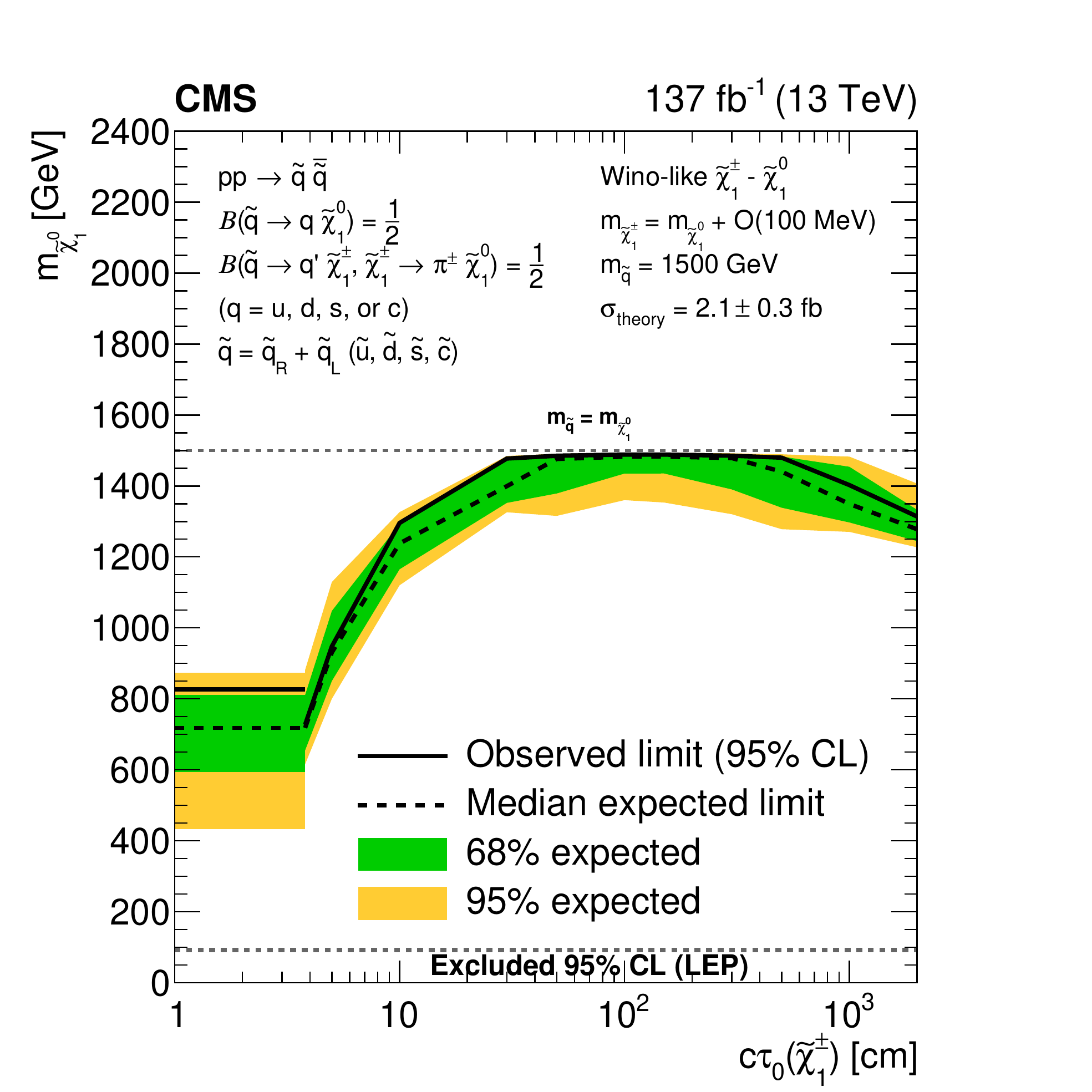}
    \caption{
    Exclusion limits at  95\% \CL on the \lsp mass, with $m_{\chargino}=m_{\lsp}+\mathcal{O}(100\MeV)$, as a function of the \chargino proper decay length,
    for (upper) direct gluino and (lower) direct light-flavor (\cPqu, \cPqd, \cPqs, \cPqc) squark pair production,
    as obtained for representative gluino and squark masses. The gluinos decay to light-flavor quarks.
    For direct squark pair production, we assume either (lower left) one--fold or (lower right) eight--fold squark degeneracy.
    The area enclosed by the solid (dashed) black curve represents the observed (median expected) exclusion region,
    while the inner green (outer yellow) band indicates the region containing 68 (95)\%
    of the distribution of limits expected under the background-only hypothesis.
    At short decay lengths, horizontal exclusion lines are obtained from the inclusive \mttwo search,
    as this is not affected by track reconstruction inefficiencies, which may arise when the \chargino decays before the CMS tracker,
    and therefore shows better sensitivity to scenarios with very small $c\tau_{0}(\chargino)$ compared to the disappearing track search,
    based on median expected limits.
    The horizontal dashed lines at (upper) $m_{\gluino}=m_{\lsp}$ and (lower) $m_{\PSQ}=m_{\lsp}$ bound the mass range in which the decays are kinematically allowed.
    If all kinematically allowed \lsp masses ($m_{\lsp} \leq m_{\gluino}$, or $m_{\lsp} \leq m_{\PSQ}$) are excluded, the curves, including 68 and 95\% expected, tend to overlap.
    The band at masses of the \lsp below 91.9\GeV represents the region of the mass plane excluded at the CERN LEP~\cite{lep_chargino}.
      Signal cross sections are calculated at approximately NNLO+NNLL order in \alpS~\cite{bib-nlo-nll-01,bib-nnll-05,bib-nlo-nll-02,bib-nlo-nll-03,bib-nlo-nll-04,bib-nnll-06,bib-nlo-nll-05,bib-nnll-02,bib-nnll-03,bib-nnll-04,bib-nnll-07,bib-nnll},
      assuming decay branching fractions ($\mathcal{B}$) as indicated in the figure.}
    \label{fig:lq_ltvsmass}
\end{figure*}

\begin{figure}[htb!]
  \centering
    \includegraphics[width=0.48\textwidth]{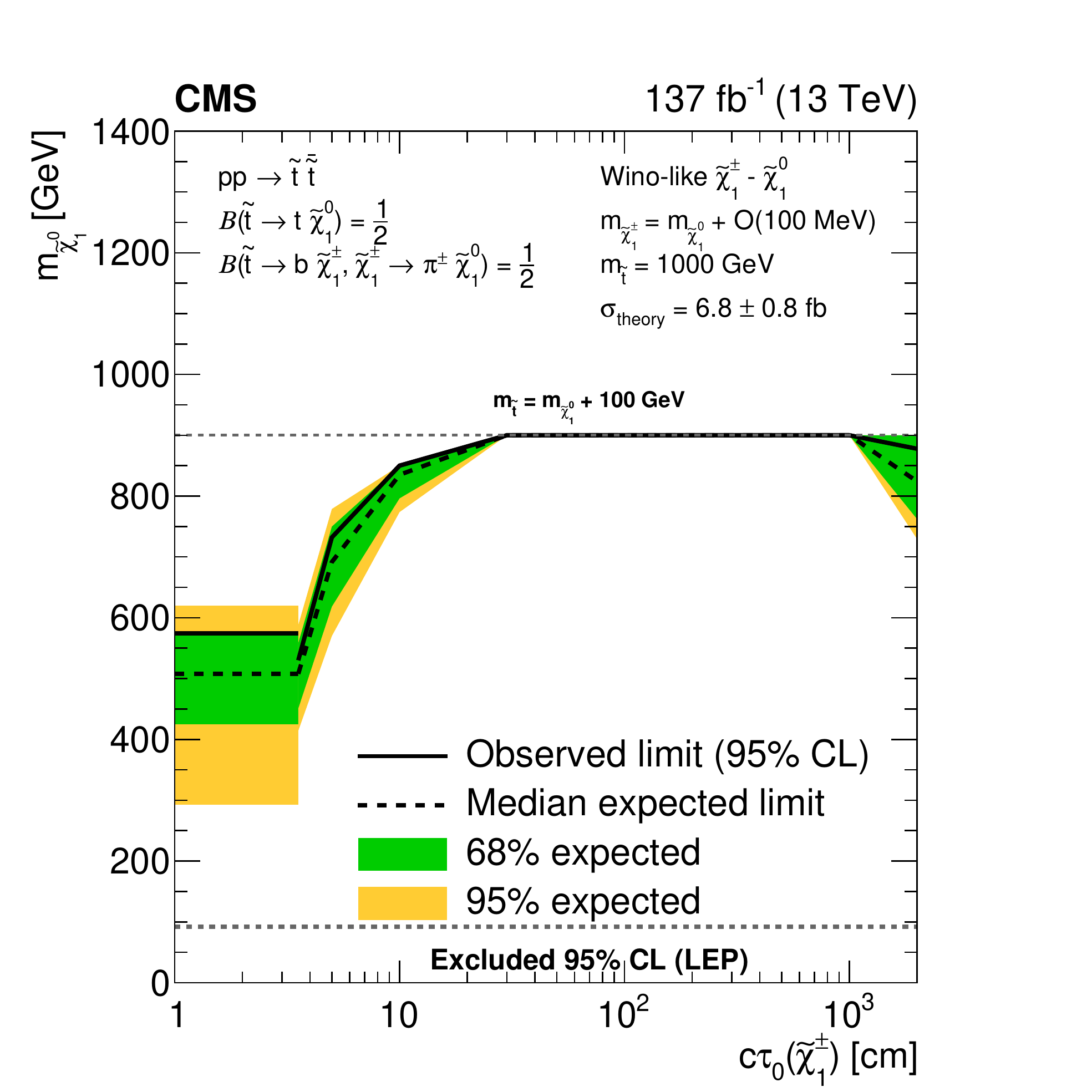}
    \caption{
    Exclusion limits at  95\% \CL on the \lsp mass, with $m_{\chargino}=m_{\lsp}+\mathcal{O}(100\MeV)$, as a function of the \chargino proper decay length,
    for  direct top squark pair production, as obtained for a representative top squark mass.
    The area enclosed by the solid (dashed) black curve represents the observed (median expected) exclusion region,
    while the inner green (outer yellow) band indicates the region containing 68 (95)\%
    of the distribution of limits expected under the background-only hypothesis.
    At short decay lengths, horizontal exclusion lines are obtained from the inclusive \mttwo search,
    as this is not affected by track reconstruction inefficiencies, which may arise when the \chargino decays before the CMS tracker,
    and therefore shows better sensitivity to scenarios with very small $c\tau_{0}(\chargino)$ compared to the disappearing track search,
    based on median expected limits.
    The horizontal dashed line at $m_{\PSQt}=m_{\lsp}+100\GeV$ indicates the minimum simulated mass difference between top squark and \lsp,
    chosen such that the decay of top quarks to on-shell $\PW$ bosons is allowed.
    If all kinematically allowed \lsp masses ($m_{\lsp} \leq m_{\PSQt}-100\GeV$) are excluded, the curves, including 68 and 95\% expected, tend to overlap.
    The band at masses of the \lsp below 91.9\GeV represents the region of the mass plane excluded at the CERN LEP~\cite{lep_chargino}.
      Signal cross sections are calculated at approximately NNLO+NNLL order in \alpS~\cite{bib-nlo-nll-01,bib-nnll-05,bib-nlo-nll-02,bib-nlo-nll-03,bib-nlo-nll-04,bib-nnll-06,bib-nlo-nll-05,bib-nnll-02,bib-nnll-03,bib-nnll-04,bib-nnll-07,bib-nnll},
      assuming decay branching fractions ($\mathcal{B}$) as indicated in the figure.
    }
    \label{fig:t2bt_ltvsmass}
\end{figure}

Two-dimensional constraints are also placed on the \chargino mass as a function of its proper decay length,
as shown in Figs.~\ref{fig:lq_ltvsmass}~and~\ref{fig:t2bt_ltvsmass},
for the pair production of gluinos and light-flavor and top squarks, respectively.
In particular, Figs.~\ref{fig:lq_ltvsmass}--\ref{fig:t2bt_ltvsmass} show
the excluded \chargino mass as a function of its proper decay length for representative gluino, light-flavor or top squark masses.
For short \chargino lifetimes, the inclusive \mttwo search is
more sensitive than the
dedicated search for disappearing tracks, based on expected exclusion limits.
As already mentioned above, the inclusive \mttwo search is not sensitive to the presence of an intermediate long-lived \chargino in the parent SUSY particle decay chain,
especially when the \chargino lifetime is short, such that the \chargino cannot be reconstructed as a stable PF candidate.
Furthermore, the signal acceptance of the inclusive \mttwo search is not affected by the track reconstruction inefficiencies
which may arise when the \chargino decays before the CMS tracker, for very short \chargino lifetimes.

\begin{figure*}[htbp]
  \centering
    \includegraphics[width=0.48\textwidth]{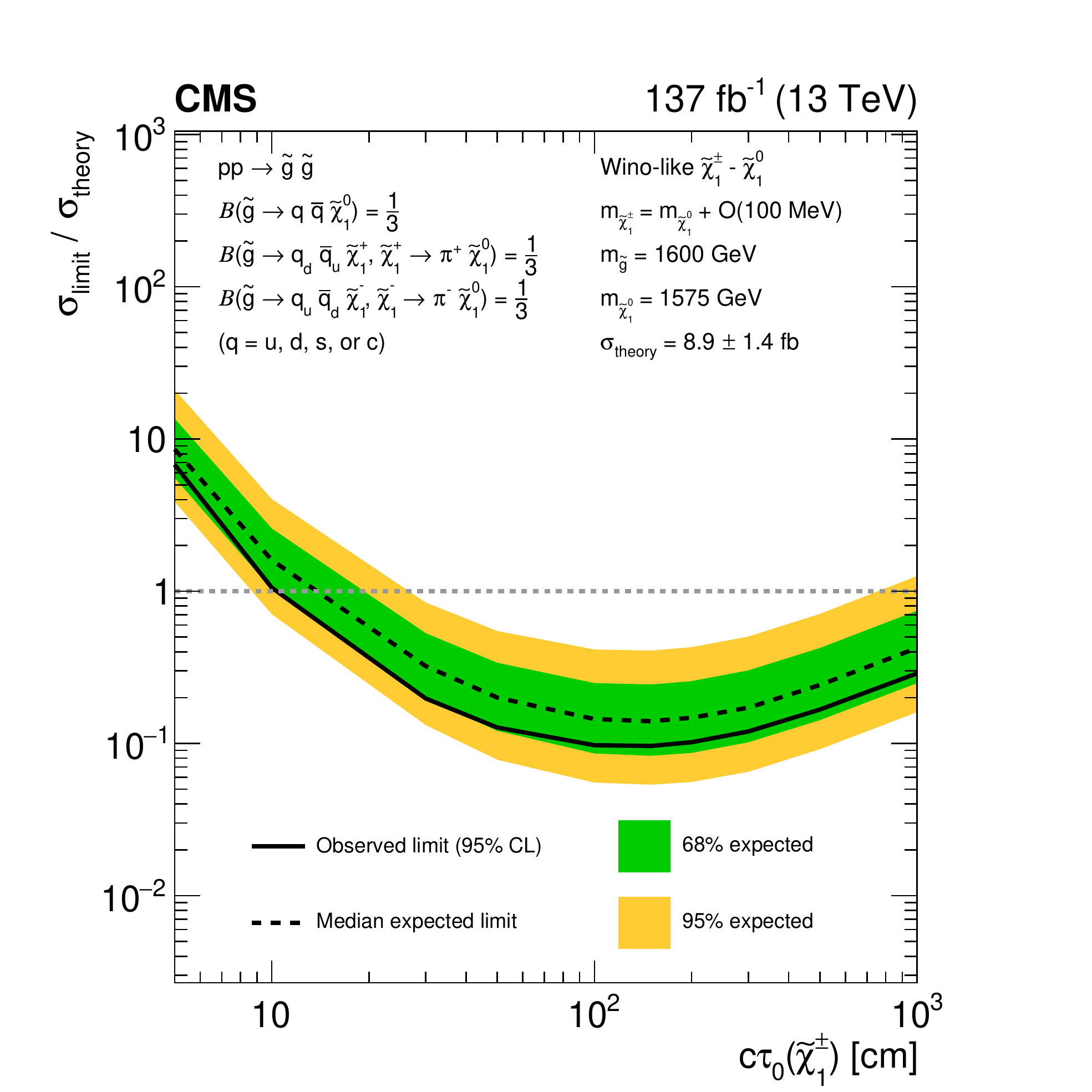}\\
    \includegraphics[width=0.48\textwidth]{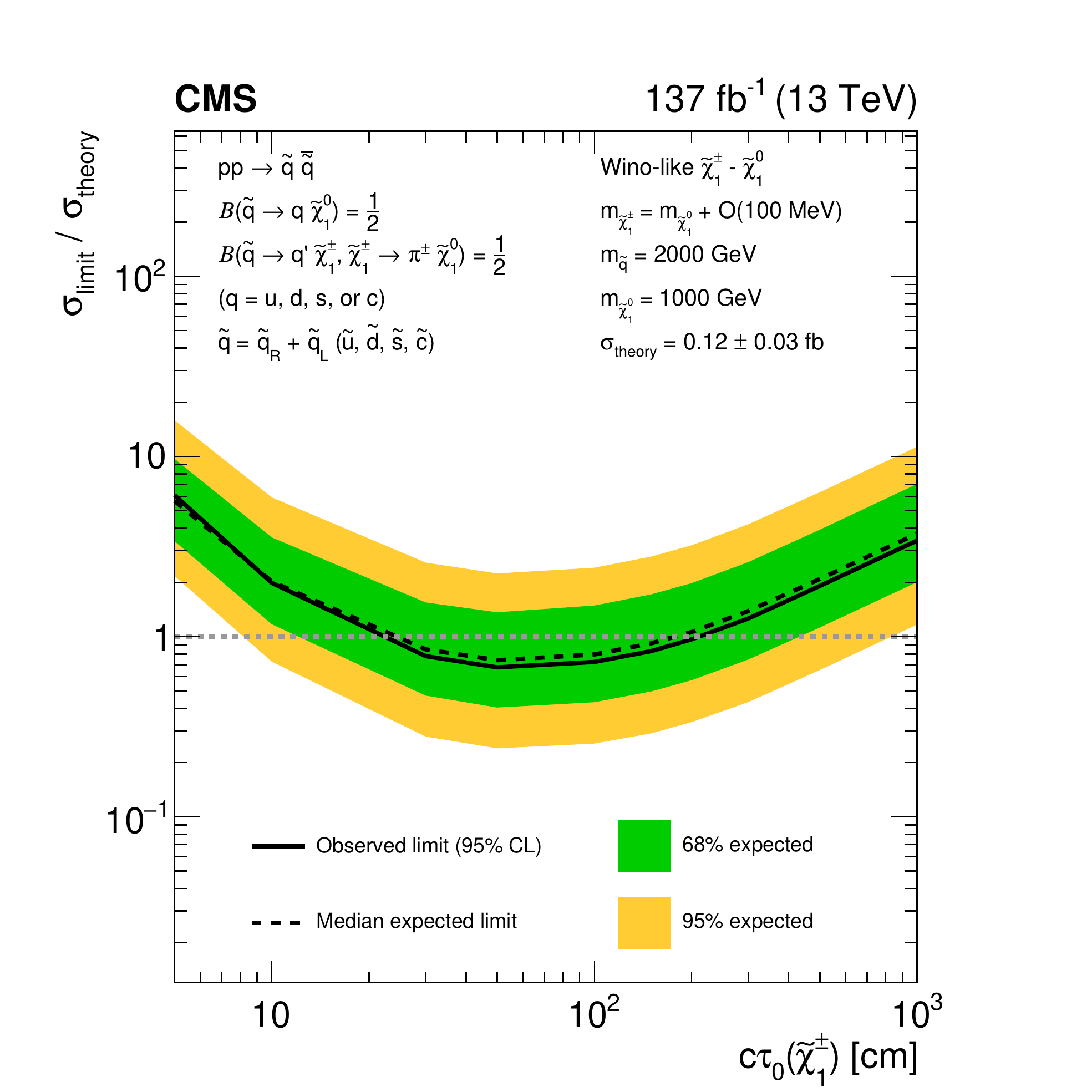}
    \includegraphics[width=0.48\textwidth]{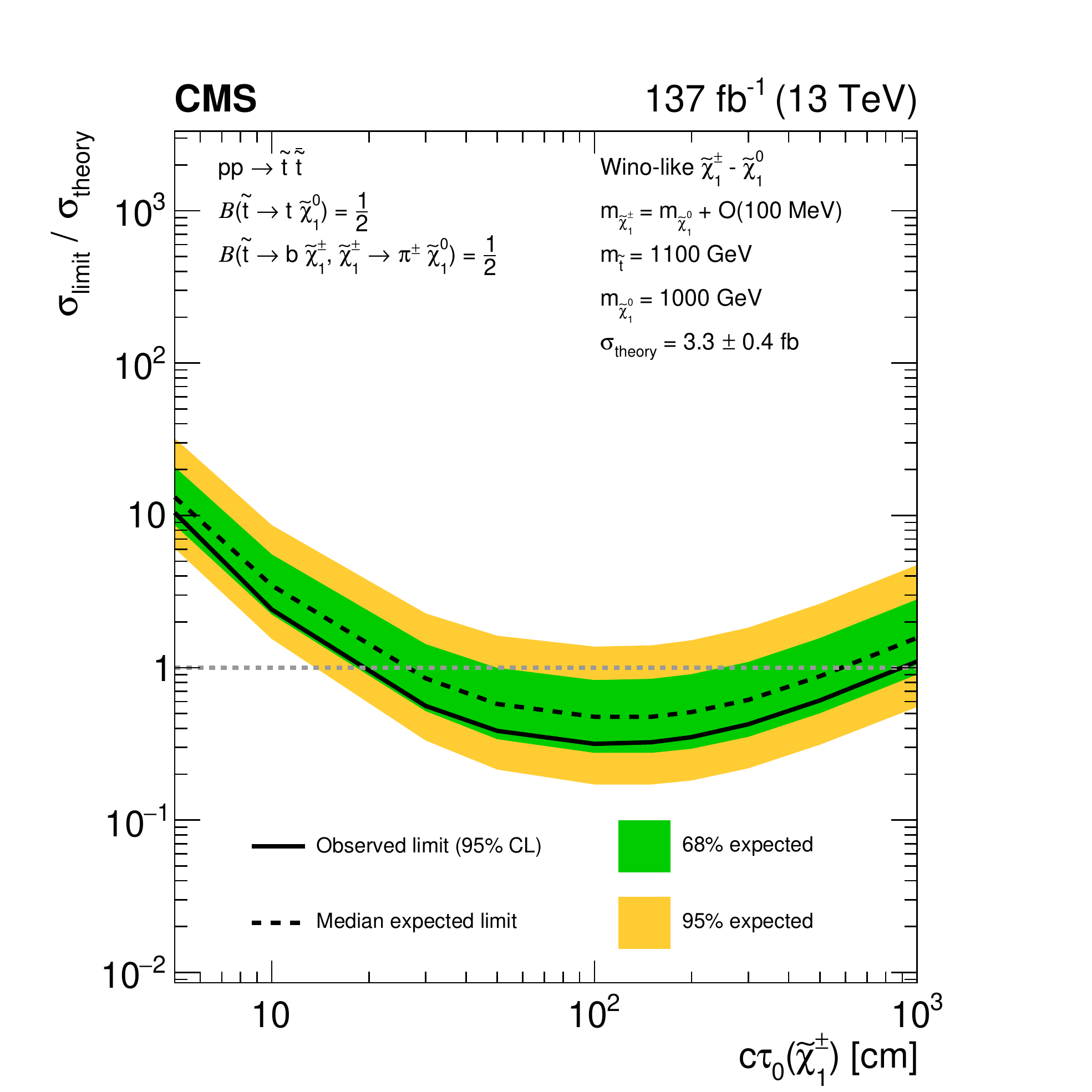}
    \caption{
    Exclusion limits at  95\% \CL on $\sigma/\sigma_{\mathrm{theory}}$ as a function of the \chargino decay length, for a choice of signal models of
    (upper) direct gluino pair production where the gluinos decay to light-flavor (\cPqu, \cPqd, \cPqs, \cPqc) quarks, (lower left) direct light-flavor squark pair production,
    and (lower right) direct top squark pair production, as obtained
    from the search for disappearing tracks.
    The area enclosed by the solid (dashed) black curve below the horizontal dashed line at $\sigma/\sigma_{\mathrm{theory}}=1$ represents the observed (median expected) exclusion region,
    while the inner green (outer yellow) band indicates the region containing 68 (95)\%
    of the distribution of limits expected under the background-only hypothesis.
      Signal cross sections are calculated at approximately NNLO+NNLL order in \alpS~\cite{bib-nlo-nll-01,bib-nnll-05,bib-nlo-nll-02,bib-nlo-nll-03,bib-nlo-nll-04,bib-nnll-06,bib-nlo-nll-05,bib-nnll-02,bib-nnll-03,bib-nnll-04,bib-nnll-07,bib-nnll},
      assuming decay branching fractions ($\mathcal{B}$) as indicated in the figure.
    }
    \label{fig:ch_rvsctau}
\end{figure*}

Figure~\ref{fig:ch_rvsctau} shows
exclusion limits on $\sigma/\sigma_{\mathrm{theory}}$ as a function of $c\tau_{0}(\chargino)$, for a choice of
signal models where gluinos and squarks can decay via a long-lived \chargino,
as obtained
from the search for disappearing tracks.
Scenarios where the mass spectrum of SUSY particles is compressed are especially constrained
across a wide range of $c\tau_{0}(\chargino)$. The exclusion limits are typically stronger at intermediate $c\tau_{0}(\chargino)$,
as a larger fraction of \chargino decay within the CMS tracker and can therefore be identified as disappearing tracks.

\section{Summary}
\label{sec:conclusions}

This paper presents the results of two related searches for phenomena beyond the standard model
using events with jets and large values of the kinematic variable \mttwo.
The first is an inclusive search, while the second requires in addition disappearing tracks.
The measurements are based on a data sample of proton-proton collisions at $\sqrt{s} =13\TeV$ collected in 2016--2018 with the CMS detector,
and corresponding to an integrated luminosity of \Lint.
No significant deviations from the standard model expectations are observed.
Limits on pair-produced gluinos and squarks are established in the context of supersymmetry models conserving $R$-parity.
The inclusive \mttwo search probes gluino masses up to 2250\GeV and the lightest neutralino \lsp masses up to 1525\GeV,
as well as light-flavor, bottom, and top squark masses up
to 1710, 1240, and 1200\GeV, respectively, and \lsp masses up to 870, 700, and 580\GeV in each respective scenario.
{\tolerance=1200
In models with a long-lived chargino \chargino, where the gluinos and squarks decay with equal probability to \lsp, \charginoplus, and \charginominus,
the search looking in addition for disappearing tracks probes gluino masses up to 2460\GeV and \lsp masses up to 2000\GeV,
as well as light-flavor (top) squark masses up to 2090~(1660)\GeV and \lsp masses up to 1650~(1210)\GeV.
\par}
A resonantly produced colored scalar state $\phi$ decaying to a massive Dirac fermion $\psi$ and a quark has recently been proposed as an explanation of an excess in data
identified in regions with low jet multiplicities, based on previous results by the ATLAS and CMS Collaborations.
From the inclusive \mttwo search, mass limits as high as 1660 and 925\GeV are obtained for $\phi$ and $\psi$, respectively,
and an upper limit on the product of the cross section and branching fraction of about 0.6\unit{pb} with a local significance of 1.1 standard deviations is observed for the
previously reported best fit point $\left(m_{\phi},m_{\psi}\right)=\left(1250,900\right)\GeV$.
The inclusive \mttwo search is also used to constrain models of scalar and vector leptoquark (LQ) pair production with the LQ decaying to a neutrino and a top, bottom, or light-flavor quark.
A vector LQ decaying with equal branching fraction to $\PQt\PGn$ and $\PQb\PGt$ has been proposed as part of an explanation of recent flavor anomalies.
In such a model, LQ masses below 1550\GeV are excluded assuming the Yang--Mills case with coupling $\kappa=1$, or 1225\GeV in the minimal coupling case $\kappa=0$.
The results presented in this paper extend the mass limits of the previous version of the CMS inclusive \mttwo search,
using a subset of the present data, by hundreds of \GeV.
In most of the cases, the results obtained are the most stringent constraints to date.

\clearpage
\begin{acknowledgments}
We congratulate our colleagues in the CERN accelerator departments for the excellent performance of the LHC and thank the technical and administrative staffs at CERN and at other CMS institutes for their contributions to the success of the CMS effort. In addition, we gratefully acknowledge the computing centers and personnel of the Worldwide LHC Computing Grid for delivering so effectively the computing infrastructure essential to our analyses. Finally, we acknowledge the enduring support for the construction and operation of the LHC and the CMS detector provided by the following funding agencies: BMBWF and FWF (Austria); FNRS and FWO (Belgium); CNPq, CAPES, FAPERJ, FAPERGS, and FAPESP (Brazil); MES (Bulgaria); CERN; CAS, MoST, and NSFC (China); COLCIENCIAS (Colombia); MSES and CSF (Croatia); RPF (Cyprus); SENESCYT (Ecuador); MoER, ERC IUT, PUT and ERDF (Estonia); Academy of Finland, MEC, and HIP (Finland); CEA and CNRS/IN2P3 (France); BMBF, DFG, and HGF (Germany); GSRT (Greece); NKFIA (Hungary); DAE and DST (India); IPM (Iran); SFI (Ireland); INFN (Italy); MSIP and NRF (Republic of Korea); MES (Latvia); LAS (Lithuania); MOE and UM (Malaysia); BUAP, CINVESTAV, CONACYT, LNS, SEP, and UASLP-FAI (Mexico); MOS (Montenegro); MBIE (New Zealand); PAEC (Pakistan); MSHE and NSC (Poland); FCT (Portugal); JINR (Dubna); MON, RosAtom, RAS, RFBR, and NRC KI (Russia); MESTD (Serbia); SEIDI, CPAN, PCTI, and FEDER (Spain); MOSTR (Sri Lanka); Swiss Funding Agencies (Switzerland); MST (Taipei); ThEPCenter, IPST, STAR, and NSTDA (Thailand); TUBITAK and TAEK (Turkey); NASU and SFFR (Ukraine); STFC (United Kingdom); DOE and NSF (USA).

\hyphenation{Rachada-pisek} Individuals have received support from the Marie-Curie program and the European Research Council and Horizon 2020 Grant, contract Nos.\ 675440, 752730, and 765710 (European Union); the Leventis Foundation; the A.P.\ Sloan Foundation; the Alexander von Humboldt Foundation; the Belgian Federal Science Policy Office; the Fonds pour la Formation \`a la Recherche dans l'Industrie et dans l'Agriculture (FRIA-Belgium); the Agentschap voor Innovatie door Wetenschap en Technologie (IWT-Belgium); the F.R.S.-FNRS and FWO (Belgium) under the ``Excellence of Science -- EOS" -- be.h project n.\ 30820817; the Beijing Municipal Science \& Technology Commission, No. Z181100004218003; the Ministry of Education, Youth and Sports (MEYS) of the Czech Republic; the Lend\"ulet (``Momentum") Program and the J\'anos Bolyai Research Scholarship of the Hungarian Academy of Sciences, the New National Excellence Program \'UNKP, the NKFIA research grants 123842, 123959, 124845, 124850, 125105, 128713, 128786, and 129058 (Hungary); the Council of Science and Industrial Research, India; the HOMING PLUS program of the Foundation for Polish Science, cofinanced from European Union, Regional Development Fund, the Mobility Plus program of the Ministry of Science and Higher Education, the National Science Center (Poland), contracts Harmonia 2014/14/M/ST2/00428, Opus 2014/13/B/ST2/02543, 2014/15/B/ST2/03998, and 2015/19/B/ST2/02861, Sonata-bis 2012/07/E/ST2/01406; the National Priorities Research Program by Qatar National Research Fund; the Ministry of Science and Education, grant no. 3.2989.2017 (Russia); the Programa Estatal de Fomento de la Investigaci{\'o}n Cient{\'i}fica y T{\'e}cnica de Excelencia Mar\'{\i}a de Maeztu, grant MDM-2015-0509 and the Programa Severo Ochoa del Principado de Asturias; the Thalis and Aristeia programs cofinanced by EU-ESF and the Greek NSRF; the Rachadapisek Sompot Fund for Postdoctoral Fellowship, Chulalongkorn University and the Chulalongkorn Academic into Its 2nd Century Project Advancement Project (Thailand); the Welch Foundation, contract C-1845; and the Weston Havens Foundation (USA).
\end{acknowledgments}

\bibliography{auto_generated}

\appendix

\section{Disappearing track selection}
\label{app:stsel}

The detailed selection of disappearing tracks (STs and STCs, as defined in Section~\ref{sec:evtsel:evtcat:distracks}) is summarized in Table~\ref{tab:stsel}.

\begin{table*}[htbp]
\setlength{\extrarowheight}{.45em}
\topcaption{Selection requirements for STs and STCs.
  For the subset of medium (M) length tracks that have just four tracking layers with a measurement, the minimum required number of layers of the pixel tracking detector with a measurement is three ($\dagger$).
  The selected tracks are required to not overlap with identified leptons. For this selection, all electrons and muons are considered, either identified as PF candidates or not.
  The selected tracks are as well required to not be identified as PF candidates,
  and to not overlap with other tracks with $\pt>15\GeV$, even if those tracks are not associated with PF candidates.
  The factor by which the selection requirement is relaxed in order to select short track candidates is also reported.
  If no factor is reported, the requirement is not relaxed for the selection of short track candidates.
  \label{tab:stsel}}
\centering
\cmsTableAlt{
\begin{tabular}{l  l  c  c}
    \hline
  Observable                            & Selection         & Track length          & STC factor\\
  \hline
  \pt [\GeVns{}]                                            & $>$15         & All                     & \\
  $\abs{\eta}$                                 & $<$2.4 and not $1.38 < \abs{\eta} < 1.6$ & All & \\
[\cmsTabSkip]
  $\sigma(\pt)$ / $\pt^2$ [\GeVns{}$^{-1}$]                          & $<$0.2; $<$0.02; $<$0.005           & P; M; L                        & 3 \\
  $d_{\mathrm{xy}}$ (from primary vertex) [cm]                                            & $<$0.02 ( $<$0.01 )           & P ( M, L )                       & 3  \\
  $d_{\mathrm{z}}$ (from primary vertex) [cm]                                             & $<$0.05           & All                     & 3  \\
[\cmsTabSkip]
  Neutral isolation ($\Delta R < 0.05$) [\GeVns{}]                          & $<$10         & All                     & 6  \\
  Neutral isolation / \pt                & $<$0.1            & All                     & 6  \\
  Isolation ($\Delta R < 0.3$) [\GeVns{}]                             & $<$10         & All                     & 6  \\
  Isolation / \pt                          & $<$0.2            & All                     & 6   \\
[\cmsTabSkip]
  Number of pixel layers             & $\geq$3 ( $\geq$2 )           & P, M$^{\dagger}$ ( M, L )                 & \\
  Number of tracker layers             & $\geq$3; $<$7; $\geq$7           & P; M; L                 & \\
  Number of lost inner hits                           & $=$0              & All                     & \\
  Number of lost outer hits                           & $\geq$2           & M, L                     & \\
[\cmsTabSkip]
  Is a PF candidate?                                    & No                 & All                     & \\
  PF lepton veto ($\Delta R < 0.1$)           & Yes   & All                     & \\
  Lepton veto ($\Delta R < 0.2$)              & Yes   & All                     & \\
  Track veto ($\Delta R < 0.1$)               & Yes   & All                     & \\
  Bad calorimeter module veto                 & Yes                 & All                     &  \\
  \Mt(track, \vMet) [\GeVns{}]                     & $>$100, if $\pt < 150\GeV$ & L  &  \\
    \hline
\end{tabular}
}
\end{table*}

\section{Definition of search regions and yields}
\label{app:srs}

\subsection{Inclusive \texorpdfstring{\mttwo}{MT2} search: search regions and yields}
\label{app:srs:mt2}

The 282 exclusive search regions defined for the inclusive \mttwo search, as described in Section~\ref{sec:evtsel:evtcat:mt2},
are summarized in Tables~\ref{tab:yieldsmonojet}--\ref{tab:yieldsUHh},
together with the pre-fit background predictions and the observed yields.

\begin{table*}[htb!]
\setlength\tabcolsep{1.5mm}
\centering
\topcaption{Predictions and observations for the 12 search regions with $\njets = 1$. For each of the background
predictions, the first uncertainty listed is statistical (from the limited size of data control samples
and Monte Carlo samples), and the second is systematic.}
\label{tab:yieldsmonojet}
\renewcommand{\arraystretch}{1.3}
\cmsTableAlt{
\begin{tabular}{c c  c c c c c}
    \hline
\njets, \nbtags & \ptj [\GeVns{}] & Lost lepton & \znunu & Multijet & Total background & Data \\
\hline
\multirow{7}{*}{1j, 0b} & 250--350 & $70\,700\pm400\pm4100$ & $167\,000\pm1000\pm11\,000$ & $530\pm20\pm160$ & $238\,000\pm1000\pm14\,000$ & 251\,941\\
 & 350--450 & $13\,440\pm130\pm790$ & $40\,100\pm500\pm3100$ & $55\pm5\pm16$ & $53\,600\pm500\pm3700$ & 54\,870\\
 & 450--575 & $3050\pm50\pm180$ & $10\,850^{+230}_{-220}\pm690$ & $5.6\pm1.1\pm1.6$ & $13\,910\pm230\pm840$ & 14\,473\\
 & 575--700 & $603^{+20}_{-19}\pm38$ & $2590^{+110}_{-100}\pm160$ & $0.38\pm0.06\pm0.11$ & $3200\pm110\pm190$ & 3432\\
 & 700--1000 & $220\pm13\pm16$ & $1076^{+70}_{-66}\pm66$ & $0.12\pm0.03\pm0.03$ & $1295^{+71}_{-67}\pm79$ & 1304\\
 & 1000--1200 & $11.7^{+4.1}_{-3.2}\pm0.9$ & $86^{+23}_{-19}\pm6$ & $<$0.01 & $98^{+24}_{-19}\pm7$ & 98\\
 & $\geq$1200 & $2.8^{+2.7}_{-1.5}\pm0.6$ & $23^{+12}_{-8}\pm2$ & $<$0.01 & $26^{+13}_{-9}\pm2$ & 30\\
[\cmsTabSkip]
\multirow{5}{*}{1j, $\geq$1b} & 250--350 & $4210\pm110\pm260$ & $9030\pm230\pm630$ & $58\pm10\pm17$ & $13\,310^{+260}_{-250}\pm820$ & 13\,549\\
 & 350--450 & $878\pm38\pm56$ & $2180^{+110}_{-100}\pm170$ & $4.6\pm0.4\pm1.3$ & $3060\pm110\pm220$ & 3078\\
 & 450--575 & $211^{+16}_{-15}\pm13$ & $651^{+57}_{-53}\pm44$ & $0.63\pm0.18\pm0.18$ & $863^{+59}_{-55}\pm53$ & 810\\
 & 575--700 & $40.3^{+6.0}_{-5.5}\pm2.5$ & $164^{+30}_{-26}\pm11$ & $0.04\pm0.02\pm0.02$ & $205^{+31}_{-26}\pm13$ & 184\\
 & $\geq$700 & $19.2^{+5.7}_{-4.6}\pm1.3$ & $74^{+21}_{-16}\pm7$ & $<$0.01 & $94^{+21}_{-17}\pm7$ & 83\\

    \hline
\end{tabular}
}
\end{table*}

\begin{table*}[htb!]
\setlength\tabcolsep{1.5mm}
\centering
\topcaption{Predictions and observations for the 30 search regions with $250 \leq \HT < 450\GeV$. For each of the background
predictions, the first uncertainty listed is statistical (from the limited size of data control samples
and Monte Carlo samples), and the second is systematic.}
\label{tab:yieldsVL}
\renewcommand{\arraystretch}{1.3}
\cmsTableAlt{
\begin{tabular}{c c  c c c c c}
    \hline
\multicolumn{7}{c}{$250 \leq \HT < 450\GeV$} \\ [\cmsTabSkip]
\njets, \nbtags & \mttwo [\GeVns{}] & Lost lepton & \znunu & Multijet & Total background & Data \\
\hline
\multirow{3}{*}{2-3j, 0b} & 200--300 & $73\,700\pm500\pm5000$ & $156\,000\pm1000\pm12\,000$ & $580\pm20\pm140$ & $231\,000\pm1000\pm16\,000$ & 240\,867\\
 & 300--400 & $12\,030\pm200\pm820$ & $31\,300\pm200\pm2500$ & $50\pm5\pm10$ & $43\,400\pm300\pm3200$ & 44\,074\\
 & $\geq$400 & $417^{+51}_{-47}\pm28$ & $1450\pm10\pm140$ & $0.44\pm0.09\pm0.09$ & $1870\pm50\pm160$ & 2022\\
[\cmsTabSkip]
\multirow{3}{*}{2-3j, 1b} & 200--300 & $12\,450\pm170\pm820$ & $18\,700\pm300\pm1500$ & $90\pm8\pm21$ & $31\,300\pm300\pm2200$ & 32\,120\\
 & 300--400 & $2380\pm80\pm160$ & $3750\pm60\pm310$ & $6.9\pm1.0\pm1.5$ & $6130\pm100\pm430$ & 6258\\
 & $\geq$400 & $97\pm8\pm39$ & $174\pm3\pm17$ & $0.01\pm0.01\pm0.00$ & $271^{+9}_{-8}\pm45$ & 275\\
[\cmsTabSkip]
\multirow{3}{*}{2-3j, 2b} & 200--300 & $2240\pm70\pm150$ & $2340^{+110}_{-100}\pm200$ & $9.7\pm1.1\pm2.3$ & $4600^{+130}_{-120}\pm320$ & 4709\\
 & 300--400 & $398^{+34}_{-32}\pm27$ & $469^{+21}_{-20}\pm39$ & $0.68\pm0.17\pm0.15$ & $868^{+40}_{-38}\pm61$ & 984\\
 & $\geq$400 & $13.3\pm2.3\pm5.4$ & $21.7^{+1.0}_{-0.9}\pm2.2$ & $<$0.01 & $35.0\pm2.5\pm6.0$ & 30\\
[\cmsTabSkip]
\multirow{3}{*}{2-6j, $\geq$3b} & 200--300 & $507^{+32}_{-31}\pm38$ & $179^{+35}_{-30}\pm27$ & $1.77\pm0.46\pm0.46$ & $688^{+47}_{-43}\pm54$ & 699\\
 & 300--400 & $69\pm6\pm15$ & $40.0^{+7.8}_{-6.6}\pm6.0$ & $0.16\pm0.12\pm0.04$ & $109^{+10}_{-9}\pm16$ & 102\\
 & $\geq$400 & $1.50\pm0.80\pm0.61$ & $1.43^{+0.28}_{-0.24}\pm0.25$ & $<$0.01 & $2.92^{+0.85}_{-0.83}\pm0.67$ & 0\\
[\cmsTabSkip]
\multirow{3}{*}{4-6j, 0b} & 200--300 & $12\,500\pm180\pm800$ & $21\,600\pm300\pm1800$ & $250\pm17\pm58$ & $34\,400\pm400\pm2400$ & 35\,187\\
 & 300--400 & $2070\pm80\pm130$ & $4660\pm70\pm410$ & $18.2\pm3.6\pm3.8$ & $6750\pm110\pm510$ & 6725\\
 & $\geq$400 & $42\pm5\pm17$ & $155\pm2\pm64$ & $0.06\pm0.03\pm0.01$ & $197\pm5\pm67$ & 170\\
[\cmsTabSkip]
\multirow{3}{*}{4-6j, 1b} & 200--300 & $5750\pm100\pm380$ & $4300\pm150\pm360$ & $61\pm7\pm15$ & $10\,120\pm180\pm680$ & 10\,564\\
 & 300--400 & $784^{+43}_{-42}\pm52$ & $928^{+32}_{-31}\pm84$ & $2.07\pm0.29\pm0.45$ & $1710\pm50\pm120$ & 1769\\
 & $\geq$400 & $14.0\pm2.5\pm5.7$ & $31\pm1\pm13$ & $0.04\pm0.02\pm0.01$ & $45\pm3\pm14$ & 40\\
[\cmsTabSkip]
\multirow{3}{*}{4-6j, 2b} & 200--300 & $2550^{+70}_{-60}\pm170$ & $921^{+68}_{-63}\pm87$ & $10.0\pm1.5\pm2.2$ & $3480\pm90\pm230$ & 3621\\
 & 300--400 & $220^{+23}_{-21}\pm15$ & $198^{+15}_{-14}\pm20$ & $0.47\pm0.15\pm0.11$ & $419^{+27}_{-25}\pm31$ & 496\\
 & $\geq$400 & $3.2\pm0.8\pm1.3$ & $6.6\pm0.5\pm2.7$ & $<$0.01 & $9.8\pm0.9\pm3.1$ & 14\\
[\cmsTabSkip]
\multirow{3}{*}{$\geq$7j, 0b} & 200--300 & $55^{+15}_{-13}\pm4$ & $61^{+23}_{-17}\pm26$ & $2.64\pm0.39\pm0.57$ & $119^{+28}_{-22}\pm27$ & 108\\
 & 300--500 & $3.8^{+2.1}_{-2.0}\pm0.8$ & $8.1^{+3.1}_{-2.3}\pm4.3$ & $0.08\pm0.04\pm0.02$ & $12.0^{+3.7}_{-3.1}\pm4.4$ & 30\\
 & $\geq$500 & $0.0^{+3.2}_{-0.0}\pm0.0$ & $0.0^{+1.2}_{-0.0}\pm0.0$ & $<$0.01 & $0.0^{+3.4}_{-0.0}\pm0.0$ & 0\\
[\cmsTabSkip]
\multirow{2}{*}{$\geq$7j, 1b} & 200--300 & $48.0^{+9.1}_{-8.2}\pm3.5$ & $19^{+19}_{-11}\pm10$ & $0.33\pm0.14\pm0.09$ & $68^{+21}_{-13}\pm11$ & 95\\
 & $\geq$300 & $3.0\pm1.4\pm1.2$ & $2.5^{+2.4}_{-1.3}\pm1.7$ & $0.03\pm0.02\pm0.01$ & $5.6^{+2.8}_{-1.9}\pm2.1$ & 12\\
[\cmsTabSkip]
\multirow{2}{*}{$\geq$7j, 2b} & 200--300 & $41.3^{+7.7}_{-7.0}\pm3.1$ & $6.0^{+5.8}_{-3.2}\pm3.7$ & $0.29\pm0.14\pm0.06$ & $47.6^{+9.7}_{-7.7}\pm5.0$ & 30\\
 & $\geq$300 & $2.15^{+0.78}_{-0.76}\pm0.87$ & $0.74^{+0.72}_{-0.40}\pm0.57$ & $<$0.01 & $2.9^{+1.1}_{-0.9}\pm1.1$ & 1\\
[\cmsTabSkip]
\multirow{2}{*}{$\geq$7j, $\geq$3b} & 200--300 & $7.3^{+1.7}_{-1.5}\pm0.9$ & $1.0^{+1.0}_{-0.6}\pm1.1$ & $0.04\pm0.04\pm0.01$ & $8.4^{+1.9}_{-1.6}\pm1.5$ & 17\\
 & $\geq$300 & $0.47\pm0.35\pm0.20$ & $0.12^{+0.11}_{-0.06}\pm0.14$ & $<$0.01 & $0.59^{+0.37}_{-0.35}\pm0.24$ & 0\\
    \hline

\end{tabular}
}
\end{table*}

\begin{table*}[htb!]
\setlength\tabcolsep{1.5mm}
\centering
\topcaption{Predictions and observations for the 28 search regions with $450 \leq \HT < 575\GeV$, and $2 \leq \njets \leq3$, $2 \leq \njets \leq 6$ and $\nbtags\geq3$, or $4 \leq \njets \leq 6$. For each of the background
predictions, the first uncertainty listed is statistical (from the limited size of data control samples
and Monte Carlo samples), and the second is systematic.}
\label{tab:yieldsLl}
\renewcommand{\arraystretch}{1.3}
\cmsTableAlt{
\begin{tabular}{c c  c c c c c}
    \hline
\multicolumn{7}{c}{$450 \leq \HT < 575\GeV$} \\ [\cmsTabSkip]
\njets, \nbtags & \mttwo [\GeVns{}] & Lost lepton & \znunu & Multijet & Total background & Data \\
\hline
\multirow{4}{*}{2-3j, 0b} & 200--300 & $8860\pm110\pm640$ & $20\,100\pm200\pm1300$ & $69\pm13\pm16$ & $29\,100\pm300\pm1900$ & 28\,956\\
 & 300--400 & $4230\pm80\pm300$ & $11\,770\pm140\pm790$ & $10.6\pm0.8\pm2.4$ & $16\,000\pm200\pm1000$ & 15\,876\\
 & 400--500 & $1510\pm60\pm110$ & $5020\pm60\pm360$ & $2.86\pm0.62\pm0.60$ & $6540\pm80\pm440$ & 6527\\
 & $\geq$500 & $121^{+24}_{-21}\pm9$ & $580\pm7\pm63$ & $0.07\pm0.03\pm0.02$ & $701^{+25}_{-22}\pm68$ & 740\\
[\cmsTabSkip]
\multirow{4}{*}{2-3j, 1b} & 200--300 & $1326\pm43\pm88$ & $2500\pm80\pm170$ & $17.0\pm8.4\pm3.8$ & $3840^{+100}_{-90}\pm240$ & 3859\\
 & 300--400 & $737\pm35\pm49$ & $1464^{+49}_{-48}\pm99$ & $1.62\pm0.20\pm0.43$ & $2200\pm60\pm140$ & 2065\\
 & 400--500 & $259^{+25}_{-23}\pm19$ & $626^{+21}_{-20}\pm45$ & $0.49\pm0.10\pm0.12$ & $885^{+32}_{-31}\pm58$ & 907\\
 & $\geq$500 & $19.1^{+2.8}_{-2.7}\pm7.8$ & $72.4\pm2.4\pm7.9$ & $0.04\pm0.02\pm0.02$ & $92\pm4\pm11$ & 79\\
[\cmsTabSkip]
\multirow{4}{*}{2-3j, 2b} & 200--300 & $201\pm15\pm13$ & $322^{+31}_{-28}\pm25$ & $1.34\pm0.62\pm0.47$ & $524^{+35}_{-32}\pm35$ & 463\\
 & 300--400 & $83.8^{+9.6}_{-9.1}\pm9.1$ & $188^{+18}_{-17}\pm15$ & $0.26\pm0.07\pm0.07$ & $272^{+21}_{-19}\pm20$ & 304\\
 & 400--500 & $31.8^{+4.1}_{-4.0}\pm6.7$ & $80.4^{+7.7}_{-7.1}\pm6.6$ & $0.02\pm0.01\pm0.01$ & $112^{+9}_{-8}\pm10$ & 120\\
 & $\geq$500 & $2.16^{+0.67}_{-0.66}\pm0.88$ & $9.3^{+0.9}_{-0.8}\pm1.1$ & $<$0.01 & $11.4\pm1.1\pm1.4$ & 15\\
[\cmsTabSkip]
\multirow{4}{*}{2-6j, $\geq$3b} & 200--300 & $232^{+17}_{-16}\pm15$ & $57^{+17}_{-13}\pm7$ & $2.20\pm0.70\pm0.80$ & $291^{+24}_{-21}\pm19$ & 297\\
 & 300--400 & $81^{+12}_{-11}\pm6$ & $33.6^{+9.9}_{-7.8}\pm4.3$ & $0.26\pm0.08\pm0.08$ & $115^{+16}_{-14}\pm8$ & 76\\
 & 400--500 & $10.7^{+2.1}_{-2.0}\pm2.3$ & $11.4^{+3.4}_{-2.7}\pm1.5$ & $<$0.01 & $22.1^{+4.0}_{-3.4}\pm2.8$ & 24\\
 & $\geq$500 & $1.08\pm0.58\pm0.44$ & $1.03^{+0.30}_{-0.24}\pm0.17$ & $<$0.01 & $2.11^{+0.65}_{-0.62}\pm0.48$ & 0\\
[\cmsTabSkip]
\multirow{4}{*}{4-6j, 0b} & 200--300 & $5660\pm90\pm370$ & $8560\pm170\pm600$ & $143\pm7\pm35$ & $14\,360\pm190\pm890$ & 15\,047\\
 & 300--400 & $2250\pm60\pm150$ & $4790^{+100}_{-90}\pm350$ & $24.3\pm2.6\pm6.2$ & $7060\pm110\pm460$ & 6939\\
 & 400--500 & $428^{+32}_{-30}\pm28$ & $1220\pm20\pm110$ & $1.42\pm0.21\pm0.52$ & $1650\pm40\pm130$ & 1817\\
 & $\geq$500 & $14.8\pm2.2\pm6.0$ & $86\pm2\pm35$ & $0.04\pm0.02\pm0.01$ & $101\pm3\pm36$ & 104\\
[\cmsTabSkip]
\multirow{4}{*}{4-6j, 1b} & 200--300 & $2810\pm60\pm190$ & $1880\pm80\pm130$ & $63\pm15\pm19$ & $4750\pm100\pm300$ & 4736\\
 & 300--400 & $937\pm36\pm63$ & $1054^{+45}_{-43}\pm78$ & $5.4\pm0.4\pm1.4$ & $2000\pm60\pm130$ & 2039\\
 & 400--500 & $138^{+17}_{-16}\pm10$ & $269\pm11\pm25$ & $0.36\pm0.10\pm0.10$ & $407^{+20}_{-19}\pm31$ & 403\\
 & $\geq$500 & $7.5\pm2.2\pm3.0$ & $19.1\pm0.8\pm7.9$ & $0.01\pm0.01\pm0.00$ & $26.5\pm2.3\pm8.5$ & 27\\
[\cmsTabSkip]
\multirow{4}{*}{4-6j, 2b} & 200--300 & $1343^{+38}_{-37}\pm89$ & $414^{+39}_{-35}\pm33$ & $11.5\pm1.0\pm3.3$ & $1770\pm50\pm110$ & 1767\\
 & 300--400 & $418^{+24}_{-23}\pm29$ & $232^{+22}_{-20}\pm19$ & $1.35\pm0.35\pm0.39$ & $651^{+32}_{-31}\pm43$ & 636\\
 & 400--500 & $45.6^{+3.9}_{-3.8}\pm9.6$ & $59.1^{+5.5}_{-5.1}\pm5.9$ & $0.03\pm0.02\pm0.01$ & $105^{+7}_{-6}\pm12$ & 120\\
 & $\geq$500 & $1.59\pm0.89\pm0.65$ & $4.2\pm0.4\pm1.7$ & $<$0.01 & $5.8\pm1.0\pm1.9$ & 7\\
    \hline

\end{tabular}
}
\end{table*}

\begin{table*}[htb!]
\setlength\tabcolsep{1.5mm}
\centering
\topcaption{Predictions and observations for the 12 search regions with $450 \leq \HT < 575\GeV$ and $\njets \geq 7$. For each of the background
predictions, the first uncertainty listed is statistical (from the limited size of data control samples
and Monte Carlo samples), and the second is systematic.}
\label{tab:yieldsLh}
\renewcommand{\arraystretch}{1.3}
\cmsTableAlt{
\begin{tabular}{c c  c c c c c}
    \hline
\multicolumn{7}{c}{$450 \leq \HT < 575\GeV$} \\ [\cmsTabSkip]
\njets, \nbtags & \mttwo [\GeVns{}] & Lost lepton & \znunu & Multijet & Total background & Data \\
\hline
\multirow{3}{*}{$\geq$7j, 0b} & 200--300 & $149^{+17}_{-16}\pm13$ & $169^{+31}_{-27}\pm34$ & $11.5\pm0.8\pm3.0$ & $329^{+36}_{-31}\pm38$ & 354\\
 & 300--400 & $38.9^{+5.8}_{-5.6}\pm8.2$ & $64^{+12}_{-10}\pm17$ & $1.24\pm0.42\pm0.32$ & $104^{+13}_{-12}\pm20$ & 110\\
 & $\geq$400 & $1.28\pm0.82\pm0.52$ & $8.8^{+1.6}_{-1.4}\pm3.8$ & $0.03\pm0.02\pm0.01$ & $10.1^{+1.8}_{-1.6}\pm3.8$ & 10\\
[\cmsTabSkip]
\multirow{3}{*}{$\geq$7j, 1b} & 200--300 & $191^{+13}_{-12}\pm15$ & $67^{+19}_{-15}\pm15$ & $4.4\pm0.5\pm1.2$ & $262^{+23}_{-19}\pm23$ & 268\\
 & 300--400 & $37.8^{+3.4}_{-3.3}\pm8.0$ & $25.3^{+7.2}_{-5.7}\pm7.3$ & $0.30\pm0.07\pm0.08$ & $63^{+8}_{-7}\pm11$ & 65\\
 & $\geq$400 & $2.31\pm0.69\pm0.94$ & $3.5^{+1.0}_{-0.8}\pm1.5$ & $0.01\pm0.01\pm0.00$ & $5.8^{+1.2}_{-1.0}\pm1.8$ & 3\\
[\cmsTabSkip]
\multirow{3}{*}{$\geq$7j, 2b} & 200--300 & $173^{+12}_{-11}\pm13$ & $19.9^{+5.7}_{-4.5}\pm5.2$ & $1.24\pm0.18\pm0.33$ & $194^{+13}_{-12}\pm15$ & 197\\
 & 300--400 & $26.8\pm2.6\pm5.7$ & $7.6^{+2.2}_{-1.7}\pm2.4$ & $0.09\pm0.04\pm0.03$ & $34.6^{+3.4}_{-3.1}\pm6.3$ & 44\\
 & $\geq$400 & $1.40\pm0.44\pm0.57$ & $1.02^{+0.29}_{-0.23}\pm0.46$ & $<$0.01 & $2.42^{+0.53}_{-0.49}\pm0.73$ & 3\\
[\cmsTabSkip]
\multirow{3}{*}{$\geq$7j, $\geq$3b} & 200--300 & $55.4^{+4.8}_{-4.7}\pm7.3$ & $2.3^{+0.7}_{-0.5}\pm1.1$ & $0.15\pm0.06\pm0.06$ & $57.8^{+4.8}_{-4.7}\pm7.4$ & 37\\
 & 300--400 & $6.4\pm1.2\pm1.5$ & $0.86^{+0.25}_{-0.20}\pm0.46$ & $0.01\pm0.01\pm0.00$ & $7.3\pm1.2\pm1.6$ & 9\\
 & $\geq$400 & $0.06\pm0.01\pm0.03$ & $0.12\pm0.03\pm0.06$ & $<$0.01 & $0.18^{+0.04}_{-0.03}\pm0.07$ & 0\\
    \hline

\end{tabular}
}
\end{table*}

\begin{table*}[htb!]
\setlength\tabcolsep{1.5mm}
\centering
\topcaption{Predictions and observations for the 21 search regions with $575 \leq \HT < 1200\GeV$ and $2 \leq \njets \leq3$. For each of the background
predictions, the first uncertainty listed is statistical (from the limited size of data control samples
and Monte Carlo samples), and the second is systematic.}
\label{tab:yieldsMll}
\renewcommand{\arraystretch}{1.3}
\cmsTableAlt{
\begin{tabular}{c c  c c c c c}
    \hline
\multicolumn{7}{c}{$575 \leq \HT < 1200\GeV$} \\ [\cmsTabSkip]
\njets, \nbtags & \mttwo [\GeVns{}] & Lost lepton & \znunu & Multijet & Total background & Data \\
\hline
\multirow{10}{*}{2-3j, 0b} & 200--300 & $5270\pm60\pm370$ & $11\,550\pm160\pm790$ & $93\pm20\pm30$ & $16\,900\pm200\pm1100$ & 17\,256\\
 & 300--400 & $2560\pm50\pm180$ & $7770^{+110}_{-100}\pm540$ & $11.9\pm1.3\pm4.4$ & $10\,340^{+120}_{-110}\pm680$ & 10\,145\\
 & 400--500 & $1101^{+32}_{-31}\pm77$ & $3900\pm50\pm280$ & $1.33\pm0.24\pm0.41$ & $5000\pm60\pm340$ & 5021\\
 & 500--600 & $502^{+24}_{-23}\pm35$ & $2250\pm30\pm170$ & $0.37\pm0.07\pm0.12$ & $2760\pm40\pm200$ & 2706\\
 & 600--700 & $180^{+16}_{-15}\pm13$ & $746\pm10\pm73$ & $0.09\pm0.03\pm0.03$ & $926^{+19}_{-18}\pm80$ & 1066\\
 & 700--800 & $52.1^{+7.3}_{-6.5}\pm5.5$ & $256\pm3\pm36$ & $0.01\pm0.01\pm0.00$ & $308^{+8}_{-7}\pm38$ & 347\\
 & 800--900 & $17.7^{+2.6}_{-2.3}\pm2.2$ & $107\pm1\pm20$ & $<$0.01 & $125\pm3\pm21$ & 111\\
 & 900--1000 & $6.0\pm0.9\pm1.3$ & $39.4\pm0.5\pm8.5$ & $0.01\pm0.01\pm0.00$ & $45.4^{+1.1}_{-1.0}\pm8.7$ & 39\\
 & 1000--1100 & $3.3^{+1.1}_{-1.0}\pm1.0$ & $13.3\pm0.2\pm3.9$ & $<$0.01 & $16.6\pm1.1\pm4.1$ & 11\\
 & $\geq$1100 & $0.31^{+0.09}_{-0.08}\pm0.12$ & $2.5\pm0.0\pm1.1$ & $<$0.01 & $2.8\pm0.1\pm1.1$ & 2\\
[\cmsTabSkip]
\multirow{6}{*}{2-3j, 1b} & 200--300 & $826^{+27}_{-26}\pm54$ & $1480^{+60}_{-50}\pm100$ & $38\pm15\pm12$ & $2340\pm60\pm140$ & 2499\\
 & 300--400 & $426^{+21}_{-20}\pm28$ & $994^{+38}_{-37}\pm69$ & $2.33\pm0.26\pm0.84$ & $1422^{+43}_{-42}\pm90$ & 1366\\
 & 400--600 & $282^{+18}_{-17}\pm20$ & $788^{+30}_{-29}\pm55$ & $0.27\pm0.06\pm0.10$ & $1071^{+35}_{-34}\pm69$ & 1057\\
 & 600--800 & $43.5^{+3.2}_{-3.1}\pm6.5$ & $129\pm5\pm12$ & $<$0.01 & $172\pm6\pm15$ & 225\\
 & 800--1000 & $4.6\pm0.7\pm1.3$ & $18.8\pm0.7\pm3.3$ & $<$0.01 & $23.4\pm1.0\pm3.6$ & 22\\
 & $\geq$1000 & $0.34\pm0.08\pm0.14$ & $2.05\pm0.08\pm0.90$ & $<$0.01 & $2.38\pm0.11\pm0.91$ & 1\\
[\cmsTabSkip]
\multirow{5}{*}{2-3j, 2b} & 200--300 & $105.1^{+9.2}_{-8.7}\pm7.6$ & $181^{+20}_{-18}\pm15$ & $3.8\pm0.5\pm1.3$ & $290^{+22}_{-20}\pm20$ & 316\\
 & 300--400 & $55.0^{+6.7}_{-6.3}\pm7.5$ & $122^{+14}_{-12}\pm10$ & $0.27\pm0.06\pm0.10$ & $177^{+15}_{-14}\pm14$ & 159\\
 & 400--600 & $36.5^{+4.6}_{-4.3}\pm5.5$ & $97^{+11}_{-10}\pm8$ & $0.08\pm0.03\pm0.03$ & $133^{+12}_{-11}\pm11$ & 107\\
 & 600--800 & $4.7\pm0.8\pm1.3$ & $15.8^{+1.8}_{-1.6}\pm1.6$ & $<$0.01 & $20.6^{+1.9}_{-1.8}\pm2.2$ & 21\\
 & $\geq$800 & $0.59\pm0.19\pm0.24$ & $2.56^{+0.29}_{-0.26}\pm0.45$ & $<$0.01 & $3.14^{+0.35}_{-0.32}\pm0.52$ & 1\\
    \hline

\end{tabular}
}
\end{table*}

\begin{table*}[htb!]
\setlength\tabcolsep{1.5mm}
\centering
\topcaption{Predictions and observations for the 26 search regions with $575 \leq \HT < 1200\GeV$, and $2\leq \njets \leq6$ and $\nbtags\geq3$, or $4\leq \njets \leq6$. For each of the background
predictions, the first uncertainty listed is statistical (from the limited size of data control samples
and Monte Carlo samples), and the second is systematic.}
\label{tab:yieldsMl}
\renewcommand{\arraystretch}{1.3}
\cmsTableAlt{
\begin{tabular}{c c  c c c c c}
    \hline
\multicolumn{7}{c}{$575 \leq \HT < 1200\GeV$} \\ [\cmsTabSkip]
\njets, \nbtags & \mttwo [\GeVns{}] & Lost lepton & \znunu & Multijet & Total background & Data \\
\hline
\multirow{5}{*}{2-6j, $\geq$3b} & 200--300 & $299^{+17}_{-16}\pm22$ & $73^{+15}_{-13}\pm10$ & $6.2\pm0.4\pm2.1$ & $379^{+22}_{-21}\pm28$ & 345\\
 & 300--400 & $100\pm10\pm7$ & $43.5^{+8.8}_{-7.4}\pm6.2$ & $0.68\pm0.09\pm0.24$ & $144^{+14}_{-12}\pm11$ & 132\\
 & 400--600 & $32.5^{+6.3}_{-5.6}\pm2.5$ & $31.2^{+6.3}_{-5.3}\pm4.4$ & $0.08\pm0.03\pm0.03$ & $63.8^{+8.9}_{-7.7}\pm5.8$ & 48\\
 & 600--800 & $3.16^{+0.95}_{-0.90}\pm0.68$ & $5.4^{+1.1}_{-0.9}\pm0.8$ & $<$0.01 & $8.6^{+1.4}_{-1.3}\pm1.1$ & 4\\
 & $\geq$800 & $0.10\pm0.03\pm0.04$ & $0.71^{+0.14}_{-0.12}\pm0.15$ & $<$0.01 & $0.81^{+0.15}_{-0.12}\pm0.16$ & 0\\
[\cmsTabSkip]
\multirow{10}{*}{4-6j, 0b} & 200--300 & $6280\pm70\pm420$ & $9470\pm160\pm650$ & $360\pm20\pm110$ & $16\,100\pm180\pm1000$ & 16\,292\\
 & 300--400 & $2700\pm50\pm180$ & $5410\pm90\pm380$ & $53\pm1\pm17$ & $8160\pm100\pm520$ & 8330\\
 & 400--500 & $927^{+28}_{-27}\pm62$ & $2420\pm40\pm180$ & $7.7\pm0.4\pm2.4$ & $3350\pm50\pm230$ & 3576\\
 & 500--600 & $324^{+17}_{-16}\pm22$ & $1171^{+20}_{-19}\pm100$ & $1.46\pm0.12\pm0.46$ & $1500\pm30\pm110$ & 1516\\
 & 600--700 & $95.4^{+9.4}_{-8.7}\pm6.4$ & $413\pm7\pm47$ & $0.33\pm0.06\pm0.10$ & $509^{+12}_{-11}\pm50$ & 543\\
 & 700--800 & $35.6^{+5.0}_{-4.5}\pm3.6$ & $171\pm3\pm27$ & $0.03\pm0.02\pm0.01$ & $206^{+6}_{-5}\pm27$ & 178\\
 & 800--900 & $13.4^{+2.0}_{-1.8}\pm1.6$ & $64\pm1\pm11$ & $0.02\pm0.01\pm0.01$ & $77\pm2\pm11$ & 62\\
 & 900--1000 & $4.39^{+0.78}_{-0.73}\pm0.93$ & $23.6\pm0.4\pm5.3$ & $<$0.01 & $28.0^{+0.9}_{-0.8}\pm5.4$ & 20\\
 & 1000--1100 & $0.64\pm0.16\pm0.20$ & $6.3\pm0.1\pm2.0$ & $<$0.01 & $6.9\pm0.2\pm2.0$ & 3\\
 & $\geq$1100 & $0.78\pm0.58\pm0.32$ & $0.89^{+0.02}_{-0.01}\pm0.40$ & $<$0.01 & $1.68\pm0.58\pm0.52$ & 1\\
[\cmsTabSkip]
\multirow{6}{*}{4-6j, 1b} & 200--300 & $2900\pm50\pm200$ & $2220^{+80}_{-70}\pm150$ & $154\pm16\pm50$ & $5270\pm90\pm330$ & 5335\\
 & 300--400 & $1066\pm29\pm74$ & $1267^{+44}_{-42}\pm89$ & $19.2\pm0.9\pm6.2$ & $2350\pm50\pm150$ & 2547\\
 & 400--600 & $504^{+22}_{-21}\pm35$ & $840^{+29}_{-28}\pm61$ & $2.98\pm0.21\pm0.93$ & $1347^{+36}_{-35}\pm88$ & 1284\\
 & 600--800 & $35.3^{+5.9}_{-5.2}\pm2.6$ & $138\pm5\pm14$ & $0.09\pm0.03\pm0.03$ & $174^{+8}_{-7}\pm16$ & 151\\
 & 800--1000 & $3.89^{+0.83}_{-0.77}\pm0.82$ & $19.3^{+0.7}_{-0.6}\pm4.3$ & $0.01\pm0.01\pm0.00$ & $23.2^{+1.1}_{-1.0}\pm4.5$ & 18\\
 & $\geq$1000 & $0.18\pm0.07\pm0.07$ & $1.57\pm0.05\pm0.65$ & $<$0.01 & $1.75\pm0.09\pm0.65$ & 1\\
[\cmsTabSkip]
\multirow{5}{*}{4-6j, 2b} & 200--300 & $1500\pm30\pm100$ & $473^{+36}_{-33}\pm36$ & $42\pm2\pm13$ & $2020\pm50\pm130$ & 1968\\
 & 300--400 & $508\pm20\pm35$ & $270^{+20}_{-19}\pm21$ & $4.9\pm0.3\pm1.6$ & $783^{+29}_{-28}\pm50$ & 788\\
 & 400--600 & $167\pm12\pm12$ & $179^{+14}_{-13}\pm14$ & $0.57\pm0.08\pm0.18$ & $346^{+18}_{-17}\pm23$ & 354\\
 & 600--800 & $11.9^{+1.3}_{-1.2}\pm2.5$ & $29.5^{+2.2}_{-2.1}\pm3.5$ & $0.02\pm0.01\pm0.01$ & $41.4^{+2.6}_{-2.4}\pm4.6$ & 37\\
 & $\geq$800 & $0.91\pm0.23\pm0.37$ & $4.4\pm0.3\pm1.8$ & $<$0.01 & $5.4\pm0.4\pm1.9$ & 7\\
    \hline

\end{tabular}
}
\end{table*}

\begin{table*}[!htbp]
\setlength\tabcolsep{1.5mm}
\centering
\topcaption{Predictions and observations for the 34 search regions with $575 \leq \HT < 1200\GeV$, and $7\leq \njets \leq9$, or $\njets \geq10$. For each of the background
predictions, the first uncertainty listed is statistical (from the limited size of data control samples
and Monte Carlo samples), and the second is systematic.}
\label{tab:yieldsMh}
\renewcommand{\arraystretch}{1.3}
\cmsTableAlt{
\begin{tabular}{c c  c c c c c}
    \hline
\multicolumn{7}{c}{$575 \leq \HT < 1200\GeV$} \\ [\cmsTabSkip]
\njets, \nbtags & \mttwo [\GeVns{}] & Lost lepton & \znunu & Multijet & Total background & Data \\
\hline
\multirow{5}{*}{7-9j, 0b} & 200--300 & $589^{+27}_{-26}\pm39$ & $573^{+47}_{-43}\pm64$ & $90\pm10\pm28$ & $1252^{+55}_{-52}\pm93$ & 1340\\
 & 300--400 & $265^{+19}_{-18}\pm18$ & $279^{+23}_{-21}\pm42$ & $14.9\pm0.5\pm4.7$ & $559^{+29}_{-28}\pm51$ & 581\\
 & 400--600 & $92^{+10}_{-9}\pm6$ & $159^{+13}_{-12}\pm28$ & $2.72\pm0.18\pm0.85$ & $253^{+16}_{-15}\pm30$ & 243\\
 & 600--800 & $8.6\pm1.2\pm1.8$ & $22.8^{+1.9}_{-1.7}\pm6.4$ & $0.10\pm0.03\pm0.03$ & $31.6^{+2.2}_{-2.1}\pm6.8$ & 32\\
 & $\geq$800 & $0.51\pm0.16\pm0.21$ & $3.0\pm0.2\pm1.3$ & $<$0.01 & $3.5\pm0.3\pm1.3$ & 2\\
[\cmsTabSkip]
\multirow{5}{*}{7-9j, 1b} & 200--300 & $733\pm21\pm52$ & $278^{+28}_{-25}\pm33$ & $48\pm3\pm16$ & $1059^{+35}_{-33}\pm73$ & 1052\\
 & 300--400 & $252^{+13}_{-12}\pm18$ & $135^{+14}_{-12}\pm21$ & $7.7\pm0.4\pm2.5$ & $395^{+19}_{-17}\pm32$ & 387\\
 & 400--600 & $71.3^{+6.9}_{-6.5}\pm5.2$ & $77^{+8}_{-7}\pm14$ & $1.36\pm0.13\pm0.45$ & $150\pm10\pm16$ & 131\\
 & 600--800 & $4.26^{+0.73}_{-0.71}\pm0.90$ & $11.0^{+1.1}_{-1.0}\pm3.1$ & $0.03\pm0.02\pm0.01$ & $15.3^{+1.3}_{-1.2}\pm3.3$ & 20\\
 & $\geq$800 & $0.11\pm0.04\pm0.05$ & $1.48^{+0.15}_{-0.13}\pm0.63$ & $<$0.01 & $1.60^{+0.15}_{-0.14}\pm0.63$ & 1\\
[\cmsTabSkip]
\multirow{5}{*}{7-9j, 2b} & 200--300 & $675\pm20\pm51$ & $82^{+8}_{-7}\pm10$ & $20.9\pm3.0\pm6.7$ & $777^{+22}_{-21}\pm56$ & 750\\
 & 300--400 & $211\pm11\pm16$ & $39.8^{+4.0}_{-3.6}\pm6.4$ & $2.42\pm0.19\pm0.79$ & $253^{+12}_{-11}\pm19$ & 259\\
 & 400--600 & $55.4^{+5.5}_{-5.2}\pm4.2$ & $22.7^{+2.3}_{-2.1}\pm4.2$ & $0.50\pm0.07\pm0.16$ & $78.6^{+5.9}_{-5.6}\pm6.6$ & 72\\
 & 600--800 & $3.00^{+0.63}_{-0.62}\pm0.64$ & $3.25^{+0.32}_{-0.30}\pm0.93$ & $0.01\pm0.01\pm0.01$ & $6.3\pm0.7\pm1.2$ & 7\\
 & $\geq$800 & $0.27\pm0.20\pm0.11$ & $0.44\pm0.04\pm0.19$ & $<$0.01 & $0.71\pm0.20\pm0.22$ & 1\\
[\cmsTabSkip]
\multirow{4}{*}{7-9j, 3b} & 200--300 & $185\pm8\pm18$ & $11.3^{+1.1}_{-1.0}\pm1.9$ & $3.6\pm0.2\pm1.2$ & $200\pm8\pm18$ & 184\\
 & 300--400 & $52.0\pm3.8\pm5.0$ & $5.5\pm0.5\pm1.2$ & $0.72\pm0.12\pm0.26$ & $58.3^{+3.9}_{-3.8}\pm5.3$ & 59\\
 & 400--600 & $13.6\pm1.8\pm1.3$ & $3.13^{+0.31}_{-0.29}\pm0.82$ & $0.05\pm0.02\pm0.02$ & $16.8\pm1.8\pm1.6$ & 14\\
 & $\geq$600 & $0.49\pm0.21\pm0.20$ & $0.51\pm0.05\pm0.21$ & $<$0.01 & $1.00\pm0.21\pm0.29$ & 2\\
[\cmsTabSkip]
\multirow{3}{*}{7-9j, $\geq$4b} & 200--300 & $38.8\pm3.1\pm7.4$ & $2.01^{+0.20}_{-0.18}\pm0.71$ & $0.55\pm0.08\pm0.19$ & $41.3^{+3.2}_{-3.1}\pm7.4$ & 38\\
 & 300--400 & $14.5^{+2.0}_{-1.9}\pm2.8$ & $0.98^{+0.10}_{-0.09}\pm0.43$ & $0.06\pm0.02\pm0.02$ & $15.6^{+2.0}_{-1.9}\pm2.8$ & 16\\
 & $\geq$400 & $3.75^{+0.98}_{-0.97}\pm0.70$ & $0.65\pm0.06\pm0.35$ & $<$0.01 & $4.40^{+0.98}_{-0.97}\pm0.79$ & 3\\
[\cmsTabSkip]
\multirow{3}{*}{$\geq$10j, 0b} & 200--300 & $11.5\pm1.6\pm1.0$ & $4.4^{+0.4}_{-0.3}\pm2.3$ & $3.1\pm0.8\pm1.1$ & $19.0\pm1.8\pm2.8$ & 27\\
 & 300--500 & $5.6\pm1.0\pm0.5$ & $3.0\pm0.2\pm1.7$ & $0.55\pm0.08\pm0.20$ & $9.1\pm1.0\pm1.8$ & 4\\
 & $\geq$500 & $0.30\pm0.11\pm0.12$ & $0.44^{+0.04}_{-0.03}\pm0.24$ & $0.02\pm0.01\pm0.01$ & $0.76\pm0.11\pm0.27$ & 3\\
[\cmsTabSkip]
\multirow{3}{*}{$\geq$10j, 1b} & 200--300 & $21.0\pm1.8\pm1.6$ & $3.5\pm0.3\pm1.9$ & $1.92\pm0.18\pm0.72$ & $26.4\pm1.8\pm2.7$ & 32\\
 & 300--500 & $7.7\pm1.0\pm0.6$ & $2.4\pm0.2\pm1.4$ & $0.45\pm0.07\pm0.17$ & $10.5\pm1.1\pm1.6$ & 15\\
 & $\geq$500 & $0.83^{+0.42}_{-0.41}\pm0.07$ & $0.36^{+0.04}_{-0.03}\pm0.20$ & $0.02\pm0.01\pm0.01$ & $1.20^{+0.42}_{-0.41}\pm0.22$ & 0\\
[\cmsTabSkip]
\multirow{3}{*}{$\geq$10j, 2b} & 200--300 & $21.8\pm1.8\pm1.6$ & $1.05\pm0.10\pm0.66$ & $0.64\pm0.08\pm0.24$ & $23.5\pm1.8\pm1.8$ & 26\\
 & 300--500 & $8.8\pm1.2\pm0.6$ & $0.69^{+0.07}_{-0.06}\pm0.45$ & $0.16\pm0.04\pm0.06$ & $9.6^{+1.3}_{-1.2}\pm0.8$ & 9\\
 & $\geq$500 & $0.22\pm0.13\pm0.02$ & $0.10\pm0.01\pm0.06$ & $<$0.01 & $0.32\pm0.13\pm0.07$ & 0\\
[\cmsTabSkip]
\multirow{2}{*}{$\geq$10j, 3b} & 200--300 & $9.9\pm1.3\pm1.2$ & $0.25\pm0.02\pm0.20$ & $0.29\pm0.05\pm0.12$ & $10.4\pm1.3\pm1.2$ & 14\\
 & $\geq$300 & $1.59\pm0.50\pm0.18$ & $0.19\pm0.02\pm0.16$ & $0.02\pm0.01\pm0.01$ & $1.80\pm0.50\pm0.25$ & 2\\
[\cmsTabSkip]
\multirow{1}{*}{$\geq$10j, $\geq$4b} & $\geq200$ & $3.9\pm1.2\pm0.8$ & $0.00^{+0.17}_{-0.00}\pm0.00$ & $0.05\pm0.02\pm0.02$ & $4.0\pm1.2\pm0.8$ & 6\\
    \hline

\end{tabular}
}
\end{table*}

\begin{table*}[!htbp]
\setlength\tabcolsep{1.5mm}
\centering
\topcaption{Predictions and observations for the 17 search regions with $1200 \leq \HT < 1500\GeV$ and $2\leq \njets \leq3$. For each of the background
predictions, the first uncertainty listed is statistical (from the limited size of data control samples
and Monte Carlo samples), and the second is systematic.}
\label{tab:yieldsHll}
\renewcommand{\arraystretch}{1.3}
\cmsTableAlt{
\begin{tabular}{c c  c c c c c}
    \hline
\multicolumn{7}{c}{$1200 \leq \HT < 1500\GeV$} \\ [\cmsTabSkip]
\njets, \nbtags & \mttwo [\GeVns{}] & Lost lepton & \znunu & Multijet & Total background & Data \\
\hline
\multirow{6}{*}{2-3j, 0b} & 200--400 & $315\pm15\pm21$ & $656^{+51}_{-47}\pm73$ & $39\pm16\pm12$ & $1009^{+55}_{-52}\pm85$ & 1128\\
 & 400--600 & $43.0^{+5.2}_{-4.7}\pm4.9$ & $185^{+14}_{-13}\pm30$ & $0.03\pm0.02\pm0.01$ & $228^{+15}_{-14}\pm31$ & 207\\
 & 600--800 & $14.1^{+2.1}_{-2.0}\pm1.7$ & $64\pm5\pm17$ & $<$0.01 & $78\pm5\pm17$ & 83\\
 & 800--1000 & $6.4^{+1.1}_{-1.0}\pm1.3$ & $32.5^{+2.5}_{-2.3}\pm7.6$ & $<$0.01 & $38.9^{+2.7}_{-2.5}\pm7.8$ & 36\\
 & 1000--1200 & $3.23^{+0.61}_{-0.59}\pm0.99$ & $17.5\pm1.3\pm5.2$ & $<$0.01 & $20.7^{+1.5}_{-1.4}\pm5.3$ & 19\\
 & $\geq$1200 & $0.87^{+0.14}_{-0.13}\pm0.35$ & $6.0^{+0.5}_{-0.4}\pm2.6$ & $<$0.01 & $6.9\pm0.5\pm2.6$ & 4\\
[\cmsTabSkip]
\multirow{6}{*}{2-3j, 1b} & 200--400 & $61.5^{+7.2}_{-6.5}\pm4.2$ & $78^{+19}_{-16}\pm10$ & $9.7\pm0.7\pm3.0$ & $149^{+21}_{-17}\pm12$ & 157\\
 & 400--600 & $10.1\pm1.4\pm1.0$ & $21.9^{+5.4}_{-4.4}\pm3.8$ & $0.03\pm0.02\pm0.01$ & $32.0^{+5.6}_{-4.6}\pm4.1$ & 27\\
 & 600--800 & $2.36^{+0.36}_{-0.35}\pm0.41$ & $7.5^{+1.9}_{-1.5}\pm2.0$ & $<$0.01 & $9.8^{+1.9}_{-1.6}\pm2.1$ & 9\\
 & 800--1000 & $0.78^{+0.16}_{-0.15}\pm0.19$ & $3.84^{+0.95}_{-0.78}\pm0.93$ & $<$0.01 & $4.62^{+0.97}_{-0.79}\pm0.96$ & 6\\
 & 1000--1200 & $0.43\pm0.08\pm0.14$ & $2.13^{+0.53}_{-0.43}\pm0.64$ & $<$0.01 & $2.56^{+0.54}_{-0.44}\pm0.66$ & 2\\
 & $\geq$1200 & $0.14^{+0.05}_{-0.04}\pm0.06$ & $0.71^{+0.18}_{-0.14}\pm0.31$ & $<$0.01 & $0.86^{+0.18}_{-0.15}\pm0.31$ & 0\\
[\cmsTabSkip]
\multirow{5}{*}{2-3j, 2b} & 200--400 & $4.8^{+2.0}_{-1.6}\pm0.3$ & $11^{+11}_{-6}\pm2$ & $1.38\pm0.13\pm0.43$ & $18^{+11}_{-6}\pm2$ & 18\\
 & 400--600 & $0.61^{+0.30}_{-0.25}\pm0.07$ & $3.2^{+3.1}_{-1.7}\pm0.7$ & $<$0.01 & $3.8^{+3.1}_{-1.8}\pm0.7$ & 5\\
 & 600--800 & $0.21^{+0.11}_{-0.09}\pm0.04$ & $1.1^{+1.1}_{-0.6}\pm0.4$ & $<$0.01 & $1.3^{+1.1}_{-0.6}\pm0.4$ & 2\\
 & 800--1000 & $0.07^{+0.04}_{-0.03}\pm0.02$ & $0.56^{+0.55}_{-0.31}\pm0.18$ & $<$0.01 & $0.63^{+0.55}_{-0.31}\pm0.18$ & 1\\
 & $\geq$1000 & $0.03\pm0.02\pm0.01$ & $0.42^{+0.41}_{-0.23}\pm0.18$ & $<$0.01 & $0.46^{+0.41}_{-0.23}\pm0.18$ & 1\\
    \hline

\end{tabular}
}
\end{table*}

\begin{table*}[!htbp]
\setlength\tabcolsep{1.5mm}
\centering
\topcaption{Predictions and observations for the 20 search regions with $1200 \leq \HT < 1500\GeV$, and $2\leq \njets \leq6$ and $\nbtags\geq3$, or $4\leq \njets \leq6$. For each of the background
predictions, the first uncertainty listed is statistical (from the limited size of data control samples
and Monte Carlo samples), and the second is systematic.}
\label{tab:yieldsHl}
\renewcommand{\arraystretch}{1.3}
\cmsTableAlt{
\begin{tabular}{c c  c c c c c}
    \hline
\multicolumn{7}{c}{$1200 \leq \HT < 1500\GeV$} \\ [\cmsTabSkip]
\njets, \nbtags & \mttwo [\GeVns{}] & Lost lepton & \znunu & Multijet & Total background & Data \\
\hline
\multirow{3}{*}{2-6j, $\geq$3b} & 200--400 & $22.6^{+4.7}_{-4.2}\pm1.8$ & $0.0^{+6.6}_{-0.0}\pm0.0$ & $4.4\pm0.2\pm1.5$ & $27.0^{+8.1}_{-4.2}\pm2.4$ & 25\\
 & 400--600 & $1.58^{+0.51}_{-0.48}\pm0.34$ & $0.0^{+1.6}_{-0.0}\pm0.0$ & $0.02\pm0.01\pm0.01$ & $1.6^{+1.7}_{-0.5}\pm0.3$ & 3\\
 & $\geq$600 & $0.47^{+0.27}_{-0.26}\pm0.19$ & $0.00^{+0.94}_{-0.00}\pm0.00$ & $<$0.01 & $0.47^{+0.98}_{-0.26}\pm0.19$ & 4\\
[\cmsTabSkip]
\multirow{6}{*}{4-6j, 0b} & 200--400 & $606^{+21}_{-20}\pm41$ & $909^{+63}_{-59}\pm90$ & $208\pm12\pm64$ & $1720^{+70}_{-60}\pm130$ & 1768\\
 & 400--600 & $84.3^{+7.4}_{-6.9}\pm5.8$ & $234^{+16}_{-15}\pm34$ & $0.88\pm0.09\pm0.27$ & $319^{+18}_{-17}\pm36$ & 301\\
 & 600--800 & $21.1^{+3.2}_{-2.9}\pm2.3$ & $75\pm5\pm17$ & $0.06\pm0.02\pm0.02$ & $96\pm6\pm17$ & 99\\
 & 800--1000 & $7.6^{+1.2}_{-1.1}\pm1.1$ & $35.2^{+2.4}_{-2.3}\pm8.0$ & $0.01\pm0.01\pm0.00$ & $42.7^{+2.7}_{-2.5}\pm8.2$ & 41\\
 & 1000--1200 & $2.23^{+0.36}_{-0.33}\pm0.61$ & $14.1^{+1.0}_{-0.9}\pm4.2$ & $<$0.01 & $16.3\pm1.0\pm4.2$ & 15\\
 & $\geq$1200 & $0.47^{+0.10}_{-0.09}\pm0.19$ & $3.0\pm0.2\pm1.3$ & $<$0.01 & $3.5\pm0.2\pm1.3$ & 5\\
[\cmsTabSkip]
\multirow{6}{*}{4-6j, 1b} & 200--400 & $278^{+15}_{-14}\pm20$ & $254^{+33}_{-30}\pm28$ & $97\pm2\pm30$ & $629^{+36}_{-33}\pm50$ & 579\\
 & 400--600 & $30.3^{+4.0}_{-3.7}\pm2.7$ & $65^{+9}_{-8}\pm10$ & $0.33\pm0.06\pm0.10$ & $96^{+9}_{-8}\pm11$ & 79\\
 & 600--800 & $8.2^{+1.4}_{-1.3}\pm1.0$ & $21.0^{+2.8}_{-2.5}\pm4.8$ & $0.02\pm0.01\pm0.01$ & $29.2^{+3.1}_{-2.8}\pm5.0$ & 16\\
 & 800--1000 & $2.36^{+0.56}_{-0.54}\pm0.50$ & $9.8^{+1.3}_{-1.1}\pm2.3$ & $0.01\pm0.01\pm0.00$ & $12.2^{+1.4}_{-1.3}\pm2.4$ & 9\\
 & 1000--1200 & $1.00\pm0.24\pm0.31$ & $4.0\pm0.5\pm1.2$ & $<$0.01 & $5.0^{+0.6}_{-0.5}\pm1.2$ & 6\\
 & $\geq$1200 & $0.07\pm0.02\pm0.03$ & $0.86^{+0.11}_{-0.10}\pm0.37$ & $<$0.01 & $0.92^{+0.11}_{-0.10}\pm0.37$ & 1\\
[\cmsTabSkip]
\multirow{5}{*}{4-6j, 2b} & 200--400 & $120.4^{+9.1}_{-8.7}\pm9.8$ & $45^{+18}_{-13}\pm5$ & $26.0\pm0.6\pm8.1$ & $191^{+20}_{-16}\pm15$ & 194\\
 & 400--600 & $11.9\pm1.4\pm1.5$ & $11.5^{+4.6}_{-3.4}\pm1.8$ & $0.11\pm0.03\pm0.04$ & $23.4^{+4.8}_{-3.7}\pm2.6$ & 27\\
 & 600--800 & $3.49\pm0.83\pm0.75$ & $3.7^{+1.5}_{-1.1}\pm1.0$ & $<$0.01 & $7.2^{+1.7}_{-1.4}\pm1.3$ & 7\\
 & 800--1000 & $0.66\pm0.16\pm0.20$ & $1.73^{+0.69}_{-0.51}\pm0.48$ & $<$0.01 & $2.38^{+0.71}_{-0.54}\pm0.53$ & 3\\
 & $\geq$1000 & $0.15\pm0.04\pm0.06$ & $0.84^{+0.34}_{-0.25}\pm0.36$ & $<$0.01 & $1.00^{+0.34}_{-0.25}\pm0.36$ & 0\\
    \hline
\end{tabular}
}
\end{table*}

\begin{table*}[htb!]
\setlength\tabcolsep{1.5mm}
\centering
\topcaption{Predictions and observations for the 31 search regions with $1200 \leq \HT < 1500\GeV$, and $7\leq \njets \leq9$, or $\njets \geq10$. For each of the background
predictions, the first uncertainty listed is statistical (from the limited size of data control samples
and Monte Carlo samples), and the second is systematic.}
\label{tab:yieldsHh}
\renewcommand{\arraystretch}{1.3}
\cmsTableAlt{
\begin{tabular}{c c  c c c c c}
    \hline
\multicolumn{7}{c}{$1200 \leq \HT < 1500\GeV$} \\ [\cmsTabSkip]
\njets, \nbtags & \mttwo [\GeVns{}] & Lost lepton & \znunu & Multijet & Total background & Data \\
\hline
\multirow{5}{*}{7-9j, 0b} & 200--400 & $120.4^{+9.8}_{-9.2}\pm9.0$ & $108^{+26}_{-21}\pm21$ & $91\pm3\pm29$ & $319^{+28}_{-24}\pm38$ & 379\\
 & 400--600 & $16.5^{+1.9}_{-1.8}\pm2.0$ & $25.8^{+6.3}_{-5.1}\pm5.7$ & $0.80\pm0.09\pm0.25$ & $43.1^{+6.5}_{-5.4}\pm6.3$ & 45\\
 & 600--800 & $2.94\pm0.42\pm0.63$ & $8.6^{+2.1}_{-1.7}\pm2.1$ & $0.06\pm0.02\pm0.02$ & $11.6^{+2.1}_{-1.8}\pm2.2$ & 17\\
 & 800--1000 & $0.77^{+0.14}_{-0.13}\pm0.24$ & $2.90^{+0.70}_{-0.58}\pm1.00$ & $0.01\pm0.01\pm0.00$ & $3.7^{+0.7}_{-0.6}\pm1.0$ & 3\\
 & $\geq$1000 & $0.11\pm0.03\pm0.05$ & $1.09^{+0.26}_{-0.22}\pm0.50$ & $<$0.01 & $1.21^{+0.27}_{-0.22}\pm0.50$ & 0\\
[\cmsTabSkip]
\multirow{5}{*}{7-9j, 1b} & 200--400 & $133.8^{+8.0}_{-7.7}\pm9.8$ & $36^{+13}_{-10}\pm8$ & $58\pm2\pm18$ & $228^{+15}_{-13}\pm23$ & 247\\
 & 400--600 & $16.6^{+2.9}_{-2.7}\pm1.3$ & $8.7^{+3.2}_{-2.4}\pm2.1$ & $0.46\pm0.07\pm0.14$ & $25.8^{+4.3}_{-3.6}\pm2.7$ & 23\\
 & 600--800 & $1.83^{+0.43}_{-0.41}\pm0.28$ & $2.9^{+1.1}_{-0.8}\pm0.8$ & $0.03\pm0.02\pm0.01$ & $4.8^{+1.1}_{-0.9}\pm0.8$ & 7\\
 & 800--1000 & $0.65^{+0.24}_{-0.23}\pm0.18$ & $0.95^{+0.34}_{-0.26}\pm0.34$ & $0.02\pm0.01\pm0.01$ & $1.62^{+0.42}_{-0.35}\pm0.39$ & 2\\
 & $\geq$1000 & $0.22\pm0.19\pm0.09$ & $0.36^{+0.13}_{-0.10}\pm0.17$ & $<$0.01 & $0.58^{+0.23}_{-0.21}\pm0.19$ & 0\\
[\cmsTabSkip]
\multirow{4}{*}{7-9j, 2b} & 200--400 & $124.0^{+7.6}_{-7.4}\pm9.1$ & $9.9^{+3.6}_{-2.7}\pm2.5$ & $21.4\pm0.5\pm6.9$ & $155\pm8\pm12$ & 162\\
 & 400--600 & $15.0^{+2.8}_{-2.6}\pm1.3$ & $2.41^{+0.87}_{-0.66}\pm0.67$ & $0.12\pm0.03\pm0.04$ & $17.5^{+3.0}_{-2.7}\pm1.5$ & 18\\
 & 600--800 & $2.47^{+0.78}_{-0.76}\pm0.53$ & $0.81^{+0.29}_{-0.22}\pm0.26$ & $0.01\pm0.01\pm0.00$ & $3.29^{+0.83}_{-0.79}\pm0.60$ & 1\\
 & $\geq$800 & $0.24\pm0.11\pm0.10$ & $0.36^{+0.13}_{-0.10}\pm0.16$ & $<$0.01 & $0.60^{+0.17}_{-0.15}\pm0.19$ & 1\\
[\cmsTabSkip]
\multirow{3}{*}{7-9j, 3b} & 200--400 & $30.0\pm2.6\pm3.2$ & $1.89^{+0.68}_{-0.52}\pm0.64$ & $5.0\pm0.3\pm1.8$ & $36.9^{+2.7}_{-2.6}\pm3.8$ & 46\\
 & 400--600 & $4.1^{+1.1}_{-1.0}\pm0.6$ & $0.45^{+0.16}_{-0.12}\pm0.18$ & $0.02\pm0.01\pm0.01$ & $4.6^{+1.1}_{-1.0}\pm0.6$ & 2\\
 & $\geq$600 & $0.92^{+0.50}_{-0.49}\pm0.38$ & $0.23^{+0.08}_{-0.06}\pm0.11$ & $<$0.01 & $1.15\pm0.50\pm0.40$ & 1\\
[\cmsTabSkip]
\multirow{2}{*}{7-9j, $\geq$4b} & 200--400 & $9.1\pm1.6\pm1.8$ & $0.26^{+0.10}_{-0.07}\pm0.23$ & $0.88\pm0.10\pm0.32$ & $10.3\pm1.6\pm1.9$ & 9\\
 & $\geq$400 & $0.44^{+0.24}_{-0.23}\pm0.08$ & $0.10^{+0.04}_{-0.03}\pm0.09$ & $<$0.01 & $0.53\pm0.24\pm0.12$ & 0\\
[\cmsTabSkip]
\multirow{3}{*}{$\geq$10j, 0b} & 200--400 & $7.7^{+1.2}_{-1.1}\pm0.8$ & $2.7^{+0.6}_{-0.5}\pm2.8$ & $8.3\pm0.9\pm3.0$ & $18.7^{+1.6}_{-1.5}\pm4.1$ & 17\\
 & 400--600 & $1.00\pm0.32\pm0.22$ & $0.56^{+0.13}_{-0.11}\pm0.62$ & $0.11\pm0.03\pm0.04$ & $1.66^{+0.35}_{-0.34}\pm0.66$ & 1\\
 & $\geq$600 & $0.10^{+0.35}_{-0.04}\pm0.04$ & $0.14^{+0.08}_{-0.03}\pm0.14$ & $0.01\pm0.01\pm0.00$ & $0.24^{+0.36}_{-0.05}\pm0.15$ & 0\\
[\cmsTabSkip]
\multirow{3}{*}{$\geq$10j, 1b} & 200--400 & $15.2\pm1.8\pm1.4$ & $1.1^{+0.4}_{-0.3}\pm1.2$ & $5.3\pm0.2\pm1.9$ & $21.6^{+1.9}_{-1.8}\pm2.7$ & 22\\
 & 400--600 & $1.27^{+0.38}_{-0.36}\pm0.11$ & $0.22^{+0.08}_{-0.06}\pm0.26$ & $0.05\pm0.02\pm0.02$ & $1.55^{+0.39}_{-0.37}\pm0.29$ & 6\\
 & $\geq$600 & $0.03\pm0.02\pm0.01$ & $0.05^{+0.10}_{-0.01}\pm0.05$ & $<$0.01 & $0.07^{+0.11}_{-0.02}\pm0.05$ & 0\\
[\cmsTabSkip]
\multirow{3}{*}{$\geq$10j, 2b} & 200--400 & $16.9\pm1.8\pm1.5$ & $0.44^{+0.16}_{-0.12}\pm0.50$ & $2.7\pm0.2\pm1.0$ & $20.1\pm1.8\pm1.9$ & 16\\
 & 400--600 & $2.62^{+0.71}_{-0.68}\pm0.30$ & $0.09\pm0.03\pm0.11$ & $0.01\pm0.01\pm0.00$ & $2.73^{+0.71}_{-0.68}\pm0.32$ & 2\\
 & $\geq$600 & $0.23\pm0.15\pm0.10$ & $0.02^{+0.08}_{-0.01}\pm0.02$ & $<$0.01 & $0.25^{+0.17}_{-0.15}\pm0.10$ & 0\\
[\cmsTabSkip]
\multirow{2}{*}{$\geq$10j, 3b} & 200--400 & $5.58^{+0.86}_{-0.85}\pm0.61$ & $0.12^{+0.11}_{-0.03}\pm0.16$ & $1.04\pm0.10\pm0.42$ & $6.74^{+0.87}_{-0.86}\pm0.76$ & 6\\
 & $\geq$400 & $0.51\pm0.22\pm0.06$ & $0.03^{+0.11}_{-0.01}\pm0.04$ & $<$0.01 & $0.54^{+0.25}_{-0.22}\pm0.08$ & 0\\
[\cmsTabSkip]
\multirow{1}{*}{$\geq$10j, $\geq$4b} & $\geq$200 & $2.59\pm0.82\pm0.62$ & $0.10^{+0.13}_{-0.03}\pm0.13$ & $0.31\pm0.06\pm0.13$ & $3.00^{+0.83}_{-0.82}\pm0.65$ & 7\\

    \hline

\end{tabular}
}
\end{table*}

\begin{table*}[htb!]
\setlength\tabcolsep{1.5mm}
\centering
\topcaption{Predictions and observations for the 30 search regions with $\HT \geq 1500\GeV$, and $2 \leq \njets \leq3$, $2 \leq \njets \leq 6$ and $\nbtags\geq3$, or $4 \leq \njets \leq 6$. For each of the background
predictions, the first uncertainty listed is statistical (from the limited size of data control samples
and Monte Carlo samples), and the second is systematic.}
\label{tab:yieldsUHl}
\renewcommand{\arraystretch}{1.3}
\cmsTableAlt{
\begin{tabular}{c c  c c c c c}
    \hline
\multicolumn{7}{c}{$\HT \geq 1500\GeV$} \\ [\cmsTabSkip]
\njets, \nbtags & \mttwo [\GeVns{}] & Lost lepton & \znunu & Multijet & Total background & Data \\
\hline
\multirow{7}{*}{2-3j, 0b} & 400--600 & $27.2^{+4.4}_{-3.9}\pm2.5$ & $150^{+14}_{-13}\pm19$ & $0.16\pm0.04\pm0.05$ & $177^{+15}_{-13}\pm20$ & 125\\
 & 600--800 & $7.8^{+1.4}_{-1.2}\pm0.8$ & $38.7^{+3.6}_{-3.3}\pm8.4$ & $<$0.01 & $46.5^{+3.9}_{-3.6}\pm8.6$ & 37\\
 & 800--1000 & $2.29^{+0.39}_{-0.34}\pm0.35$ & $17.2^{+1.6}_{-1.5}\pm3.4$ & $<$0.01 & $19.5^{+1.7}_{-1.5}\pm3.4$ & 19\\
 & 1000--1200 & $1.20^{+0.21}_{-0.19}\pm0.26$ & $9.0\pm0.8\pm1.8$ & $<$0.01 & $10.2^{+0.9}_{-0.8}\pm1.9$ & 14\\
 & 1200--1400 & $0.80^{+0.16}_{-0.14}\pm0.22$ & $4.9^{+0.5}_{-0.4}\pm1.3$ & $<$0.01 & $5.7^{+0.5}_{-0.4}\pm1.4$ & 4\\
 & 1400--1800 & $0.43^{+0.09}_{-0.08}\pm0.15$ & $2.80^{+0.26}_{-0.24}\pm0.98$ & $<$0.01 & $3.23^{+0.28}_{-0.26}\pm0.99$ & 3\\
 & $\geq$1800 & $0.05\pm0.02\pm0.02$ & $0.41^{+0.04}_{-0.03}\pm0.19$ & $<$0.01 & $0.46\pm0.04\pm0.19$ & 0\\
[\cmsTabSkip]
\multirow{5}{*}{2-3j, 1b} & 400--600 & $5.2^{+1.1}_{-1.0}\pm0.6$ & $13.4^{+4.9}_{-3.7}\pm1.9$ & $0.09\pm0.03\pm0.03$ & $18.7^{+5.0}_{-3.8}\pm2.1$ & 23\\
 & 600--800 & $1.52^{+0.43}_{-0.41}\pm0.27$ & $3.5^{+1.3}_{-1.0}\pm1.0$ & $<$0.01 & $5.0^{+1.3}_{-1.0}\pm1.0$ & 3\\
 & 800--1000 & $0.38\pm0.09\pm0.10$ & $1.53^{+0.55}_{-0.42}\pm0.35$ & $<$0.01 & $1.90^{+0.56}_{-0.43}\pm0.37$ & 3\\
 & 1000--1200 & $0.10\pm0.03\pm0.03$ & $0.81^{+0.29}_{-0.22}\pm0.24$ & $<$0.01 & $0.91^{+0.29}_{-0.22}\pm0.24$ & 4\\
 & $\geq$1200 & $0.19\pm0.06\pm0.08$ & $0.73^{+0.26}_{-0.20}\pm0.31$ & $<$0.01 & $0.92^{+0.27}_{-0.21}\pm0.32$ & 0\\
[\cmsTabSkip]
\multirow{1}{*}{2-3j, 2b} & $\geq$400 & $0.63^{+0.49}_{-0.36}\pm0.26$ & $0.0^{+3.0}_{-0.0}\pm0.0$ & $<$0.01 & $0.6^{+3.0}_{-0.4}\pm0.3$ & 2\\
[\cmsTabSkip]
\multirow{2}{*}{2-6j, $\geq$3b} & 400--600 & $1.72^{+0.73}_{-0.68}\pm0.42$ & $1.1^{+2.4}_{-0.9}\pm0.3$ & $0.03\pm0.02\pm0.01$ & $2.8^{+2.5}_{-1.1}\pm0.6$ & 1\\
 & $\geq$600 & $0.37^{+0.19}_{-0.18}\pm0.16$ & $0.5^{+1.2}_{-0.4}\pm0.2$ & $<$0.01 & $0.9^{+1.2}_{-0.5}\pm0.2$ & 0\\
[\cmsTabSkip]
\multirow{7}{*}{4-6j, 0b} & 400--600 & $46.4^{+5.6}_{-5.1}\pm3.6$ & $176^{+15}_{-14}\pm23$ & $1.62\pm0.13\pm0.46$ & $224^{+16}_{-15}\pm24$ & 207\\
 & 600--800 & $10.6^{+2.3}_{-1.9}\pm1.2$ & $45.5^{+4.0}_{-3.7}\pm9.9$ & $0.07\pm0.03\pm0.02$ & $56^{+5}_{-4}\pm10$ & 62\\
 & 800--1000 & $4.5^{+1.1}_{-1.0}\pm0.5$ & $20.3^{+1.8}_{-1.6}\pm3.9$ & $<$0.01 & $24.8^{+2.1}_{-1.9}\pm4.1$ & 31\\
 & 1000--1200 & $1.35^{+0.30}_{-0.26}\pm0.24$ & $10.6\pm0.9\pm2.1$ & $<$0.01 & $11.9^{+1.0}_{-0.9}\pm2.2$ & 12\\
 & 1200--1400 & $0.89^{+0.27}_{-0.25}\pm0.23$ & $5.7\pm0.5\pm1.5$ & $<$0.01 & $6.6^{+0.6}_{-0.5}\pm1.6$ & 9\\
 & 1400--1600 & $0.20\pm0.05\pm0.07$ & $2.64^{+0.23}_{-0.21}\pm0.92$ & $<$0.01 & $2.84^{+0.24}_{-0.22}\pm0.92$ & 3\\
 & $\geq$1600 & $0.09\pm0.03\pm0.04$ & $1.18\pm0.10\pm0.51$ & $<$0.01 & $1.27^{+0.11}_{-0.10}\pm0.51$ & 2\\
[\cmsTabSkip]
\multirow{5}{*}{4-6j, 1b} & 400--600 & $21.0^{+3.7}_{-3.3}\pm2.0$ & $32.6^{+7.0}_{-5.8}\pm5.5$ & $0.81\pm0.09\pm0.23$ & $54.5^{+7.9}_{-6.7}\pm6.3$ & 72\\
 & 600--800 & $4.79^{+0.91}_{-0.83}\pm0.62$ & $8.4^{+1.8}_{-1.5}\pm2.3$ & $0.02\pm0.01\pm0.01$ & $13.2^{+2.0}_{-1.7}\pm2.5$ & 20\\
 & 800--1000 & $1.27^{+0.26}_{-0.24}\pm0.27$ & $3.71^{+0.79}_{-0.66}\pm0.92$ & $0.03\pm0.02\pm0.01$ & $5.01^{+0.84}_{-0.71}\pm0.97$ & 8\\
 & 1000--1400 & $0.89^{+0.21}_{-0.20}\pm0.28$ & $3.00^{+0.64}_{-0.54}\pm0.93$ & $<$0.01 & $3.89^{+0.68}_{-0.57}\pm0.98$ & 6\\
 & $\geq$1400 & $0.40^{+0.34}_{-0.33}\pm0.16$ & $0.72^{+0.15}_{-0.13}\pm0.31$ & $<$0.01 & $1.12^{+0.37}_{-0.36}\pm0.36$ & 3\\
[\cmsTabSkip]
\multirow{3}{*}{4-6j, 2b} & 400--600 & $7.2^{+1.2}_{-1.1}\pm1.1$ & $4.3^{+2.9}_{-1.9}\pm1.4$ & $0.17\pm0.04\pm0.05$ & $11.7^{+3.2}_{-2.2}\pm1.9$ & 11\\
 & 600--800 & $1.66^{+0.41}_{-0.40}\pm0.46$ & $1.12^{+0.76}_{-0.48}\pm0.55$ & $0.01\pm0.01\pm0.00$ & $2.79^{+0.86}_{-0.63}\pm0.73$ & 3\\
 & $\geq$800 & $0.32\pm0.13\pm0.13$ & $0.99^{+0.67}_{-0.43}\pm0.52$ & $<$0.01 & $1.31^{+0.68}_{-0.45}\pm0.54$ & 4\\

    \hline

\end{tabular}
}
\end{table*}

\begin{table*}[htb!]
\setlength\tabcolsep{1.5mm}
\centering
\topcaption{Predictions and observations for the 21 search regions with $\HT \geq 1500\GeV$, and $7\leq \njets \leq9$, or $\njets \geq10$. For each of the background
predictions, the first uncertainty listed is statistical (from the limited size of data control samples
and Monte Carlo samples), and the second is systematic.}
\label{tab:yieldsUHh}
\renewcommand{\arraystretch}{1.3}
\cmsTableAlt{
\begin{tabular}{c c  c c c c c}
    \hline
\multicolumn{7}{c}{$\HT \geq 1500\GeV$} \\ [\cmsTabSkip]
\njets, \nbtags & \mttwo [\GeVns{}] & Lost lepton & \znunu & Multijet & Total background & Data \\
\hline
\multirow{5}{*}{7-9j, 0b} & 400--600 & $14.3^{+1.8}_{-1.7}\pm1.7$ & $32.3^{+7.5}_{-6.2}\pm4.3$ & $1.50\pm0.13\pm0.44$ & $48.1^{+7.7}_{-6.4}\pm5.0$ & 36\\
 & 600--800 & $3.77^{+0.56}_{-0.55}\pm0.69$ & $8.3^{+1.9}_{-1.6}\pm2.2$ & $0.18\pm0.04\pm0.05$ & $12.3^{+2.0}_{-1.7}\pm2.3$ & 9\\
 & 800--1000 & $1.16^{+0.18}_{-0.17}\pm0.30$ & $3.70^{+0.86}_{-0.71}\pm0.83$ & $0.01\pm0.01\pm0.00$ & $4.86^{+0.88}_{-0.73}\pm0.90$ & 6\\
 & 1000--1400 & $0.58\pm0.11\pm0.19$ & $2.96^{+0.69}_{-0.57}\pm0.86$ & $0.01\pm0.01\pm0.00$ & $3.55^{+0.69}_{-0.58}\pm0.89$ & 4\\
 & $\geq$1400 & $0.05\pm0.01\pm0.02$ & $0.71^{+0.17}_{-0.14}\pm0.30$ & $<$0.01 & $0.76^{+0.17}_{-0.14}\pm0.30$ & 2\\
[\cmsTabSkip]
\multirow{3}{*}{7-9j, 1b} & 400--600 & $12.8^{+2.5}_{-2.3}\pm1.6$ & $9.2^{+4.2}_{-3.0}\pm1.4$ & $0.82\pm0.09\pm0.24$ & $22.9^{+4.9}_{-3.8}\pm2.3$ & 25\\
 & 600--800 & $3.49^{+0.94}_{-0.89}\pm0.76$ & $2.4^{+1.1}_{-0.8}\pm1.0$ & $0.06\pm0.02\pm0.02$ & $5.9^{+1.4}_{-1.2}\pm1.2$ & 7\\
 & $\geq$800 & $1.09^{+0.34}_{-0.32}\pm0.45$ & $2.10^{+0.96}_{-0.69}\pm0.93$ & $<$0.01 & $3.2^{+1.0}_{-0.8}\pm1.0$ & 2\\
[\cmsTabSkip]
\multirow{3}{*}{7-9j, 2b} & 400--600 & $8.1^{+1.8}_{-1.6}\pm1.0$ & $2.4^{+1.1}_{-0.8}\pm0.4$ & $0.35\pm0.06\pm0.10$ & $10.9^{+2.1}_{-1.8}\pm1.2$ & 10\\
 & 600--800 & $1.78^{+0.54}_{-0.52}\pm0.40$ & $0.62^{+0.28}_{-0.20}\pm0.25$ & $0.02\pm0.01\pm0.01$ & $2.41^{+0.61}_{-0.56}\pm0.49$ & 5\\
 & $\geq$800 & $0.40^{+0.19}_{-0.18}\pm0.17$ & $0.55^{+0.25}_{-0.18}\pm0.25$ & $0.01\pm0.01\pm0.00$ & $0.96^{+0.31}_{-0.26}\pm0.30$ & 0\\
[\cmsTabSkip]
\multirow{2}{*}{7-9j, 3b} & 400--800 & $2.40^{+0.74}_{-0.72}\pm0.29$ & $0.32^{+0.15}_{-0.10}\pm0.12$ & $0.10\pm0.03\pm0.03$ & $2.82^{+0.76}_{-0.72}\pm0.32$ & 2\\
 & $\geq$800 & $0.16\pm0.09\pm0.07$ & $0.08^{+0.04}_{-0.03}\pm0.04$ & $<$0.01 & $0.24\pm0.09\pm0.08$ & 0\\
[\cmsTabSkip]
\multirow{1}{*}{7-9j, $\geq$4b} & $\geq$400 & $0.52^{+0.23}_{-0.22}\pm0.08$ & $0.07^{+0.03}_{-0.02}\pm0.06$ & $0.02\pm0.01\pm0.01$ & $0.61^{+0.23}_{-0.22}\pm0.10$ & 1\\
[\cmsTabSkip]
\multirow{2}{*}{$\geq$10j, 0b} & 400--800 & $1.41\pm0.38\pm0.33$ & $1.52^{+0.35}_{-0.29}\pm0.34$ & $0.23\pm0.05\pm0.08$ & $3.17^{+0.52}_{-0.48}\pm0.49$ & 11\\
 & $\geq$800 & $0.05\pm0.02\pm0.02$ & $0.37^{+0.09}_{-0.07}\pm0.17$ & $0.01\pm0.01\pm0.00$ & $0.43^{+0.09}_{-0.08}\pm0.17$ & 0\\
[\cmsTabSkip]
\multirow{2}{*}{$\geq$10j, 1b} & 400--800 & $2.16^{+0.71}_{-0.69}\pm0.25$ & $0.56^{+0.25}_{-0.18}\pm0.16$ & $0.14\pm0.04\pm0.05$ & $2.85^{+0.76}_{-0.71}\pm0.31$ & 3\\
 & $\geq$800 & $0.55\pm0.30\pm0.22$ & $0.13^{+0.06}_{-0.04}\pm0.07$ & $<$0.01 & $0.68^{+0.31}_{-0.30}\pm0.23$ & 0\\
[\cmsTabSkip]
\multirow{1}{*}{$\geq$10j, 2b} & $\geq$400 & $1.98^{+0.69}_{-0.67}\pm0.24$ & $0.30^{+0.14}_{-0.10}\pm0.12$ & $0.05\pm0.02\pm0.02$ & $2.33^{+0.70}_{-0.68}\pm0.28$ & 0\\
[\cmsTabSkip]
\multirow{1}{*}{$\geq$10j, 3b} & $\geq$400 & $0.77\pm0.35\pm0.09$ & $0.00^{+0.45}_{-0.00}\pm0.00$ & $0.05\pm0.03\pm0.02$ & $0.82^{+0.57}_{-0.35}\pm0.09$ & 1\\
[\cmsTabSkip]
\multirow{1}{*}{$\geq$10j, $\geq$4b} & $\geq$400 & $0.09\pm0.05\pm0.01$ & $0.00^{+0.45}_{-0.00}\pm0.00$ & $<$0.01 & $0.09^{+0.45}_{-0.05}\pm0.01$ & 0\\
    \hline

\end{tabular}
}
\end{table*}

\subsection{Search for disappearing tracks: search regions and yields}
\label{app:srs:distr}

The 68 search regions defined for the disappearing track search, as described in Section~\ref{sec:evtsel:evtcat:distracks},
are summarized in Tables~\ref{tab:sr1_distracks}--\ref{tab:sr3_distracks},
together with the pre-fit background predictions and the observed yields.

\begin{table*}[htbp]

\centering

\topcaption{Summary of the 28 signal regions of the search for disappearing tracks, for the 2016 data set, together with the corresponding background predictions and observations. For the background
predictions, the first uncertainty listed is statistical (from the limited size of control samples), and the second is systematic. The systematic uncertainty is not shown when it is negligible. \label{tab:sr1_distracks}}
\renewcommand{\arraystretch}{1.3}
\cmsTableAlt{
\begin{tabular}{cccclcc}

    \hline

Track length & \njets & \HT range [\GeVns{}] & Track \pt [\GeVns{}] & Label & Background & Data\\

\hline

\multirow{12}{*}{P}

& \multirow{6}{*}{2--3} & \multirow{2}{*}{[ 250, 450 )} & [ 15, 50 ) & P LL lo & $15.5^{+3.0}_{-2.7}\pm3.2$ & 16\\

& & & [ 50, $\infty$ ) & P LL hi & $9.8^{+2.6}_{-2.2}\pm2.5$ & 3\\

& & \multirow{2}{*}{[ 450, 1200 )} & [ 15, 50 ) & P LM lo & $4.2^{+1.0}_{-0.9}\pm1.2$ & 2\\

& & & [ 50, $\infty$ ) & P LM hi & $2.02^{+0.66}_{-0.55}\pm0.63$ & 1\\

& & \multirow{2}{*}{[ 1200, $\infty$ )} & [ 15, 50 ) & P LH lo & $0.19^{+0.26}_{-0.13}\pm0.13$ & 0\\

& & & [ 50, $\infty$ ) & P LH hi & $0.06^{+0.14}_{-0.05}\pm0.03$ & 0\\

& \multirow{6}{*}{$\geq$4} & \multirow{2}{*}{[ 250, 450 )} & [ 15, 50 ) & P HL lo & $3.3^{+0.7}_{-0.6}\pm1.4$ & 1\\

& & & [ 50, $\infty$ ) & P HL hi & $1.98^{+0.43}_{-0.38}\pm0.57$ & 1\\

& & \multirow{2}{*}{[ 450, 1200 )} & [ 15, 50 ) & P HM lo & $4.7^{+0.8}_{-0.7}\pm1.9$ & 6\\

& & & [ 50, $\infty$ ) & P HM hi & $2.37^{+0.50}_{-0.44}\pm0.55$ & 1\\

& & \multirow{2}{*}{[ 1200, $\infty$ )} & [ 15, 50 ) & P HH lo & $0.43^{+0.24}_{-0.17}\pm0.27$ & 0\\

& & & [ 50, $\infty$ ) & P HH hi & $0.17^{+0.10}_{-0.07}\pm0.04$ & 0\\ [\cmsTabSkip]

\multirow{12}{*}{M}

& \multirow{6}{*}{2--3} & \multirow{2}{*}{[ 250, 450 )} & [ 15, 50 ) & M LL lo & $3.9^{+1.5}_{-1.2}\pm1.3$ & 3\\

& & & [ 50, $\infty$ ) & M LL hi & $14^{+3.7}_{-3.2}\pm4.0$ & 8\\

& & \multirow{2}{*}{[ 450, 1200 )} & [ 15, 50 ) & M LM lo & $2.1^{+0.89}_{-0.71}\pm1.1$ & 3\\

& & & [ 50, $\infty$ ) & M LM hi & $0.68^{+0.90}_{-0.45}\pm0.35$ & 4\\

& & \multirow{2}{*}{[ 1200, $\infty$ )} & [ 15, 50 ) & M LH lo & $0.0^{+0.25}_{-0.0}\pm0.0$ & 0\\

& & & [ 50, $\infty$ ) & M LH hi & $0.0^{+0.7}_{-0.0}$ & 0\\

& \multirow{6}{*}{$\geq$4} & \multirow{2}{*}{[ 250, 450 )} & [ 15, 50 ) & M HL lo & $1.8^{+0.6}_{-0.5}\pm0.9$ & 0\\

& & & [ 50, $\infty$ ) & M HL hi & $2.1^{+0.8}_{-0.6}$ $^{+2.3}_{-2.1}$ & 2\\

& & \multirow{2}{*}{[ 450, 1200 )} & [ 15, 50 ) & M HM lo & $2.2^{+0.7}_{-0.6}\pm1.3$ & 1\\

& & & [ 50, $\infty$ ) & M HM hi & $2.9^{+0.9}_{-0.8}\pm2.3$ & 0\\

& & \multirow{2}{*}{[ 1200, $\infty$ )} & [ 15, 50 ) & M HH lo & $0.23^{+0.23}_{-0.13}\pm0.11$ & 0\\

& & & [ 50, $\infty$ ) & M HH hi & $0.30^{+0.40}_{-0.20}\pm0.29$ & 1\\ [\cmsTabSkip]

\multirow{4}{*}{L}

& \multirow{2}{*}{2--3} & [ 250, 1200 ) & [ 15, $\infty$ ) & L LLM  & $0.046^{+0.050}_{-0.034}$ $^{+0.057}_{-0.046}$ & 0\\

&  & [ 1200, $\infty$ ) & [ 15, $\infty$ ) & L LH  & $0.015^{+0.036}_{-0.015}$ $^{+0.022}_{-0.015}$ & 0\\

& \multirow{2}{*}{$\geq$4} & [ 250, 1200 ) & [ 15, $\infty$ ) & L HLM  & $0.092^{+0.136}_{-0.085}$ $^{+0.130}_{-0.092}$ & 0\\

&  & [ 1200, $\infty$ ) & [ 15, $\infty$ ) & L HH  & $0.0^{+0.1}_{-0.0}$ & 0\\

    \hline

\end{tabular}
}
\end{table*}

\begin{table*}[htbp]

\centering

\topcaption{Summary of the 24 signal regions of the search for disappearing tracks for pixel tracks, for the 2017--2018 data set,
together with the corresponding background predictions and observations.
For the background predictions, the first uncertainty listed is statistical (from the limited size of control samples), and the second is systematic.
The systematic uncertainty is not shown when it is negligible. \label{tab:sr2_distracks}}
\renewcommand{\arraystretch}{1.3}
\cmsTableAlt{
\begin{tabular}{cccclcc}

    \hline

Track length & \njets & \HT range [\GeVns{}] & Track \pt [\GeVns{}] & Label & Background & Data\\

\hline

\multirow{12}{*}{P3}

& \multirow{6}{*}{2--3} & \multirow{2}{*}{[ 250, 450 )} & [ 15, 50 ) & P3 LL lo & $78^{+9}_{-9}\pm34$ & 73\\

& & & [ 50, $\infty$ ) & P3 LL hi & $43.9^{+6.7}_{-6.2}\pm8.1$& 41\\

& & \multirow{2}{*}{[ 450, 1200 )} & [ 15, 50 ) & P3 LM lo & $30^{+5}_{-5}\pm16$ & 21\\

& & & [ 50, $\infty$ ) & P3 LM hi & $13^{+3}_{-3}\pm13$ & 16\\

& & \multirow{2}{*}{[ 1200, $\infty$ )} & [ 15, 50 ) & P3 LH lo & $0.0^{+1.0}_{-0.0}$ & 1\\

& & & [ 50, $\infty$ ) & P3 LH hi & $0.43^{+0.98}_{-0.36}\pm0.34$ & 0\\

& \multirow{6}{*}{$\geq$4} & \multirow{2}{*}{[ 250, 450 )} & [ 15, 50 ) & P3 HL lo & $25.8^{+3.8}_{-3.4}\pm7.9$ & 17\\

& & & [ 50, $\infty$ ) & P3 HL hi & $10.8^{+2.1}_{-1.8}\pm3.5$ & 7\\

& & \multirow{2}{*}{[ 450, 1200 )} & [ 15, 50 ) & P3 HM lo & $28.9^{+4.0}_{-3.7}\pm5.7$ & 37\\

& & & [ 50, $\infty$ ) & P3 HM hi & $12.3^{+2.2}_{-1.9}\pm6.8$ & 11\\

& & \multirow{2}{*}{[ 1200, $\infty$ )} & [ 15, 50 ) & P3 HH lo & $3.1^{+1.5}_{-1.1}\pm0.5$ & 5\\

& & & [ 50, $\infty$ ) & P3 HH hi & $0.49^{+0.65}_{-0.32}\pm0.12$ & 3\\ [\cmsTabSkip]

\multirow{12}{*}{P4}

& \multirow{6}{*}{2--3} & \multirow{2}{*}{[ 250, 450 )} & [ 15, 50 ) & P4 LL lo & $24^{+5}_{-5}\pm11$ & 10\\

& & & [ 50, $\infty$ ) & P4 LL hi & $4.1^{+1.9}_{-1.5}\pm3.7$ & 0\\

& & \multirow{2}{*}{[ 450, 1200 )} & [ 15, 50 ) & P4 LM lo & $8.7^{+2.7}_{-2.2}\pm4.6$ & 8\\

& & & [ 50, $\infty$ ) & P4 LM hi & $1.1^{+0.7}_{-0.5}$ $^{+1.4}_{-1.1}$  & 0\\

& & \multirow{2}{*}{[ 1200, $\infty$ )} & [ 15, 50 ) & P4 LH lo & $0.40^{+0.91}_{-0.33}\pm0.40$  & 0\\

& & & [ 50, $\infty$ ) & P4 LH hi & $0.0^{+0.39}_{-0.0}$  & 0\\

& \multirow{6}{*}{$\geq$4} & \multirow{2}{*}{[ 250, 450 )} & [ 15, 50 ) & P4 HL lo & $6.3^{+1.6}_{-1.3}\pm2.2$  & 7\\

& & & [ 50, $\infty$ ) & P4 HL hi & $0.62^{+0.35}_{-0.25}\pm0.43$  & 0\\

& & \multirow{2}{*}{[ 450, 1200 )} & [ 15, 50 ) & P4 HM lo & $6.9^{+1.6}_{-1.4}\pm6.2$  & 2\\

& & & [ 50, $\infty$ ) & P4 HM hi & $1.32^{+0.54}_{-0.43}\pm0.63$  & 2\\

& & \multirow{2}{*}{[ 1200, $\infty$ )} & [ 15, 50 ) & P4 HH lo & $0.42^{+0.56}_{-0.28}\pm0.12$  & 0\\

& & & [ 50, $\infty$ ) & P4 HH hi & $0.08^{+0.18}_{-0.07}\pm0.03$  & 0\\ [\cmsTabSkip]

    \hline

\end{tabular}
}
\end{table*}

\begin{table*}[htbp]

\centering

\topcaption{Summary of the 16 signal regions of the search for disappearing tracks for medium (M) length and long (L) tracks, for the 2017--2018 data set,
together with the corresponding background predictions and observations.
For the background predictions, the first uncertainty listed is statistical (from the limited size of control samples), and the second is systematic.
The systematic uncertainty is not shown when it is negligible. \label{tab:sr3_distracks}}
\renewcommand{\arraystretch}{1.3}
\cmsTableAlt{
\begin{tabular}{cccclcc}

    \hline

Track length & \njets & \HT range [\GeVns{}] & Track \pt [\GeVns{}] & Label & Background & Data\\

\hline

\multirow{12}{*}{M}

& \multirow{6}{*}{2--3} & \multirow{2}{*}{[ 250, 450 )} & [ 15, 50 ) & M LL lo & $8.4^{+2.4}_{-2.0}\pm3.4$  & 8\\

& & & [ 50, $\infty$ ) & M LL hi & $5.4^{+2.2}_{-1.8}\pm2.6$  & 2\\

& & \multirow{2}{*}{[ 450, 1200 )} & [ 15, 50 ) & M LM lo & $1.90^{+0.85}_{-0.66}\pm0.92$  & 6\\

& & & [ 50, $\infty$ ) & M LM hi & $1.12^{+0.77}_{-0.54}\pm0.97$  & 1\\

& & \multirow{2}{*}{[ 1200, $\infty$ )} & [ 15, 50 ) & M LH lo & $0.00^{+0.36}_{-0}$  & 0\\

& & & [ 50, $\infty$ ) & M LH hi & $0.00^{+0.46}_{-0}$  & 0\\

& \multirow{6}{*}{$\geq$4} & \multirow{2}{*}{[ 250, 450 )} & [ 15, 50 ) & M HL lo & $1.6^{+0.6}_{-0.5}$ $^{+3.0}_{-1.6}$  & 3\\

& & & [ 50, $\infty$ ) & M HL hi & $1.11^{+0.57}_{-0.42}\pm0.58$  & 1\\

& & \multirow{2}{*}{[ 450, 1200 )} & [ 15, 50 ) & M HM lo & $1.9^{+0.6}_{-0.5}$ $^{+3.5}_{-1.9}$  & 3\\

& & & [ 50, $\infty$ ) & M HM hi & $1.5^{+0.7}_{-0.5}\pm1.1$  & 0\\

& & \multirow{2}{*}{[ 1200, $\infty$ )} & [ 15, 50 ) & M HH lo & $0.38^{+0.31}_{-0.19}$ $^{+0.70}_{-0.38}$  & 1\\

& & & [ 50, $\infty$ ) & M HH hi & $0.12^{+0.29}_{-0.10}\pm0.04$  & 0\\ [\cmsTabSkip]

\multirow{4}{*}{L}

& \multirow{2}{*}{2--3} & [ 250, 1200 ) & [ 15, $\infty$ ) & L LLM  & $0.46^{+0.26}_{-0.20}$ $^{+0.53}_{-0.46}$  & 0\\

&  & [ 1200, $\infty$ ) & [ 15, $\infty$ ) & L LH  & $0.00^{+0.14}_{-0}$  & 0\\

& \multirow{2}{*}{$\geq$4} & [ 250, 1200 ) & [ 15, $\infty$ ) & L HLM  & $0.013^{+0.015}_{-0.014}$ $^{+0.018}_{-0.013}$  & 0\\

&  & [ 1200, $\infty$ ) & [ 15, $\infty$ ) & L HH  & $0.000^{+0.008}_{-0}$ & 0\\

    \hline

\end{tabular}
}
\end{table*}

\section{Detailed results}
\label{app:results}

\subsection{Inclusive \texorpdfstring{\mttwo}{MT2} search}
\label{app:results:mt2}

\begin{figure*}[htb]
  \centering
    \includegraphics[width=0.85\textwidth]{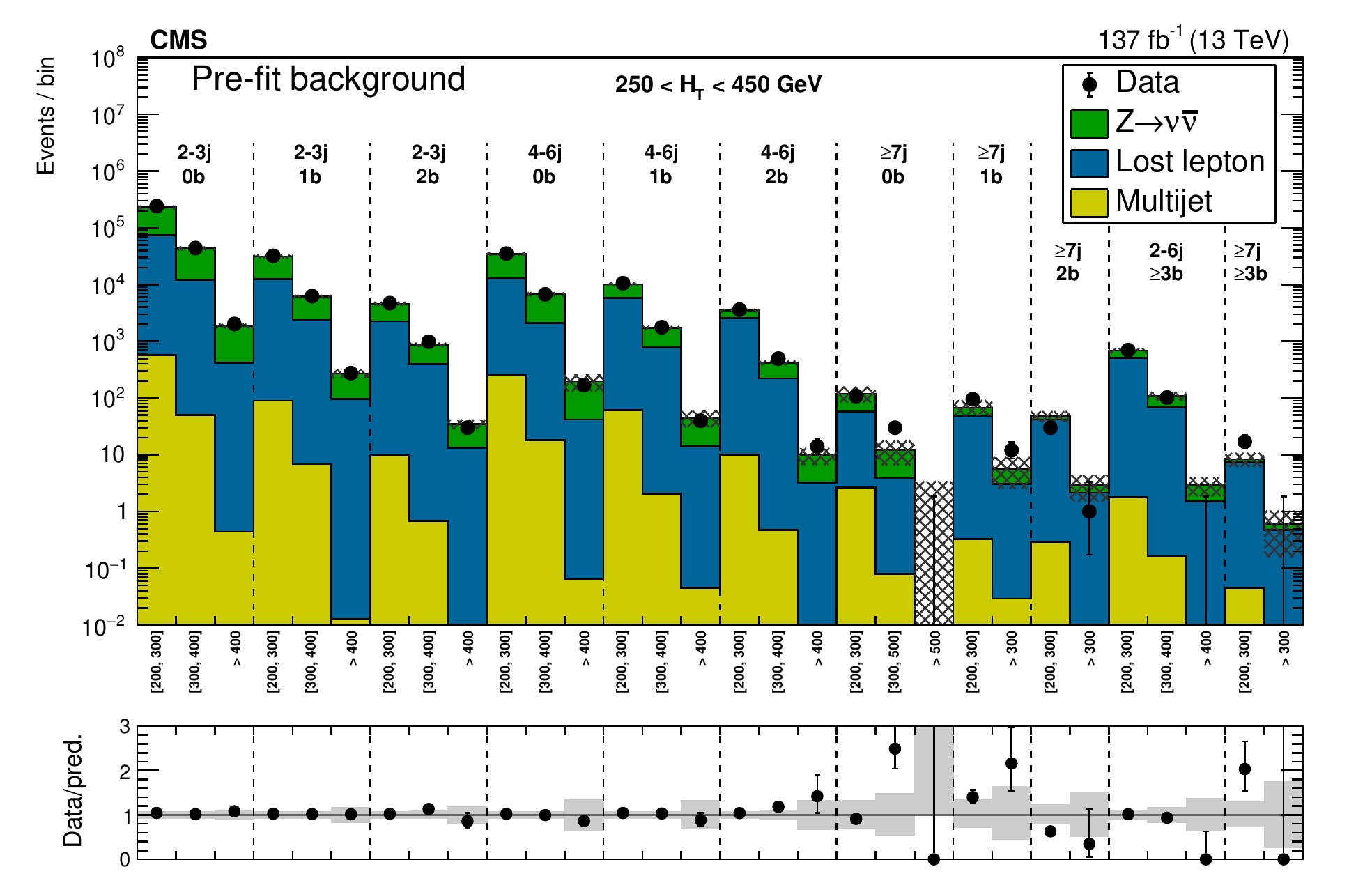}\\
    \includegraphics[width=0.85\textwidth]{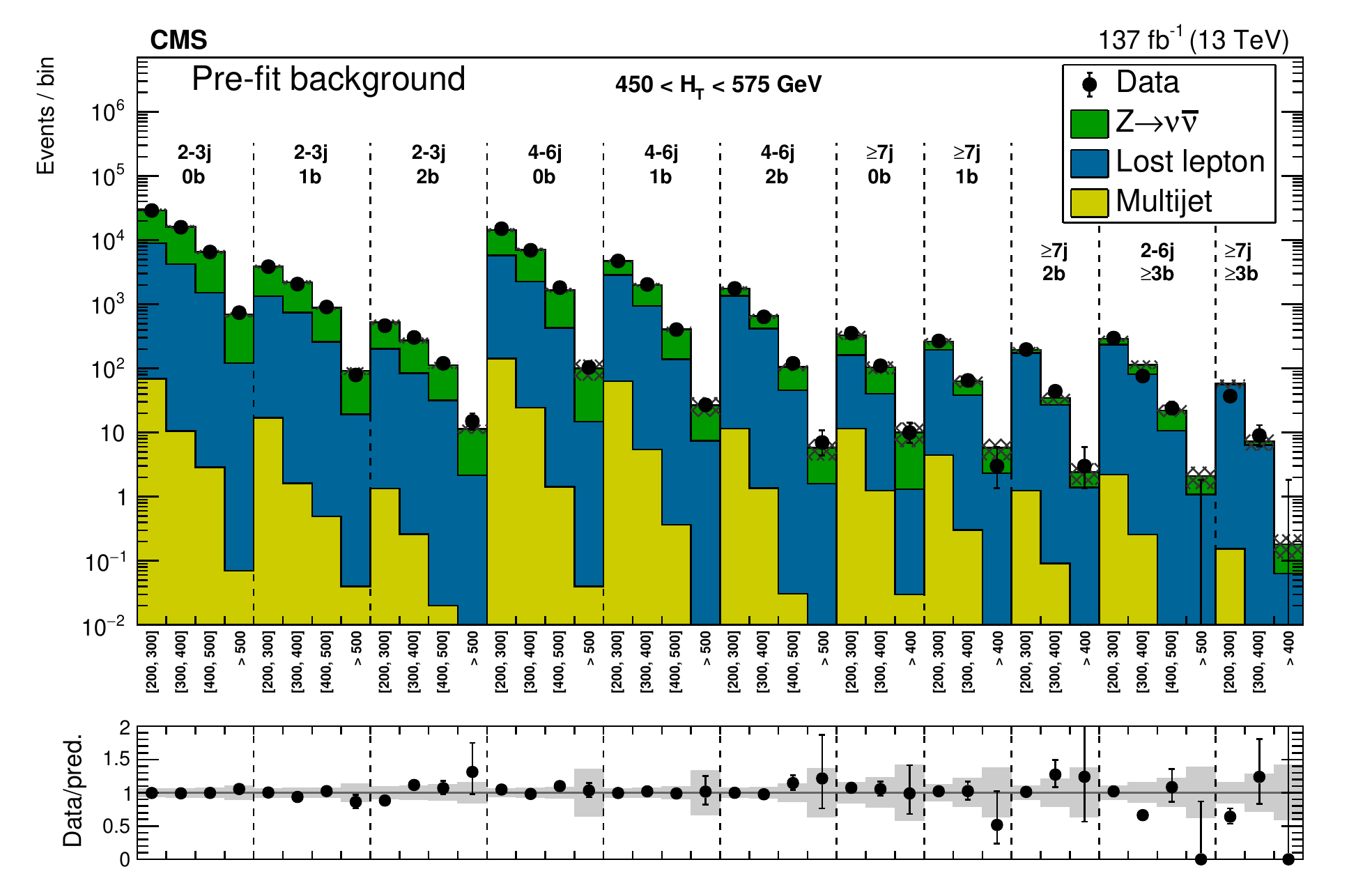}\\
    \caption{(Upper) Comparison of the estimated background and observed data events in each signal bin in the very-low-\HT region.
      The hatched bands represent the full uncertainty in the background estimate.
      The notations j, b indicate \njets, \nbtags labeling.
    (Lower) Same for the low-\HT region.  On the $x$ axis, the \mttwo binning is shown in units of GeV.}
    \label{fig:otherResults1}
\end{figure*}

\begin{figure*}[htb]
  \centering
    \includegraphics[width=0.85\textwidth]{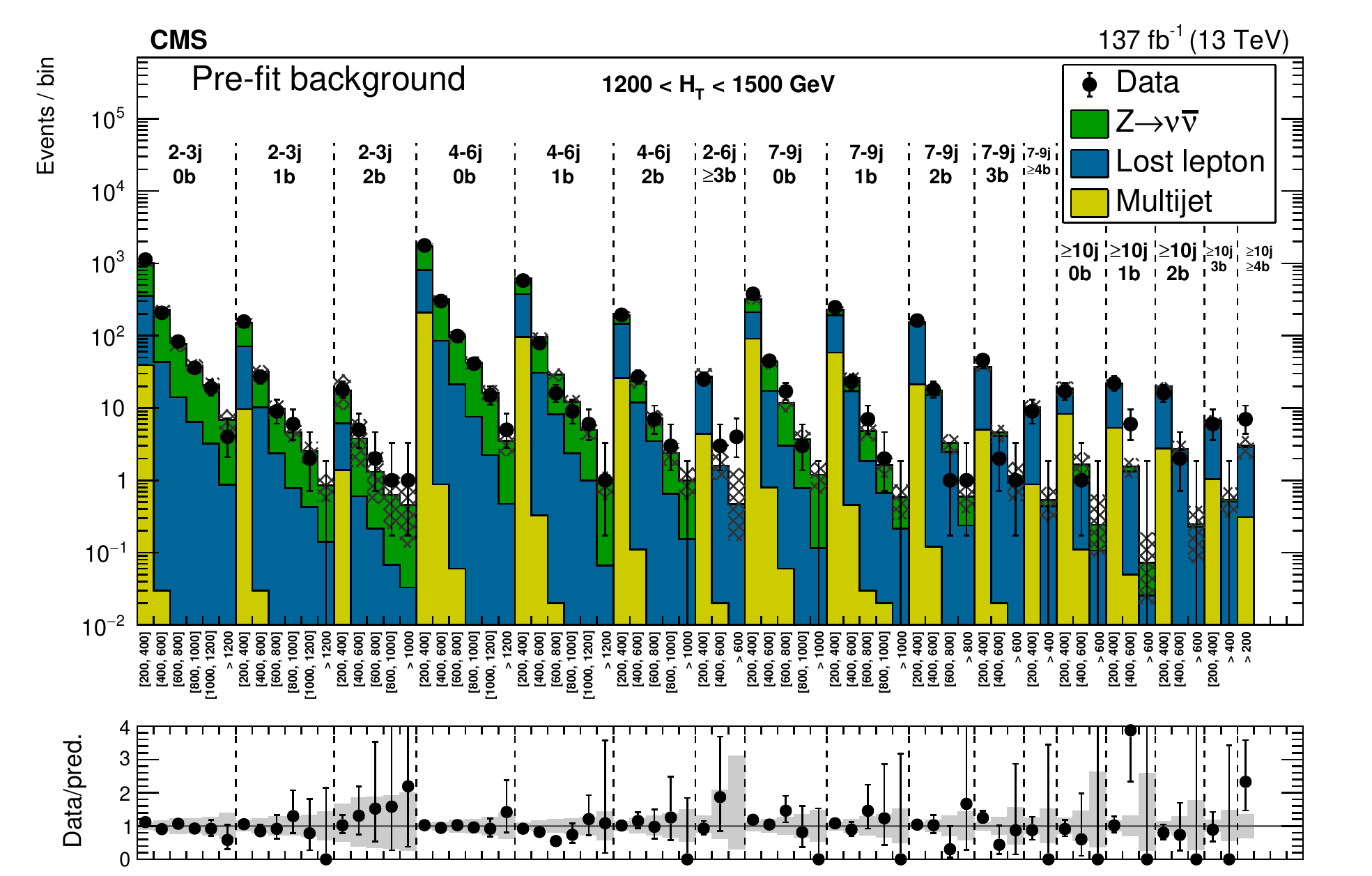}\\
    \includegraphics[width=0.85\textwidth]{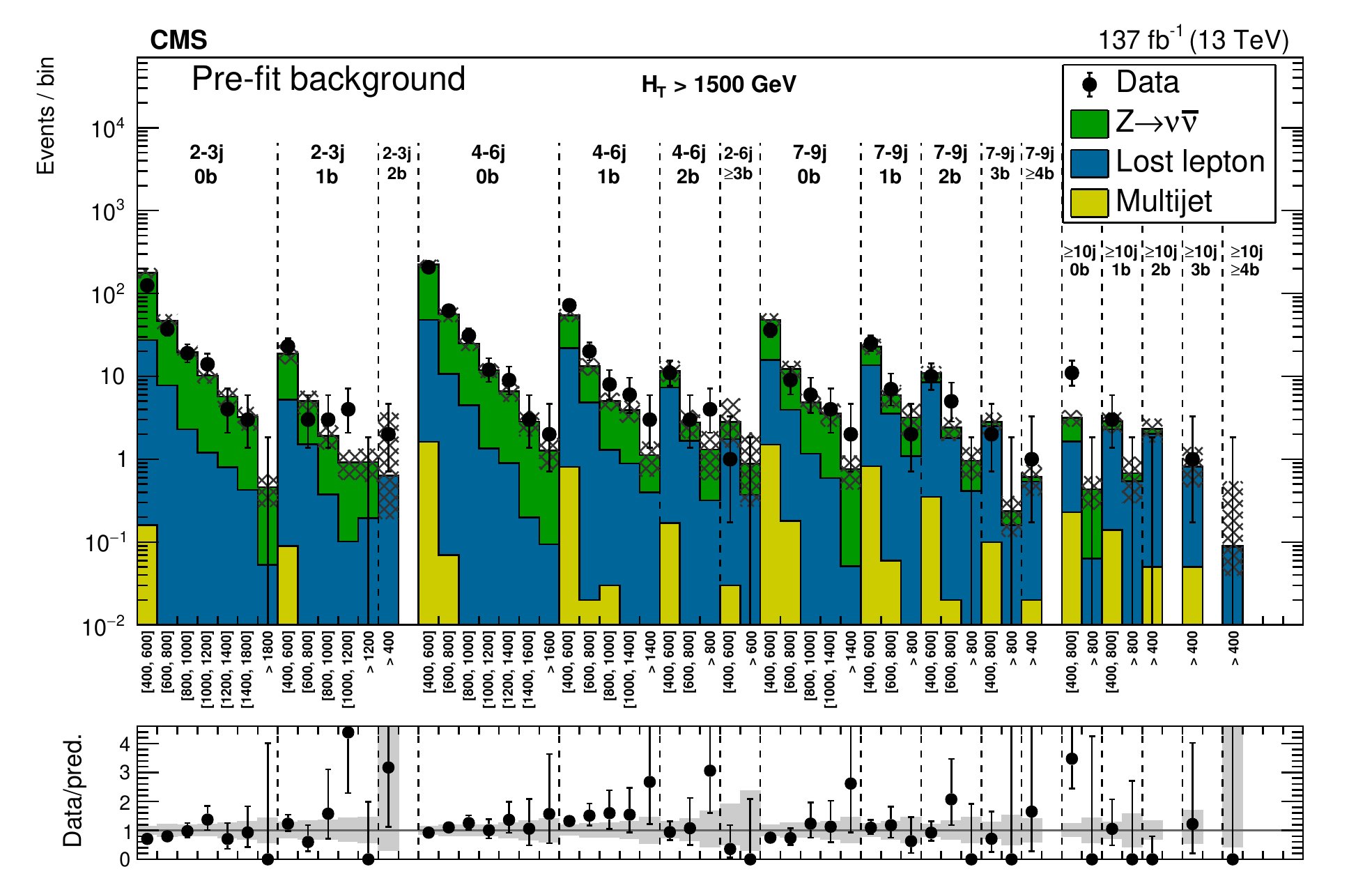}\\
    \caption{(Upper) Comparison of the estimated background and observed data events in each signal bin in the high-\HT region.
      The hatched bands represent the full uncertainty in the background estimate.
    The notations j, b indicate \njets, \nbtags labeling.
    (Lower) Same for the extreme-\HT region. On the $x$ axis, the \mttwo binning is shown in units of GeV.}
    \label{fig:otherResults2}
\end{figure*}

Figures~\ref{fig:otherResults1}--\ref{fig:otherResults2} show
the background estimates and observed data yields in the regions $250 < \HT <450$, $450 < \HT <575$, $1200 < \HT <1500$, and $\HT >1500$\GeV.
Each bin corresponds to a single \mttwo bin, and vertical lines identify (\HT, \njets, \nbtags) topological regions.

\cleardoublepage \section{The CMS Collaboration \label{app:collab}}\begin{sloppypar}\hyphenpenalty=5000\widowpenalty=500\clubpenalty=5000\vskip\cmsinstskip
\textbf{Yerevan Physics Institute, Yerevan, Armenia}\\*[0pt]
A.M.~Sirunyan$^{\textrm{\dag}}$, A.~Tumasyan
\vskip\cmsinstskip
\textbf{Institut f\"{u}r Hochenergiephysik, Wien, Austria}\\*[0pt]
W.~Adam, F.~Ambrogi, T.~Bergauer, J.~Brandstetter, M.~Dragicevic, J.~Er\"{o}, A.~Escalante~Del~Valle, M.~Flechl, R.~Fr\"{u}hwirth\cmsAuthorMark{1}, M.~Jeitler\cmsAuthorMark{1}, N.~Krammer, I.~Kr\"{a}tschmer, D.~Liko, T.~Madlener, I.~Mikulec, N.~Rad, J.~Schieck\cmsAuthorMark{1}, R.~Sch\"{o}fbeck, M.~Spanring, D.~Spitzbart, W.~Waltenberger, C.-E.~Wulz\cmsAuthorMark{1}, M.~Zarucki
% ...
\vskip\cmsinstskip
\dag: Deceased\\
1:  Also at Vienna University of Technology, Vienna, Austria\\
% ...
\end{sloppypar}
\end{document}